\documentclass[aip, jcp,11pt,reprint]{revtex4-1}

\usepackage{graphicx}   
\usepackage{dcolumn}    
\usepackage{bm}         
\usepackage{amsmath}         
\usepackage[T1]{fontenc}       
\usepackage{color,soul}
\usepackage{multirow}

\newcommand{\eqn}[1]{Eq.~(\ref{#1})}
\newcommand{\eqnn}[2]{Eqs.~(\ref{#1}) and (\ref{#2})}
\newcommand{\eqnnn}[3]{Eqs.~(\ref{#1}), (\ref{#2}) and (\ref{#3})}

\usepackage{bm}         
\usepackage{color,soul}

\begin{document}

\title{Tunneling splittings of vibrationally excited states
using general instanton paths}


\author{Mihael Erakovi\'{c}}
\affiliation{Department of Physical Chemistry, Ru{\dj}er Bo\v{s}kovi\'{c} Institute,
Bijeni\v{c}ka Cesta 54, 10000 Zagreb, Croatia}

\author{Marko T.~Cvita{\v s}}
\email[Author to whom correspondence should be addressed: ]{mcvitas@irb.hr}
\affiliation{Department of Physical Chemistry, Ru{\dj}er Bo\v{s}kovi\'{c} Institute,
Bijeni\v{c}ka Cesta 54, 10000 Zagreb, Croatia}

\date{\today}

\begin{abstract}
A multidimensional semiclassical method for calculating tunneling splittings
in vibrationally excited states of molecules using Cartesian coordinates is
developed. It is an extension of the theory by Mil'nikov and Nakamura
[{\em J.~Chem.~Phys.}~{\bf 122}, 124311 (2005)] to asymmetric paths that are
necessary for calculating tunneling splitting patterns in multi-well systems,
such as water clusters. Additionally, new terms are introduced in the description
of the semiclassical wavefunction that drastically improve the splitting estimates
for certain systems. The method is based on the instanton theory and builds 
the semiclassical wavefunction of the vibrationally excited states from
the ground-state instanton wavefunction along the minimum action path and
its harmonic neighborhood.
The splittings of excited states are thus obtained at a negligible
added numerical effort. The cost is concentrated, as for the ground-state
splittings, in the instanton path optimization and the hessian evaluation along
the path. The method can thus be applied without modification to many mid-sized
molecules in full dimensionality and in combination with on-the-fly evaluation
of electronic potentials. The tests were performed on several model potentials
and on the water dimer.
\\
\\
The following article has been submitted to Journal of Chemical Physics. After it is published, it will be found at https://aip.scitation.org/journal/jcp
\end{abstract}

\maketitle

\section{INTRODUCTION}
Tunneling splittings of molecular energy levels are spectroscopic
signatures of rearrangements that take place between degenerate
symmetric wells via tunneling motion
\cite{BellBook,Coudert1988dimer,Walsh1996rearrangements}. These splittings can
be detected in high-precision spectroscopic measurements
\cite{Xu1997,Keutsch2001water} and carry information about the molecular
structure and dynamics along the accessible tunneling paths
\cite{BellBook,Liu1996pentamer}.
Dynamical theories, in combination with potential
energy surfaces (PES) or first principles electronic structure calculations,
aim to reach an agreement with the measurements and provide a physical
interpretation \cite{BellBook,Waterclust}.

Computational studies of tunneling splittings concentrated initially on
the symmetric tunneling systems. Proton transfer in
malonaldehyde \cite{Hammer2009malonaldehyde}, collective migration of
hydrogen atoms in ammonia \cite{Fabri2019} or concerted monomer motion in
the HF dimer \cite{Felker2019}
are some examples of extensively studied systems.
More recently, the splitting patterns in water clusters \cite{Waterclust}
have also come into focus,
motivated by the development of a universal water model that is capable of
predicting properties of liquid water from first principles
\cite{Ceriotti2016water,Wang2011water,Reddy2016MBpol}.
Water clusters are multi-well systems and exhibit multiple tunneling pathways.
These tunneling paths are often asymmetric, whereby tunneling atoms take on
different roles in the minima they connect \cite{Walsh1996rearrangements}.

The splittings vary over many orders of magnitude even in a single system.
In water dimer, for instance, they vary over three orders of magnitude \cite{water}
depending on which of the five tunneling pathways is taken, all of which reflect
on the appearance of the splitting pattern in the spectrum.
Likewise, the experiments on water trimer \cite{Keutsch2001} and pentamer
\cite{Cole2017pentamer} show that the splittings of vibrationally excited states
differ by up to three orders of magnitude in comparison to the ground-state
splittings, depending on which normal mode is excited.
The interplay of different rearrangement pathways can lead to an increase in
the width of a vibrational manifold and a reduction in another
\cite{water,Cvitas2019}, as contributions from different pathways enter
the splitting pattern with the same or opposite signs, respectively.
Qualitatively different tunneling splitting patterns in water hexamer spectrum
distinguish the prism and cage structures \cite{Mhin1994hexamer,Perez2012hexamer}
of almost equal energy.
The contributions of different tunneling pathways can be disentangled,
by computation, to reveal the experimental evidence of unexpected mechanisms,
such as the simultaneous double hydrogen-bond breaking \cite{hexamerprism} in
the water hexamer prism.
The investigations of tunneling splitting patterns thus provide a sensitive test
of both the dynamical theories and the potentials at geometries along
which the hydrogen bonds rearrange.

Tunneling splittings can be determined by solving the Schr\"{o}dinger equation.
Variational methods have been used to determine the tunneling splittings in, 
e.g., HF dimer \cite{Felker2019},
ammonia \cite{Fabri2019,Leonard2002,Neff2014},
vinyl radical \cite{Smydke2019}, malonaldehyde \cite{Wu2016}
and water dimer \cite{Leforestier2012,Wang2018}, using time-independent methods, and, e.g.,
malonaldehyde
\cite{Schroeder2011malonaldehyde,Schroder2014,Hammer2009malonaldehyde,Hammer2011malonaldehyde},
using time-dependent methods. Both, ground- and excited-state
splittings are obtained in this way, however, the cost of these methods scales
prohibitively with the basis set size and a different approach is needed for
larger systems. Diffusion Monte Carlo in combination with the projection operator techniques
has been used to calculate tunneling splittings in water trimer \cite{Blume2000} and
malonaldehyde \cite{Viel2007,Wang2008malonaldehydePES}.
The recently-developed path-integral molecular dynamics method has been used
to obtain the splittings in water trimer and hexamer \cite{Vaillant2019pimd}
in full dimensionality.
However, the tunneling splittings of vibrationally excited states, which are the
the topic of our investigations here, cannot be obtained using these approaches.
The remaining options include resorting to dynamical approximations \cite{Nesbitt2008,Qu2016},
reduced-dimensionality approaches \cite{Althorpe1995dimer,Matanovic2008,Kamarchik2009reduced}
or semiclassical methods \cite{Sewell1995malonaldehyde,Tautermann2002semiclassical,Ceotto2012instanton,Burd2020,Makri1989}.

The development in this paper belongs to the class of semiclassical methods
based on the instanton theory \cite{Uses_of_Instantons,ABCofInstantons,Miller1975semiclassical}.
In the standard instanton formulation \cite{Benderskii}, tunneling splitting
is calculated from the zero-temperature limit of the quantum partition function
in the path-integral formalism.
The dominant contribution to the partition function comes from
the minimum action path (MAP) that connects the symmetry-related minima.
The contribution from all other paths is estimated analytically using
the parameters in a harmonic expansion of the potential
in the directions perpendicular to the MAP.
Instanton theories of tunneling splittings come in several variants.
Some approaches use approximate MAPs \cite{Siebrand1999AIM,Smedarchina1995AIM},
determined from the stationary points on the PES, and
approximate hamiltonians \cite{Smedarchina2012rainbow,Benderskii1997excited},
in which analytic expressions for vibrational couplings are fitted to
the PES. The present contribution belongs to the category that is
based on the numerically exact MAPs.
Mil'nikov and Nakamura \cite{Milnikov2001,Milnikov2005} use the exact MAP
and Hessians along the MAP to obtain splittings via the integration of
Jacobi fields (henceforth reffered to as the JFI method). They employ internal
coordinates in their treatment in order to separate the overall rotational motion.
Ring-polymer instanton (RPI) method \cite{tunnel,water} likewise uses
the numerically exact MAP and evalulates the splitting from
the eigenvalues of the discretized functional determinant of
the action Hessian. This approach is therefore computationally more
demanding than the JFI method \cite{Milnikov2001} and recovering the rotational
dependence of the splittings, when it is significant, becomes
elaborate \cite{Vaillant2018rotation}.
Its advantage is that it can be applied without modification to any molecule of
interest, as it works in Cartesian coordinates, and it can readily be applied
to systems that exhibit asymmetric MAPs.
The RPI method featured prominently in the recent calculations of
tunneling splitting patterns in water clusters. It was used to obtain
the ground-state tunneling splitting pattern and reveal machanisms responsible
for its formation in asymmetric systems such as the water dimer,
trimer \cite{water}, hexamer \cite{hexamerprism}
and octamer \cite{octamer} in full dimensionality.

Standard instanton approaches for calculating tunneling splittings
suffer from the same drawback as the Monte-Carlo and path-integral based method
mentioned above in that they cannot provide the splittings of vibrationally excited
states from the outset.
It is well-known though that the instanton expression for the ground-state
tunneling splitting can be obtained using a variant of the WKB theory \cite{Garg2000}
and Herring formula \cite{Herring1962,Landau+Lifshitz}.
This link thus provides a consistent route for calculating tunneling splittings of
vibrationally excited states \cite{Milnikov2001,Milnikov2005},
where this paper aims to contribute. In fact, the semiclassical methods based on
the wavefunction along the classical trajectory that connects the minima on
the inverted potential energy surface (PES), i.e., along the MAP, are regularly referred to
as the instanton methods in literature \cite{Milnikov2005,Benderskii1997two,Benderskii2000}.
Tunneling splittings of vibrationally excited states have been
obtained using the related methods in symmetric systems such as
malonaldehyde \cite{Benderskii2000}, tropolone \cite{Smedarchina1996},
9-hydeoxyphenalenone \cite{Fernandez-Ramos1998},
HO$_2$ \cite{Milnikov2005}, formic acid dimer \cite{Milnikov2005formic} and
the vinyl radical \cite{Milnikov2006}.

In our recent work \cite{Erakovic2020}, we generalized the JFI approach
of Mil'nikov and Nakamura \cite{Milnikov2001} to obtain the ground-state
tunneling splittings for asymmetric paths in Cartesian coordinates.
We obtained an almost perfect agreement between the JFI and RPI splittings
\cite{Erakovic2020} for systems in which rotations do not couple
strongly to the internal degrees of freedom, like water trimer or
malonaldehyde. The development enabled us to treat large asymmetric systems
that exhibit slow motion of a heavy-atom skeleton, such as
the water pentamer \cite{Cvitas2019}, in full dimensionality. We were able
to calculate the 320-level ground-state splitting pattern of the pentamer,
including the state symmetries, and to identify rearrangement motions
responsible for its formation, in a treatment which would become extremely
cumbersome in the RPI approach due to the large imaginary time periods
involved.

Motivated by the effectiveness of our JFI approach, the present work aims
to derive the tunneling splittings of vibrationally excited states for
general, symmetric and asymmetric paths, in a consistent approach.
This is accomplished by a WKB construction of wavefunction that reproduces
our JFI result in the ground state. In essence, our approach below follows
the work of Mil'nikov and Nakamura \cite{Milnikov2005} in which they extend
their ground-state instanton theory of Ref.~\onlinecite{Milnikov2001} to
treat the low-lying vibrationally excited states.
Distinctly, in our approach we can readily treat asymmetric paths, that
are regularly encountered in the studies of clusters, and we again work
in Cartesian coordinates in order to make our approach general.
Unlike Ref.~\onlinecite{Milnikov2005}, we treat the `longitudinal' modes,
that are parallel to the MAP at minima, and `transversal' modes, that
are perpendicular to the MAP at minima, on an equal footing.
We achieve this by using a different form of the  matching wavefunction
near minima, which allows for a displacement of the wavefunction node
away from the MAP.
In particular, this means that we can treat the asymmetric paths in which
the excited mode is the longitudinal mode at one minimum and is
a transversal mode near the other end of the MAP. 
The straightforward generalization of Ref.~\onlinecite{Milnikov2005} to
asymmetric paths would give a zero splitting in that case.
The theory thus includes newly added terms which for certain cases
dramatically improve the splitting estimates even in symmetric systems.
It is applicable to low vibrationally excited states.

Instanton method evaluates the splittings with a modest number of
potential evaluations (on the order of a thousand) in comparison with
the exact methods \cite{Cvitas2016instanton,Cvitas2018instanton}.
This means that the computations can be performed on larger systems
or using more accurate electronic potentials.
In certain circumstances, it can probably provide the best possible
splittings in a compromise between the accuracy of the dynamical theory
and the level of electronic structure theory that the dynamical treatment
allows. Numerical effort is concentrated in the MAP optimization and
the Hessian evaluation along the MAP \cite{Cvitas2016instanton,Cvitas2018instanton}.
Since the calculations of splittings in vibrationally excited states do
not require any additional information about the molecular system,
they too enjoy the same advantages over the exact methods.

The paper is organized as follows. In Section II, we use a semiclassical
expansion to approximate the wavefunction about the MAP. The wavefunctions
that start from the `left' and from the `right' symmetry-connected minima
along the MAP are constructed and used in Herring formula at the dividing
surface to obtain the ground-state tunneling splitting, which is identical
in form to the JFI instanton expression from our previous work \cite{Erakovic2020}.
The derivation follows Ref.~\onlinecite{Milnikov2001}, but does not assume
the mirror symmetry of the potential along the MAP.
We prove explicitly that the expression for the splitting does not depend
on the position of the connection point between the left- and right-localized
wavefunctions along the MAP.
Section II thus lays the groundwork for constructing the wavefunctions
of the excited states in Section III. Section III follows the work of
Ref.~\onlinecite{Milnikov2005}, but arrives at a different expression for
the tunneling splittings of vibrationally excited states. As stated above,
our formulation treats longitudinal and transversal excitations in
a unified approach. In certain cases, as the numerical exercises on symmetric
and asymmetric model potentials in Section IV show, the contribution from
the newly added terms can dominate the splittings.
The deuterated water dimer provides a real-life test system that exhibits
asymmetric paths, including the path featuring the longitudinal-transversal
excitation mode and the vibrational modes that do not line up either in
either parallel or perpendicular direction with respect to the MAP near minima. 
The importance of different terms in the semiclassical expansion is discussed
in terms of the accuracy improvements that they bring to the splittings
and the stability with regards to the position of the dividing surface.
Conclusions and outlook are given in Section V. Atomic units $(\hbar=1)$
are used throughout unless indicated otherwise.

\section{GROUND-STATE TUNNELING SPLITTING}
Tunneling splittings in molecular systems with multiple symmetry-related
minima  can be expressed as the eigenvalues of a tunneling matrix \cite{water}
in which rows and columns are numbered by the indices of the minima, using group
theoretic arguments. The tunneling matrix element $h$ connecting two minima,
termed L and R for convenience, is the transition amplitude
between the degenerate states $\phi^{({\rm L})}$ and $\phi^{({\rm R})}$,
localized in their respective wells, that neglect the presence of
tunneling motion.
The tunneling splitting of the isolated double-well system connecting
minima L and R is thus $\Delta=-2h$, the difference between the tunneling matrix
eigenvalues. The tunneling matrix eigenvectors are comprised of the coefficients
of the energy eigenstates in the $\phi^{({\rm L/R})}$ basis. For a double-well system,
they form the symmetric and antisymmetric linear combinations of
$\phi^{({\rm L})}$ and $\phi^{({\rm R})}$.

In our previous work \cite{Erakovic2020}, we derived the tunneling matrix 
element $h$, or equivalently the tunneling splitting $\Delta$, using 
the JFI theory. The splitting is dominated by the Euclidean action of
the MAP, while the contributions from all other paths in the harmonic neighborhood
of the MAP are collected into the fluctuation prefactor. The fluctuation
prefactor is then evaluated via integration of Jacobi fields \cite{Kleinert,Milnikov2001}.
We now proceed along the lines of Refs.~\onlinecite{Garg2000,Benderskii1995excited,Milnikov2005}
to derive an identical expression using the semiclassical WKB approach to construct
the localized states $\phi^{({\rm L/R})}$.


Whenever the energy eigenstates are well approximated by the symmetric and
antisymmetric combinations of the localized state functions, $\phi^{({\rm L/R})}$,
the tunneling splitting can be calculated using Herring formula
\cite{Herring1962,Landau+Lifshitz},
\begin{equation}
\Delta = \frac{\int \left( \phi^{(\rm R)} \frac{\partial}{\partial S} \phi^{(\rm L)}-\phi^{(\rm R)} \frac{\partial}{\partial S} \phi^{(\rm L)} \right) 
\delta(f(\mathbf x)) \rm d \mathbf x}{\int \left| \phi^{(\rm L)} \right|^2 \rm d \mathbf x},
\label{herring}
\end{equation}
where $\mathbf x$ is the molecular geometry in mass-scaled Cartesian coordinates and
$f(\mathbf x)=0$ is an implicit equation of an arbitrary dividing plane, which separates
the two minima. Variable $S$ corresponds to the position on a local normal to the dividing plane.

We now construct the localized states $\phi^{({\rm L/R})}$ in the familiar WKB form as
\begin{equation}
\phi = {\rm e} ^{-\frac{1}{\hbar}(W_0+W_1\hbar)},
\label{WKB_wavefunction}
\end{equation}
where we drop the labels (L/R) from this point onwards as the equations are valid in both
wells. In \eqn{WKB_wavefunction}, $W_0$ satisfies Hamilton-Jacobi equation 
\begin{equation}
\frac{\partial W_0}{\partial x_i}  \frac{\partial W_0}{\partial x_i} = 2V(\mathbf x),
\label{Hamilton-Jacobi}
\end{equation}
where $V(\mathbf x)$ is the PES, and $W_1$ satisfies the transport equation,
\begin{equation}
\frac{\partial W_0}{\partial x_i} \frac{\partial W_1}{\partial x_i} - \frac{1}{2} \frac{\partial^2 W_0}{\partial x_i \partial x_i}+E = 0 .
\label{Transport}
\end{equation}
We note here that $E$ is approximated by the ground-state energy of the quantum harmonic oscillator
and is of the order $\hbar^1$. The whole energy dependence is moved to the transport equation,
\eqn{Transport}, following Ref.~\onlinecite{Milnikov2001,Garg2000}.

Hamilton-Jacobi equation, \eqn{Hamilton-Jacobi}, can be solved using the method of characteristics
that we briefly describe in Appendix A. The characteristics of Hamilton-Jacobi equation are given by
\begin{equation}
\ddot{\mathbf x} (\tau) = \nabla V ({\mathbf x} (\tau)),
\label{characteristic}
\end{equation}
with $\tau$ as parameter. The form of \eqn{characteristic} suggests that
the characteristics represent classical trajectories on the inverted PES
and that $\tau$ represents time.
As shown in Appendix A, these trajectories must have zero energy in order to satisfy
\eqn{Hamilton-Jacobi}. On a characteristic, $W_0$ can be obtained by a simple integration,
\begin{equation}
W_0({\rm x}(\tau_2)) = W_0({\rm x}(\tau_1)) + \int_{\tau_1}^{\tau_2} p_0^2(\tau) {\rm d} \tau,
\label{two_point_W}
\end{equation}
where $p_0 = \sqrt{2V}$ corresponds to the mass-scaled momentum on the classical trajectory.
It is convenient to choose one point to correspond to the minimum of the PES
and define $W_0({\mathbf x}_{\rm min}) = 0$. The reason behind this choice is that in the vicinity
of the minimum, the wavefunction can then be matched to that of the harmonic oscillator,
which will be used later on to determine its norm. With that choice, since the minimum on the PES
is a maximum on the inverted PES, all other points along the characteristic correspond to
time $\tau > \tau_{\rm min}$ and the integral in \eqn{two_point_W} remains positive.
However, by choosing the first point at the minimum, the time to any other point will be
infinite, since it takes infinite time to move away from the minimum with zero energy.
This presents a problem in a numerical implementation, which can conveniently be fixed
by reparametrizing the characteristics using the arc length distance $S$ from the minimum
along the characteristic,
\begin{equation}
\frac{{\rm d} S}{{\rm d} \tau}=\sqrt{\frac{{\rm d} x_i}{{\rm d} \tau} \frac{{\rm d} x_i}{{\rm d} \tau}} = p_0.
\label{arc}
\end{equation}
Using this transformation, \eqn{two_point_W} reduces to
\begin{equation}
W_0({\mathbf x}) = \int_{0}^{S(\mathbf{x})} p_0(S') {\rm d} S'.
\label{W_integral}
\end{equation}
We observe that $W_0$ equals Jacobi action between the minimum and the point $S$ on
the characteristic. The characteristic between the minimum and a point $\mathbf{x}$,
as well as $W_0$, can both be determined by a Jacobi action minimization.
The gradient of $W_0$ is therefore parallel to the characteristic.

In order to describe $W_0$ in the vicinity of a given characteristic, we assume that the Hessian
of the potential, $\mathbf H(S)$, along the characteristic is known.
The equation for the Hessian of $W_0$, $A_{ij}=\frac{\partial^2 W_0}{\partial x_i \partial x_j}$,
along a characteristic is then obtained, by differentiating \eqn{Hamilton-Jacobi} twice, as
\begin{equation}
p_0 \frac{\partial}{\partial S} \mathbf A (S) = \mathbf H (S) - {\mathbf A}^2(S).
\label{hessian}
\end{equation}
Riccatti equation in \eqn{hessian} is identical to the equation that emerges in
the JFI method \cite{Milnikov2001,Erakovic2020} as the equation for
the log-derivative of a Jacobi field. The initial condition for \eqn{hessian} at the minimum,
where $p_0=0$, is $\mathbf{A}_0=\mathbf{H}(0)^{1/2}$. This identification later serves
to match the semiclassical wavefunction $\phi$ in \eqn{WKB_wavefunction} to that of
the harmonic oscillator at the minimum.

We can now expand $W_0$ around the characteristic as
\begin{equation}
W_0(S, \Delta \mathbf x) = \int_{0}^{S} p_0(S') {\rm d} S' + \frac{1}{2} \Delta \mathbf x^{\top} \mathbf A \Delta \mathbf x,
\label{taylor_W}
\end{equation}
where $\{S, \Delta x_i \}$ is a set of local coordinates \cite{Milnikov2005} for an arbitrary
point $\mathbf x$. Coordinate $S$ corresponds to the position of the point ${\mathbf x}_0$ on 
the characteristic which satisfies $(x_i-x_{0 i})p_{0i}=0$. The coordinates $\Delta x_i$ define
an orthogonal shift from ${\mathbf x}_0$ to $\mathbf x$, so that $\Delta x_i = x_i-x_{0i}$.
Jacobian of the transformation is derived in Appendix A. The first term in the expansion is 
missing, since $\nabla W_0$ is tangent to the classical trajectory.
\eqn{taylor_W} serves to describe $W_0$ in the neighborhood of the characteristic without
the need to compute new characteristics.

Transport equation in \eqn{Transport} can be solved on a characteristic by a simple integration
\begin{equation}
W_1(S) = \frac{1}{2} \int_{0}^{S} \frac{{\rm Tr} \left( {\mathbf A}(S')- {\mathbf A}_0 \right)}{p_0} {\rm d} S',
\label{W1_integral}
\end{equation}
where we inserted the energy of harmonic oscillator $E=\frac{1}{2} {\rm Tr} {\mathbf A}_0$
into the expression.
Using \eqnn{taylor_W}{W1_integral},
the localized wavefunctions in \eqn{WKB_wavefunction} take the following forms in
their respective wells,
\begin{align}
\nonumber
\phi^{(\rm L)}(S) = &{\rm e} ^{-\int_{0}^{S} p_0(S') {\rm d} S'-\frac{1}{2} \int_{0}^{S} \frac{{\rm Tr} ( {\mathbf A}^{(\rm L)}(S')- {\mathbf A}^{(\rm L)}_0 )}{p_0} {\rm d} S'} \\
\nonumber
&{\rm e}^{-\frac{1}{2} \Delta \mathbf x^\top {\mathbf A}^{(\rm L)} \Delta \mathbf x} \\
\nonumber
\phi^{(\rm R)}(\tilde{S}) = &{\rm e} ^{-\int_{0}^{\tilde{S}} p_0(\tilde{S}') {\rm d} \tilde{S}'-\frac{1}{2} \int_{0}^{\tilde{S}} \frac{{\rm Tr} ( {\mathbf A}^{(\rm R)}(\tilde{S}')- {\mathbf A}^{(\rm R)}_0 )}{p_0} {\rm d} \tilde{S}'} \\
&{\rm e}^{- \frac{1}{2} \Delta \mathbf x^\top {\mathbf A}^{(\rm R)} \Delta \mathbf x},
\label{ground_state_wavefunction}
\end{align}
where $S$ is the distance from the left minimum along the characteristic,
while $\tilde{S}$ denotes the corresponding distance from the right minimum.
In the harmonic regions near minima, these wavefunctions are matched to that of
the quantum harmonic oscillator, as we describe in Appendix B.
From that identification, we obtain their norm as
\begin{equation}
\int \left| \phi \right|^2 {\rm d} {\mathbf x} = \sqrt{\frac{\pi^N}{{\rm det} {\mathbf A}_0}}.
\label{ground_state_norm}
\end{equation}

Having obtained the localized wavefunctions, \eqnn{ground_state_wavefunction}{ground_state_norm},
we are ready to compute the tunneling splitting via Herring formula in \eqn{herring}.
One could take an arbitrary dividing surface and compute the surface integral in \eqn{herring}
numerically. However, this requires computing the characteristics that connect the minima with
every point at which the integrand is evaluated on the dividing surface.
An economical way to compute the integral is to choose one point on the dividing surface
and use Taylor expansion of $W_0$ around it to evaluate the integrand at other points.
If the dividing surface is chosen to be a hyperplane and the gradient of
$W_0$ taken to be constant, the integral can be computed analytically.
Since the integrand in Herring formula is proportional to the product
$\phi^{(\rm L)}\phi^{(\rm R)}$, the integral will be best approximated if the point
on the dividing surface is chosen so that it maximizes this product.
This is equivalent to the minimization of
\begin{equation}
\int_{0}^{S^{({\rm L})}} p_0^{({\rm L})}(S') {\rm d} S' + \int_{0}^{\tilde{S}^{({\rm R})}} p_0^{({\rm R})}(\tilde{S}') {\rm d} \tilde{S}',
\label{total_W_0}
\end{equation}
which is accomplished when the point lies on the classical trajectory that connects the two minima.
In that case, the characteristics that originate at two minima are smoothly joined at
the connection point $S=S_{\rm cp}$ and $\tilde{S} = S_{\rm tot} - S_{\rm cp}$, where
$S_{\rm tot}$ is the total length of the MAP that connects the two minima. The two joined
characteristics coincide with the instanton trajectory \cite{ABCofInstantons,Milnikov2001}.
The sum of $W^{({\rm L})}_0$ and $W^{({\rm R})}_0$ then becomes the Jacobi action of
the instanton trajectory,
$W^{({\rm L})}_0+W^{({\rm R})}_0=\int_{0}^{S_{\rm tot}} p_0 {\rm d} S$.
The dividing surface is taken to be orthogonal to the trajectory at the connection point and
Herring formula gives the ground-state tunneling splitting as
\begin{align}
\nonumber
&\Delta_0 = \sqrt{\frac{{\rm det} {\mathbf A}_0}{\pi^N}} {\rm e}^{-\int_{0}^{S_{\rm tot}} p_0 {\rm d} S-W^{({\rm L})}_1-W^{({\rm R})}_1} \\
&\int \left( \frac{\partial W^{({\rm L})}_0}{\partial S}-\frac{\partial W^{({\rm R})}_0}{\partial S} \right) 
{\rm e}^{-\Delta \mathbf x^\top \frac{{\mathbf A}^{({\rm L})}+{\mathbf A}^{({\rm R})}}{2} \Delta \mathbf x} \delta (f({\mathbf x})) {\rm d} {\mathbf x},
\label{delta_0_integral}
\end{align}
where $\frac{\partial W^{({\rm R})}_0}{\partial S} = - \frac{\partial W^{({\rm R})}_0}{\partial \tilde{S}}$
evaluates to $p_0$ at the connection point, and is kept constant in the surface integral.

In order to solve the integral in \eqn{delta_0_integral}, we note that the matrix
\begin{equation}
\bar{{\mathbf A}} = \frac{{\mathbf A}^{({\rm L})}+{\mathbf A}^{({\rm R})}}{2}
\label{A_bar}
\end{equation}
possesses a zero eigenvalue, which corresponds to the tangent vector.
This is easily proved by differentiating Hamilton-Jacobi equation, \eqn{Hamilton-Jacobi},
which yields ${\mathbf A}^{(\rm L / R)} {\mathbf p}^{(\rm L / R)}_0 = \nabla V$.
Subtracting these two equations and using the fact that ${\mathbf p}^{(\rm L)}_0 = -{\mathbf p}^{(\rm R)}_0$,
it follows that $({\mathbf A}^{(\rm L)}+{\mathbf A}^{(\rm R)}) {\mathbf p}^{(\rm L)}_0 = 0$.
The eigenvectors of $\bar{{\mathbf A}}$ which correspond to its non-zero eigenvalues $\lambda_i$ then
span the dividing surface. Transforming to the eigenvector basis reduces this integral to
\begin{eqnarray}
\nonumber
\Delta_0 &=& 2 p_0 \sqrt{\frac{{\rm det} {\mathbf A}_0}{\pi^N}} {\rm e}^{-\int_{0}^{S_{\rm tot}} p_0 {\rm d} S-W^{({\rm L})}_1-W^{({\rm R})}_1} \int 
{\rm e}^{-\lambda_i \xi_i^2} {\rm d} {\mathbf \xi} \\
&=& 2 p_0 \sqrt{\frac{{\rm det} {\mathbf A}_0}{\pi {\rm det'} \bar{\mathbf A}}} {\rm e}^{-\int_{0}^{S_{\rm tot}} p_0 {\rm d} S-W^{({\rm L})}_1-W^{({\rm R})}_1},
\label{delta_0}
\end{eqnarray}
where $\rm det'$ denotes the product of non-zero $\lambda_i$'s, and $W_1^{({\rm L/R})}$
at $S=S_{\rm cp}$ are calculated using \eqn{W1_integral}.
The ground-state tunneling splitting formula in \eqn{delta_0}
is identical to the instanton formula, Eq.~(33) in Ref.~\onlinecite{Erakovic2020}.
The splitting in \eqn{delta_0} does not depend on the position of the connection point
on the instanton trajectory. This is evident from the derivation of Ref.~\onlinecite{Erakovic2020},
but the present treatment does not guarantee it and we prove it in Appendix D.

\section{EXCITED-STATE TUNNELING SPLITTING}
The calculation of tunneling splittings in vibrationally excited states is approached in
a consistent manner, following Ref.~\onlinecite{Milnikov2005}. We assume one quantum
of vibrational  excitation in the mode with frequency $\omega_{\rm e}$ and construct
the WKB wavefunctions in \eqn{WKB_wavefunction} by solving the Hamilton-Jacobi
and transport equations,
\eqnn{Hamilton-Jacobi}{Transport}, and finally insert them into Herring formula,
\eqn{herring}, which remains valid for the excited states.

Only the transport equation depends on the energy and is different for the excited
state. We decompose $W_1$ in form
\begin{equation}
W_1=W_1^{(0)}+w,
\label{Excited_W1}
\end{equation}
where $W_1^{(0)}$ is the ground-state function given by \eqn{W1_integral}, and
insert \eqn{Excited_W1} in \eqn{Transport}. We then find that $w$ satisfies
\begin{equation}
\frac{\partial W_0}{\partial x_i} \frac{\partial w}{\partial x_i}+\omega_{\rm e}=0.
\label{w_equation}
\end{equation}
In a crucial difference from Ref.~\onlinecite{Milnikov2005}, we seek the solution
of \eqn{w_equation} along the characteristic in the following form
\begin{equation}
w=-{\rm ln} \left( {\mathbf U}^\top \Delta {\mathbf x} + F \right).
\label{w_form}
\end{equation}
The above form, when used in \eqn{WKB_wavefunction}, allows the matching to a
harmonic oscillator wavefunction in the neighborhood of minima for both,
the longitudinally and transversally excited modes with respect to the MAP,
in a unified approach.
We insert \eqn{w_form} into \eqn{w_equation}, multiply through
with ${\mathbf U}^\top \Delta {\mathbf x} + F$ and equate the terms 
of order $\Delta {\mathbf x}^0$ and $\Delta {\mathbf x}^1$ to obtain
equations for $F$ and ${\mathbf U}$ as
\begin{equation}
p_0 \frac{\rm d}{{\rm d} S} F = \omega_{\rm e} F,
\label{F_equation}
\end{equation}
\begin{equation}
p_0 \frac{\rm d}{{\rm d} S} {\mathbf U} = \omega_{\rm e} {\mathbf U} - {\mathbf A}{\mathbf U} + 
2 \left( {\mathbf U}^\top{\mathbf p}_0 -\omega_{\rm e} F \right) \frac{{\mathbf A}{\mathbf p}_0}{p_0^2}.
\label{U_equation_full}
\end{equation}
\eqn{U_equation_full} can be simplified by noting that, by definition, components of
${\mathbf U}$ equal to
\begin{equation}
U_i=\frac{\partial}{\partial x_i} {\rm e}^{-w} = \frac{\partial}{\partial x_i} F,
\label{definition_U_i}
\end{equation}
where the second equality is due to the fact that the partial derivative is taken
on the characteristic. This means that the projection of ${\mathbf U}$ onto the tangent is
\begin{equation}
{\mathbf U}^\top {\mathbf p}_0 = \frac{\partial F}{\partial x_i} \frac{\partial W_0}{\partial x_i} = p_0 \frac{\rm d}{{\rm d} S} F = \omega_{\rm e} F,
\label{U_projection}
\end{equation}
where \eqn{F_equation} was used. Combining \eqnn{U_projection}{U_equation_full}, reduces
the equation for $\mathbf U$ to 
\begin{equation}
p_0 \frac{\rm d}{{\rm d} S} {\mathbf U} = \omega_{\rm e} {\mathbf U} - {\mathbf A}{\mathbf U}.
\label{U_equation}
\end{equation}
This is the same equation that Mil'nikov and Nakamura \cite{Milnikov2005} obtained
in their treatment of transversal excitations. Here, however, we use it for both,
longitudinal and transversal excitations. As our test calculations below demonstrate,
it is important to propagate both components of ${\mathbf U}$ simultaneously for best
accuracy.

\eqnn{F_equation}{U_equation} have singularities at the minima of PES.
In order to avoid them, we need to start the propagation a small distance $\varepsilon$
away from the minimum along the characteristic. If this distance is sufficiently
small to fall into the harmonic region around the minimum, the initial conditions
at $\varepsilon$ can be taken in form 
\begin{equation}
F(\varepsilon) = {\mathbf U}_0^\top \left( \mathbf{x}_0(\varepsilon) - \mathbf{x}_0(0)\right),
\label{initial_F}
\end{equation}
as justified in Appendix B, and
\begin{equation}
{\mathbf U}(\varepsilon) = {\mathbf U}_0,
\label{initial_U}
\end{equation}
where ${\mathbf U}_0$ is the excited normal mode at the minimum.

Alternatively, we can solve \eqn{U_equation} in the region $[0, \varepsilon]$
using the same procedure that was used for solving \eqn{hessian} in
Refs.~\onlinecite{Milnikov2001,Erakovic2020}.
We expand $p_0$, $\mathbf A$ and $\mathbf U$ around minimum as
\begin{align}
\nonumber
p_0 &= p_0^{(1)} S, \\
\nonumber
{\mathbf A} &= {\mathbf A}_0 + {\mathbf A}_1 S, \\
{\mathbf U} &= \sum_i {\mathbf C}^{(i)} S^i.
\label{Taylor_U}
\end{align}
We then insert \eqn{Taylor_U} into \eqn{U_equation} and equate
the terms of the same order in $S^i$ to obtain the recurrence relation for ${\mathbf C}^{(i)}$,
\begin{align}
\nonumber
{\mathbf A}_0 {\mathbf C}^{(0)} &= \omega_{\rm e} {\mathbf C}^{(0)}, \\
\left( {\mathbf A}_0 + (i p_0^{(1)} - \omega_{\rm e}) {\mathbf I} \right) {\mathbf C}^{(i)} &= - {\mathbf A}_1 {\mathbf C}_{i-1}.
\label{recurrence_U}
\end{align}
Once ${\mathbf U}$ has been determined, $F$ can be obtained from \eqn{U_projection} as
\begin{equation}
F(S)=\int_0^S {\mathbf U}^{\top}(S')\mathbf{t}(S'){\rm d}S'=\frac{{\mathbf U}^\top (S) {\mathbf p}_0(S) }{\omega_{\rm e}},
\label{F_from_U}
\end{equation}
where $\mathbf{t}=\mathbf{p}_0/p_0$ is the tangent vector at instanton trajectory.
In this way, the anharmonicity of the PES near minima is accounted for by ${\mathbf A}_1$.
Having obtained ${\mathbf U}(\varepsilon)$, \eqn{U_equation} is readily solved by a simple integrator,
such as the Runge-Kutta method \cite{NumRep}.

At the dividing plane, the wavefunction of the excited state in \eqn{WKB_wavefunction} takes the form
\begin{align}
\nonumber
\phi^{(\rm L)} = &\left( {\mathbf U}^{(\rm L) \top} \Delta {\mathbf x} + F^{(\rm L)} \right) {\rm e} ^{-\int_{0}^{S} p_0(S') {\rm d} S'} \\
\nonumber
&{\rm e}^{-\frac{1}{2} \int_{0}^{S} \frac{{\rm Tr} ( {\mathbf A}^{(\rm L)}(S')- {\mathbf A}^{(\rm L)}_0 )}{p_0} {\rm d} S'- \frac{1}{2} \Delta \mathbf x^\top {\mathbf A}^{(\rm L)} \Delta \mathbf x}, \\
\nonumber
\phi^{(\rm R)} = &\left( {\mathbf U}^{(\rm R) \top} \Delta {\mathbf x} + F^{(\rm R)} \right) {\rm e} ^{-\int_{0}^{\tilde{S}} p_0(\tilde{S}') {\rm d} \tilde{S}'} \\
&{\rm e}^{-\frac{1}{2} \int_{0}^{\tilde{S}} \frac{{\rm Tr} ( {\mathbf A}^{(\rm R)}(\tilde{S}')- {\mathbf A}^{(\rm R)}_0 )}{p_0} {\rm d} \tilde{S}'- \frac{1}{2} \Delta \mathbf x^\top {\mathbf A}^{(\rm R)} \Delta \mathbf x}.
\label{excited_state_wavefunction}
\end{align}
By matching the above wavefunction to that of the harmonic oscillator at a minimum, one obtains the norm as
\begin{equation}
\int \left| \phi \right|^2 {\rm d} {\mathbf x} = \sqrt{\frac{\pi^N}{{\rm det} {\mathbf A}_0}} \frac{1}{2 \omega_{\rm e}}.
\label{excited_state_norm}
\end{equation}
Wavefunctions in \eqn{excited_state_wavefunction} are then inserted into Herring formula
and the surface integral evaluated in a similar manner to the ground-state case.
This gives the tunneling splitting of vibrationally excited states as
\begin{equation}
\Delta_1 = \Delta_0 (2\omega_{\rm e}) \left( F^{(\rm L)}F^{(\rm R)} + \frac{1}{2}{\mathbf U}^{(\rm L)} \bar{\mathbf A}^{-1} {\mathbf U}^{(\rm R)} \right).
\label{excited_state_splitting}
\end{equation}
Since $\bar{\mathbf A}$ possesses a zero eigenvalue, $\bar{\mathbf A}^{-1}$ in \eqn{excited_state_splitting}
denotes a pseudoinverse of $\bar{\mathbf A}$, defined by
$\bar{\mathbf A} \bar{\mathbf A}^{-1} = \bar{\mathbf A}^{-1} \bar{\mathbf A} = {\mathbf P}$,
where ${\mathbf P} = {\mathbf I} - {\mathbf t}{\mathbf t}^{\top}$
is a projector onto the orthogonal plane. The pseudoinverse has the same eigenvectors as
$\bar{\mathbf A}$, while its nonzero eigenvalues are reciprocals of the eigenvalues
of $\bar{\mathbf A}$.

It turns out, the tunneling splitting formula in \eqn{excited_state_splitting} is dependent on
the position of the connection point at which the dividing surface and
the instanton trajectory cross. This undesirable behavior, which was not present in
the ground-state formula in \eqn{delta_0}, arises from 
the ${\mathbf U}^{(\rm L)} \bar{\mathbf A}^{-1} {\mathbf U}^{(\rm R)}$ term,
as shown in Appendix D. It can further be shown, by a similar analysis, that 
the terms which cause this dependency cancel out if the next order term is
included in the Taylor expansion of $\exp(-w)$,
\begin{equation}
w=-{\rm ln} \left( F + U_i\Delta x_i + \frac{1}{2} Z_{ij} \Delta x_i \Delta x_j \right).
\label{w_quadratic}
\end{equation}
However, the inclusion of $\mathbf Z$ in \eqn{w_quadratic} brings new terms that
are again do depend on the connection point and to eliminate their dependence on
$S_{\rm cp}$, it would be necessary to include higher order terms in
the wavefunction expansion \eqn{WKB_wavefunction}, such as the $W_2$ term.
The root of the problem is that the expansion of $\exp(-w)$ is inconsistent
with the expansion of $W_1$, as it gives rise to terms of all orders
in $\Delta \mathbf{x}$ in the expansion of $w$. Excluding the higher order terms
of $w$ in \eqn{w_quadratic}, on the other hand, would degrade the quality of
matching with the harmonic oscillator near minima.

In fact, any improvement of the accuracy of the WKB wavefunction through
the inclusion of extra terms in $W_{0}$ and $W_1$ necessarily requires
the calculation of higher order derivatives of potential along the path.
Calculation of the tensor of third derivatives of potential along the
path allows us to expand $W_0$ in \eqn{taylor_W} up to the $\Delta \mathbf{x}^3$ term,
$W_1^{(0)}$ in \eqn{W1_integral} up to $\Delta \mathbf{x}^1$, and to include
the $\Delta \mathbf{x}^2$ term in \eqn{w_quadratic}. The tensor of fourth derivatives
of potential allows for the correction of the vibrational energy,
the inclusion of the $\Delta \mathbf{x}^0$ term of $W_2$ and the higher order terms
in $W_0$, $W_1^{(0)}$ and $w$. The calculation of higher order derivatives of
the potential quickly becomes computationally unfeasable for realistic
potential energy functions and, in most cases, does not improve
the results significantly.

In order to study the effect of anharmonicity that originates from
the inclusion of third derivatives of potential on
the tunneling splittings in numerical tests below, we derive
the equation for $\mathbf Z$ along a characteristic in Appendix C.
It turns out that from all terms that can be computed using the
third derivatives of potential, this is the only term that is meaningful
to include in the tunneling splitting formula, \eqn{excited_state_splitting_Z},
below. The inclusion of $\nabla W_1^{(0)}$ does not appreciably
influence the results, whereas the inclusion of the $\Delta \mathbf{x}^3$
term in $W_0$ in \eqn{taylor_W} does not result in convergent
integrals on the dividing surface.

It can be shown, by using the $\mathbf Z$ contribution to the splitting, derived
in  Appendix C, that the connection point is best placed in
the middle of the instanton path for symmetric systems, i.e., at the top
of the barrier, because, at this place, the $\mathbf Z$ contribution
is the smallest. We found no such justification for the placement
of the connection point in asymmetric systems, so the safest place
to set it is at the barrier maximum as well.

Alternatively, we can discard the terms that are responsible for
the connection point dependence of the splittings in order to obtain
an unambiguous formulation.
For this purpose, we decompose the vector $\mathbf U$ into longitudinal
and transversal parts as
\begin{equation}
{\mathbf U} = {\mathbf U}_{\perp} + F' {\mathbf t},
\label{U_separated}
\end{equation}
where ${\mathbf U}_{\perp}$ is the component of ${\mathbf U}$ that is perpendicular
to the path.
Since only the ${\mathbf U}_{\perp}$ components contribute to the splitting
in the ${\mathbf U}^{(\rm L)} \bar{\mathbf A}^{-1} {\mathbf U}^{(\rm R)}$ term
in \eqn{excited_state_splitting}, due to the fact that the tangent vector
is an eigenvector of $\bar{\mathbf A}^{-1}$ with zero eigenvalue, it can be used
instead of the vector ${\mathbf U}$.
We carry out the separation in \eqn{U_separated} at $S=\varepsilon$ and
propagate ${\mathbf U}_{\perp}$ and $F$ independently towards the connection point
from both minima.
It can be shown that ${\mathbf U}_{\perp}$ satisfies the following equation
\begin{equation}
p_0 \frac{{\rm d}}{{\rm d}S}{\mathbf U}_\perp = \omega_{\rm e} {\mathbf U}_\perp - {\mathbf A}{\mathbf U}_\perp
-2 \omega_{\rm e} F \frac{{\rm d}{\mathbf t}}{{\rm d}S}.
\label{U_orthogonal_propagation}
\end{equation}
If we neglect the last term in \eqn{U_orthogonal_propagation}, ${\mathbf U}_\perp$
satisfies the same equation as ${\mathbf U}$. Vector ${\mathbf U}_\perp$ remains
perpendicular to the instanton path \cite{Milnikov2001}, when it is propagated using
\eqn{U_equation}, and, as Appendix D shows, the splitting becomes independent of
the position of the connection point.
Since the neglected term is proportional to the curvature of the instanton path,
it can safely be neglected for paths with small curvatures. For paths with a large curvature,
it turns out in Section IV, it is better to work with the full vector ${\mathbf U}$,
as the deviations in the splittings, when the connection point is moved along the instanton
path, are smaller than the error introduced by the above approximation.

\section{NUMERICAL TESTS}
We now perform tests of the above theory on a two-dimensional (2D) symmetric system, a 2D
asymmetric system and the deuterated water dimer. Each calculation of the tunneling splitting
in a vibrationally excited state is preceded by a calculation of the ground-state
tunneling splitting using the JFI method of Ref.~\onlinecite{Erakovic2020}.
A JFI calculation starts by an action minimization, using the string or quadratic string
method \cite{Cvitas2016instanton,Cvitas2018instanton}, followed by the evaluation of Hessians
along the MAP, and, finally, it ends with the computation of $\mathbf{A}$ by solving
the Riccatti equation in \eqn{hessian} along the MAP. Excited-state calculations
additionally require a propagation of $\mathbf{U}$ along the MAP using \eqn{U_equation}
for each vibrationally excited state of interest.
In our tests below, we also evaluate $\mathbf{Z}$ along the MAP in order to check
the accuracy and convergence of the obtained results.
To accomplish this, we first compute the tensor of third derivatives of
potential along the MAP, we then use it to propagate \eqn{B_propagation},
and, finally, use $\mathbf{B}$, as well as $\mathbf{A}$ and
$\mathbf{U}$, to propagate $\mathbf{Z}$ along the MAP using \eqn{Z_propagation}.
The splittings are evaluated using
\eqnnn{delta_0}{excited_state_splitting}{excited_state_splitting_Z}.

In the tests, we discretized all instanton paths using 600 equally spaced beads
(or points) in mass-scaled Cartesian coordinates and used the string method of
Ref.~\onlinecite{Cvitas2016instanton} for the optimization of MAP.
In the tests on water dimer, the orientations of end beads were adjusted during
optimization by minimizing the distance to the first neighbor bead at every iteration
\cite{Cvitas2016instanton}.
Convergence criterion was taken to be the maximum value of the action gradient
orthogonal to the string as $\max \left\{ S^{\perp}_i \right\} < 10^{-8} {\rm a.u.}$.
A large number of beads and a tight convergence criterion were used to ensure that
the results do not depend on the accuracy of the MAP.
Hessians and third-derivative tensors were computed at all beads
using fourth-order finite difference method with the grid spacing of $10^{-3} {\rm a.u.}$.
In water dimer calculations, we projected out the overall translations and rotations,
as described in Ref.~\onlinecite{Kawatsu2014RPI}. Molecular geometries, potential,
Hessian matrix elements and third derivative tensor elements were all interpolated with
respect to the mass-scaled arc length distance $S$ along the MAP using
natural cubic splines. \eqnnn{hessian}{B_propagation}{Z_propagation} were solved on
the interval $[0, \varepsilon]$ by linearization, as described previously in
Ref.~\onlinecite{Milnikov2001,Erakovic2020} and in Appendix C,
while on the interval $[\varepsilon, S_{\rm cp}]$, they were integrated using
Runge-Kutta method \cite{NumRep} with the fixed step length of
$10^{-3} m_{\rm e}^{1/2} a_0$.
The parameter $\varepsilon$ was taken as $\varepsilon = 0.1 m_{\rm e}^{1/2} a_0$
in all test systems.

The normal modes were calculated at one minimum and obtained at the other minimum
by utilizing the symmetry operation that connects them in order to avoid sign
ambiguity. \eqn{U_equation} was then solved on the interval
$[0, \varepsilon]$ using the recurrence relation, \eqn{recurrence_U}.
Taylor series of $\mathbf{U}$ in \eqn{Taylor_U} was cut when the change in the norm
of ${\mathbf U}(\varepsilon)$ fell below the threshold value of $10^{-12}$.
On the interval $[\varepsilon, S_{\rm cp}]$, we used the exponential propagator to
solve \eqn{U_equation},
\begin{equation}
{\mathbf U}(S+h) = {\rm e}^{(\omega_{\rm e}{\mathbf I}-\mathbf A)\frac{h}{p_0}} {\mathbf U}(S),
\label{exponential_propagator}
\end{equation}
with fixed step length $h=10^{-3} m_{\rm e}^{1/2} a_0$.
$F$ values were computed from the tangent projection of the $\mathbf U$ vector,
using \eqn{U_projection}. That procedure was found to be less sensitive to
the value of $F(\varepsilon)$ than the direct integration of \eqn{F_equation},
in \eqn{analytic_F}.

\subsection{SYMMETRIC DOUBLE-WELL 2D POTENTIAL}
We first test the theory on a model 2D double-well system. We call the system symmetric, since
the potential along the MAP connecting two minima has a left-right mirror symmetry with the maximum
of the potential in the middle of the path. The potential is given by the following equations,
\begin{align}
\nonumber
V({\mathbf x}) &= \frac{V_1V_2}{V_1+V_2}, \\
\nonumber
V_1({\mathbf x}) &= \frac{1}{2} \left( {\mathbf x} - {\mathbf x}^{(1)} \right) ^{\top} {\mathbf U}_1
\begin{pmatrix} \alpha_1^2 & 0 \\ 0 & \alpha_2^2 \end{pmatrix}
{\mathbf U}_1^{\top} \left( {\mathbf x} - {\mathbf x}^{(1)} \right), \\
\nonumber
V_2({\mathbf x}) &= \frac{1}{2} \left( {\mathbf x} - {\mathbf x}^{(2)} \right) ^{\top} {\mathbf U}_2
\begin{pmatrix} \alpha_1^2 & 0 \\ 0 & \alpha_2^2 \end{pmatrix}
{\mathbf U}_2^{\top} \left( {\mathbf x} - {\mathbf x}^{(2)} \right), \\
\nonumber
{\mathbf U}_1 &= \begin{pmatrix} \cos{\theta} & -\sin{\theta} \\ \sin{\theta} & \cos{\theta} \end{pmatrix}, \\
\nonumber
{\mathbf U}_2 &= \begin{pmatrix} -\cos{\theta} & \sin{\theta} \\ \sin{\theta} & \cos{\theta} \end{pmatrix}, \\
{\mathbf x}^{(1,2)} &= \left( 0, \pm \beta \right)^\top,
\label{sym_pot}
\end{align}
where ${\mathbf x}$ are not mass scaled.
It has two minima, located at ${\mathbf x}^{(1,2)}$, with normal modes given by matrices
${\mathbf U}_{1,2}$. The parameters were set to $\beta=2$, $\alpha_1=1.265$, $\alpha_2=2$
and $m=27$. Changing the angle $\theta$ changes the angle between the normal modes
of the two minima, as can be seen in Figure 1.
\begin{figure} [htbp]
\begin{center}
\rotatebox{0}{ \resizebox{8cm}{!}
{\includegraphics[width=8cm]{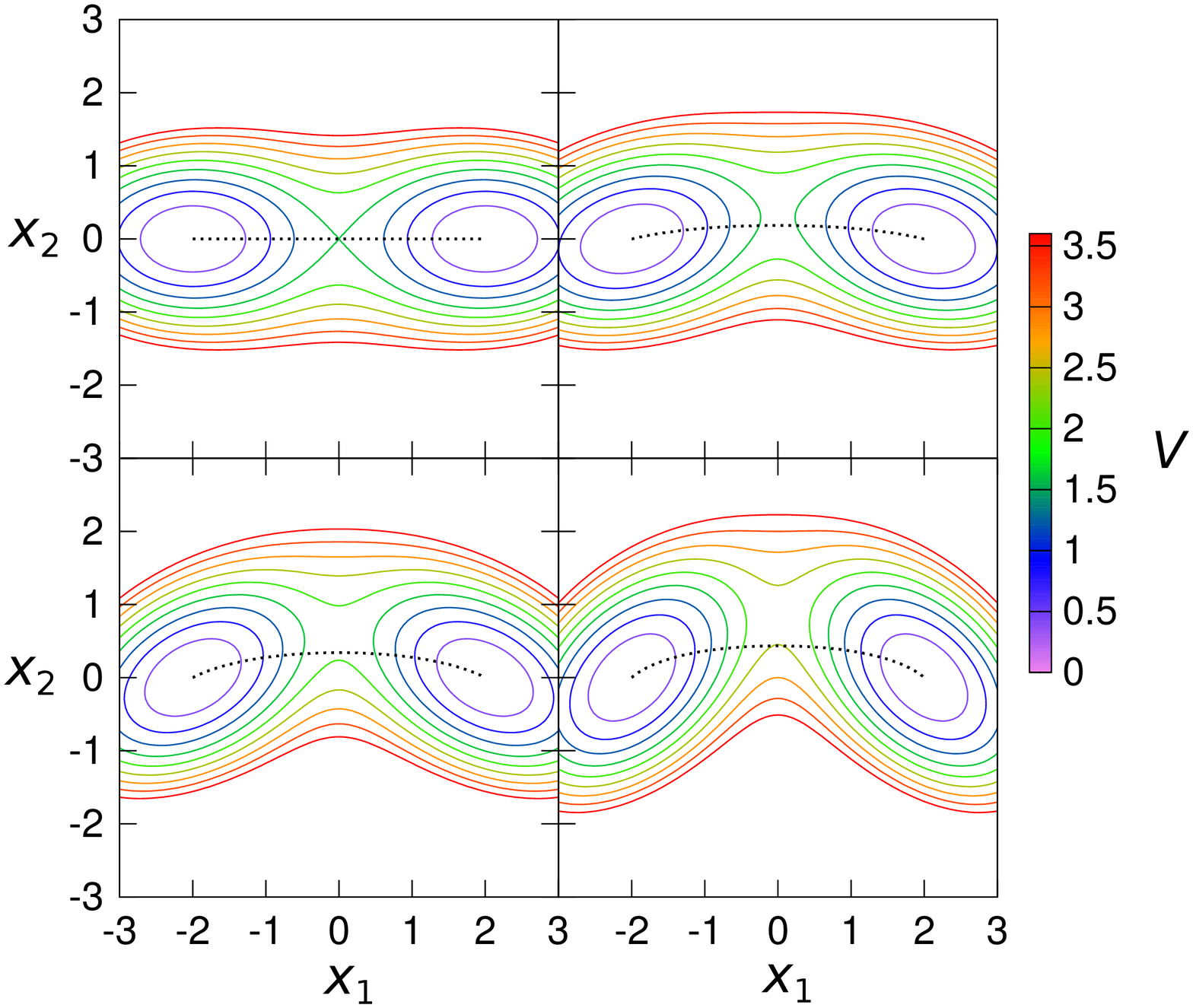}}}
\caption{Potential energy surfaces for model potential
in \eqn{sym_pot} ($\alpha_1=1.265$, $\alpha_2=2$, $\beta=2$)
for angles $\theta$ of, left to right and top to bottom,
0, $\pi/12$, $ \pi/6$ and $\pi/4$. Superposed on each potential
energy surface are the corresponding instanton pathways.}
\label{fig_1}
\end{center}
\end{figure}
With $\theta=0$, the instanton path is a straight line which connects the two minima
and, near minima, the path direction coincides with the lowest normal mode.
As values of $\theta$ increase and normal modes rotate, the instanton path does not
rotate as quickly near minima. Instead, it picks up a non-zero displacement
along the higher normal mode. It turns out that this small displacement can
significantly affect the splitting. Combining \eqnn{F_equation}{initial_F},
we obtain $F$ at the dividing plane in the form
\begin{equation}
F(S_{\rm cp})={\mathbf U}^\top({\mathbf x}(\varepsilon)-{\mathbf x}(0)) {\rm e}^{\omega_{\rm e} \int_\varepsilon^{S_{\rm cp}} \frac{1}{p_0} {\rm d}S'}.
\label{analytic_F}
\end{equation}
The exponential growth of the $F$ term in \eqn{analytic_F} is responsible for this behavior.
Even small displacements along the excited mode near minima can be magnified and result in
an important contribution to the splitting. A useful parameter for quantifying
the displacement near minima is
\begin{equation}
\eta={\mathbf U}_0^{\top}({\mathbf x}(\varepsilon)-{\mathbf x}(0))/\varepsilon,
\label{initial_projection}
\end{equation}
where the division with $\varepsilon$ is made to cancel out the dependence on
the step length $\varepsilon$, where it is observed.
The dependence of the displacement $\eta$ on the angle $\theta$ is given in Table I.
It can be seen that the displacement is predominantly along the lower mode for
all angles $\theta$ in Table I.

\begin{table}[hbtp]
\begin{center}
\begin{tabular}{ccc} \cline{1-3}
$\theta$ & $\eta(1,0)$ & $\eta(0,1)$ \\
\cline{1-3}
$0$ & $1.00000$ $(1.000)$ & $0.00000$ $(0.000)$ \\
\cline{1-3}
$\pi / 12$ & $0.99998$ $(0.999)$ & $0.00664$ $(0.681)$ \\
\cline{1-3}
$\pi / 6$ & $0.99989$ $(0.995)$ & $0.01488$ $(0.919)$ \\
\cline{1-3}
$\pi / 4$ & $0.99962$ $(0.987)$ & $0.02774$ $(0.976)$ \\
\hline
\end{tabular}
\caption{Displacement $\eta$ in \eqn{initial_projection}
at $\varepsilon = 0.1$ for the two normal modes, (1,0) and (0,1),
of the 2D symmetric potential in \eqn{sym_pot}.
Fractional contribution of the $F^{(\rm L)}F^{(\rm R)}$ term to the tunneling
splitting in \eqn{excited_state_splitting}, when the mode is excited, is given in
parentheses.}
\end{center}
\end{table}

Table II shows the tunneling splittings in the ground state and in the first two excited
states, with the lower, $(1,0)$, and the higher mode $(0,1)$ excited with one quantum of
vibration.
Convergence of the excited-state splittings with the addition of
$F$, $\mathbf U$ and $\mathbf Z$ terms in the $\exp (-w)$ expansion
is also shown. The exact quantum-mechanical results are obtained by the diagonalization of
Hamiltonian in the sine DVR basis \cite{Light1985DVR} with grid boundaries at $[-6.0, 6.0]$
in both coordinates and 150 basis functions for each degree of freedom. They are
given in Table II in parentheses for comparison.
It can be seen that the $\mathbf Z$ term contribution is small for all the test
cases. The contribution of $F$ term is dominant for the longitudinal excitation
of the mode $(1,0)$. On the other hand, when the higher mode $(0,1)$ is excited,
the relative contribution of $F$ and $\mathbf U$ terms changes with angle $\theta$.
Displacement $\eta$ suggests that the excitation of $(0,1)$ is in the transversal mode.
Indeed, at $\theta=0$, $F$ term does not contribute and the $\mathbf U$ term determines
the splitting, as in the theory of Ref.~\onlinecite{Milnikov2005}.
But with an increase of $\theta$, the $F$ contribution quickly rises to
account for more than $90 \%$ of the splitting at $\theta=\pi/6$, while
the displacement remains small at $\eta=0.015$.
This demonstrates that it is crucial to include the $F$ term in the expansion
of $\exp(-w)$ even when the excited mode appears to be transversal.
The contribution from a small displacement can exponentially grow and finally
dominate the splitting.

\begin{table}[hbtp]
\begin{center}
\begin{tabular}{cccc} \cline{1-4}
$\theta$ & $\Delta_0$ & $\Delta_1(1,0)$ & $\Delta_1(0,1)$ \\
\cline{1-4}
\multirow{4}{*}{0} &   & $1.830(-8)$ & $0.000$ \\
& $2.630(-10)$ & $1.830(-8)$ & $5.026(-10)$ \\
& $(2.639(-10))$ & $1.838(-8)$ & $5.026(-10)$ \\
&   & $(1.811(-8))$ & $(5.155(-10))$ \\
\cline{1-4}
\multirow{4}{*}{$\pi / 12$} &   & $9.870(-9)$ & $5.492(-10)$ \\
& $1.463(-10)$ & $9.882(-9)$ & $8.066(-10)$ \\
& $(1.472(-10))$ & $9.927(-9)$ & $8.062(-10)$ \\
&   & $(9.858(-9))$ & $(8.089(-10))$ \\
\cline{1-4}
\multirow{4}{*}{$\pi / 6$} &   & $1.563(-9)$ & $4.029(-10)$ \\
& $2.573(-11)$ & $1.571(-9)$ & $4.383(-10)$ \\
& $(2.599(-11))$ & $1.578(-9)$ & $4.390(-10)$ \\
&   & $(1.583(-9))$ & $(4.477(-10))$ \\
\cline{1-4}
\multirow{4}{*}{$\pi / 4$} &   & $7.729(-11)$ & $5.932(-11)$ \\
& $1.606(-12)$ & $7.827(-11)$ & $6.077(-11)$ \\
& $(1.620(-12))$ & $7.863(-11)$ & $6.097(-11)$ \\
&   & $(7.879(-11))$ & $(6.224(-11))$ \\
\hline
\end{tabular}
\caption{Tunneling splittings in the ground and first two vibrationally excited states for
the potential in \eqn{sym_pot} at various angles $\theta$ obtained using instanton theory.
The excited-state splittings are, top to bottom, obtained using the expansion of
$\exp(-w)$ to $F$, $F+U_i \Delta x_i$ and $F+U_i \Delta x_i+\frac{1}{2} Z_{ij} \Delta x_i\Delta x_j$
terms, respectively. The exact quantum-mechanical results are given in parentheses.}
\end{center}
\end{table}

The tunneling splittings are invariant with respect to the position of
the dividing plane when only $F$ terms are considered, in accord with
the analysis of Appendix D. The same is true for the splitting obtained with
the inclusion of the $\mathbf U$ terms at $\theta=0$.
In this case, the instanton path is a straight line and vector $\mathbf U$
remains perpendicular to the path. We can see that in \eqn{U_orthogonal_propagation},
the last term disappears in that case, since the path curvature is zero.
However, we observed in all other cases that the splittings decrease as
the position of the dividing plane changes from
$0.5S_{\rm tot}$ to $0.25S_{\rm tot}$. This decrease varies from
$0.02 \% $ to $0.2 \% $ for the excitation in the lower, longitudinal, mode
and from $3 \% $ to $2 \% $ for the excitation in the higher, transversal, mode.
This variation can be eliminated by using ${\mathbf U}_{\perp}$ instead of ${\mathbf U}$,
in other words, by ignoring the last term in \eqn{U_orthogonal_propagation}.
In this approach, the $F$ term is still included, e.g., by using \eqn{analytic_F}, while
the ${\mathbf U}^{(\rm L)}\bar{{\mathbf A}}^{-1}{\mathbf U}^{(\rm R)}$ contribution
in \eqn{excited_state_splitting} is computed with ${\mathbf U}_{\perp}$.
This approach thus eliminates the dependence of the splitting on
the position of the dividing plane, as discussed in Appendix D.
However, we noticed an increase in all computed splittings by as much
as $8 \%$, which resulted in an overestimation of quantum-mechanical results.
Since the error introduced is larger than the variation of splitting with
the connection point position, using the full expression seems to be 
the preferable option.

\begin{table*}[hbtp]
\begin{center}
\begin{tabular}{cccccc} \cline{1-6}
$m$ & $\Delta_0$ & $\Delta_1(1,0)$ & $V_{\rm eff}^{(1,0)}$ & $\Delta_1(0,1)$ & $V_{\rm eff}^{(0,1)}$ \\
\cline{1-6}
\multirow{4}{*}{$27.0$} &   & $9.870(-9)$ && $5.492(-10)$ \\
& $1.463(-10)$ & $9.882(-9)$ &  & $8.066(-10)$ & \\
& $(1.472(-10))$ & $9.927(-9)$ & $1.273$ & $8.062(-10)$ & $1.428$ \\
&   & $(9.858(-9))$ && $(8.089(-10))$ \\
\cline{1-6}
\multirow{4}{*}{$5.0$} &   & $4.156(-3)$ && $2.312(-4)$ \\
& $1.431(-4)$ & $4.168(-3)$ &  & $4.831(-4)$ & \\
& $(1.435(-4))$ & $4.212(-3)$ & $0.731$ & $4.827(-4)$ & $1.091$ \\
&   & $(3.921(-3))$ && $(4.979(-4))$ \\
\cline{1-6}
\multirow{4}{*}{$1.7$} &   & $0.214$ && $1.191(-2)$ \\
& $1.264(-2)$ & $0.215$ &  & $3.416(-2)$ &  \\
& $(1.231(-2))$ & $0.219$ & $0.051$ & $3.413(-2)$ & $0.668$ \\
&& $(0.146)$ && $(3.080(-2))$ \\
\cline{1-6}
\multirow{4}{*}{$1.5$} &   & $0.297$ && $1.649(-2)$ \\
& $1.865(-2)$ & $0.298$ &  & $4.932(-2)$ &  \\
& $(1.802(-2))$ & $0.304$ & $-0.055$ & $4.927(-2)$ & $0.603$ \\
&& $(0.188)$ && $(4.157(-2))$ \\
\cline{1-6}
\multirow{4}{*}{$1.0$} &   & $0.740$ && $0.041$ \\
& $5.696(-2)$ & $0.745$ &  & $0.141$ &  \\
& $(5.300(-2))$ & $0.763$ & $-0.445$ & $0.141$ & $0.360$ \\
&& $(0.361)$ && $(0.102)$ \\
\hline
\end{tabular}
\caption{Tunneling splittings in the ground ($\Delta_0$)
and first two vibrationally excited states ($\Delta_1$)
for the potential in \eqn{sym_pot} at $\theta=\pi / 12$ and various masses $m$ obtained using
instaton theory.
The excited-state splittings are, top to bottom, obtained using the expansion of $\exp(-w)$
to $F$, $F+U_i \Delta x_i$ and $F+U_i \Delta x_i+\frac{1}{2} Z_{ij} \Delta x_i\Delta x_j$
terms, respectively. The exact quantum-mechanical results are given in parentheses.
For each excitation, the effective barrier heights $V_{\rm eff}$ on the instanton path
are also given.}
\end{center}
\end{table*}

In Table III, we studied the dependence of splittings on the reduction
of the mass of the system.
Convergence of the excited-state splittings with the addition of
$F$, $\mathbf U$ and $\mathbf Z$ terms in the $\exp (-w)$ expansion
is again shown, as well as the exact quantum-mechanical results in parentheses.
The reduction of mass causes an increase in the energy of vibrational states,
which provides an insight into the limits of theory as the energy approaches
the barrier height. In the ground state, the effective barrier height can be computed as
\begin{equation}
V_{\rm eff}^{(0,0)}=V_0+\frac{1}{2}\left(\lambda_2-\omega_1-\omega_2 \right),
\end{equation}
where $V_0$ is the potential energy and $\lambda_2$ is the nonegative eigenvalue
of matrix $\mathbf A$ at the position of the barrier, whereas $\omega_1$ and
$\omega_2$ are vibrational frequencies at the minimum. If lower, longitudinal mode
is excited, the effective barrier is lowered by $\omega_1$ and becomes
\begin{equation}
V_{\rm eff}^{(1,0)}=V_{\rm eff}^{(0,0)}-\omega_1,
\end{equation}
while if the higher, transversal mode is excited, the effective barrier changes as
\begin{equation}
V_{\rm eff}^{(0,1)}=V_{\rm eff}^{(0,0)}-\omega_2+\lambda_2.
\end{equation}
As we reduce the effective barrier height, by varying the mass in Table III,
the instanton method starts to overestimate the tunneling splittings.
When  $V_{\rm eff} \approx 0$, the excited-state splitting is overestimated
by about a factor of 2, similarly to the earlier observations in
the ground state \cite{tunnel}.
This is mainly caused by the overestimation of the state energy
in the harmonic approximation, which is then used in the transport equation.
Furthermore, a significant effect comes from the underestimation of
the norm of the localized wavefunction in the harmonic approximation,
as it extends further on the other side of the barrier.
Therefore, in the case of a 'shallow' splitting or the 'over-the-barrier' splitting,
the estimates obtained using the instanton method should only serve as an upper limit.

\subsection{ASYMMETRIC DOUBLE-WELL 2D POTENTIAL}
We next perform tests on an asymmetric model 2D system. The potential profile along the MAP
connecting any two minima does not have the left-right symmetry and the maximum does not,
in general, lie at the midpoint. The MAP can approach two minima along different normal
modes in an asymmetric system. The asymmetric potential that we use in our tests
is given by the following equations,
\begin{align}
\nonumber
V_1=\frac{1}{2}\alpha_1^2(x_1+\beta)^2+\frac{1}{2}\alpha_2^2(x_2+\beta)^2, \\
\nonumber
V_2=\frac{1}{2}\alpha_2^2(x_1-\beta)^2+\frac{1}{2}\alpha_1^2(x_2+\beta)^2, \\
\nonumber
V_3=\frac{1}{2}\alpha_1^2(x_1-\beta)^2+\frac{1}{2}\alpha_2^2(x_2-\beta)^2, \\
\nonumber
V_4=\frac{1}{2}\alpha_2^2(x_1+\beta)^2+\frac{1}{2}\alpha_1^2(x_2-\beta)^2, \\
V=\frac{V_1V_2V_3V_4}{V_1V_2V_3+V_1V_2V_4+V_1V_3V_4+V_2V_3V_4},
\label{asymmetric_potential}
\end{align}
where $x_i$ are not mass scaled.
The potential parameters in \eqn{asymmetric_potential} are taken as $\beta = 2$,
$\alpha_1 = 1.265$  $\alpha_2 = 2$ and $m = 27$.
The potential has four minima, and possesses a $C_4$ symmetry axis,
as shown in Figure 2. Instanton paths connect the neighboring minima
as indicated in the figure. The 'diagonal' instanton paths have large
actions and are negligible. Energy levels split due to tunneling into
a triplet, in which the middle level is doubly degenerate.
\begin{figure} [htbp]
\begin{center}
\rotatebox{0}{ \resizebox{8cm}{!}
{\includegraphics[width=8cm]{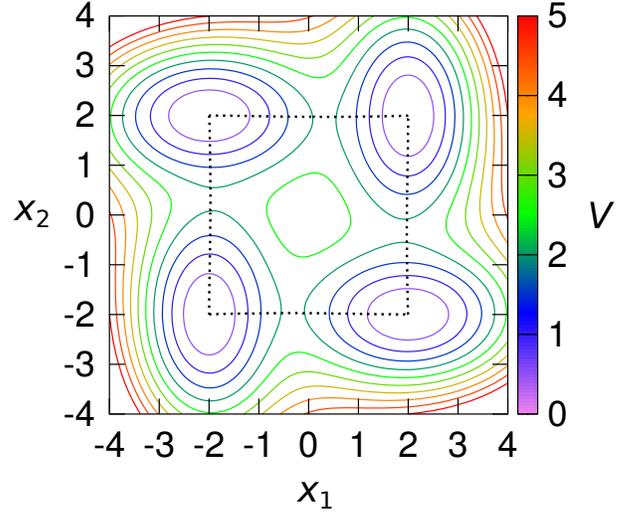}}}
\caption{Potential energy surface for model potential
in \eqn{asymmetric_potential} ($\alpha_1=1.265$, $\alpha_2=2$, $\beta=2$)
Superposed on potential energy surface are the instanton pathways
that are responsible for the formation of the tunneling splitting pattern.}
\label{fig_2}
\end{center}
\end{figure}
The tunneling splitting pattern consists of energy levels $E_1=E_0-\Delta$,
$E_2=E_3=E_0$ and $E_4=E_0+\Delta$,
where $\Delta$ corresponds to the tunneling splitting between the neighboring minima
and $E_0$ is the harmonic energy.
We now label the minimum at $(-\beta, -\beta)$ as 'left' and the minimum
at $(\beta, -\beta)$ as 'right'.
Each instanton path is almost a straight line between two minima, however, because of
the anharmonicity, the path is slightly deflected near minima.
As a result of this deflection, it enters the left minimum along the lower mode, instead
of the higher one, as explained in Appendix B. However, it also possesses a large
displacement $\eta$ in \eqn{initial_projection} along the higher mode.
The higher mode is therefore longitudinal at the left minimum, while the lower mode
is longitudinal near the right minimum. As a result, when either of the modes is excited,
it cannot be described as a longitudinal or a transversal excitation with respect to
the instanton path. It represents the case of longitudinal-transversal excitation, where
the excited mode is longitudinal at one minimum and trasversal to the path at the other
minimum. This case cannot be treated with the method of Ref.~\onlinecite{Milnikov2005}.
The localized wavefunction that corresponds to the longitudinal excitation is of
the form $p_0\exp{(-1/2 \Delta{\mathbf x}^{\top}{\mathbf A}\Delta{\mathbf x})}$, which means
that it is even in the dividing plane. On the other hand, the wavefunction that corresponds
to the transversal excitation is of the form
$({\mathbf U}^{\top}\Delta {\mathbf x})\exp{(-1/2 \Delta{\mathbf x}^{\top}{\mathbf A}\Delta{\mathbf x})}$,
which is odd in the dividing plane. As a result, the surface integral in Herring formula
is odd and identically equal zero.
\begin{figure} [htbp]
\begin{center}
\rotatebox{0}{ \resizebox{8cm}{!}
{\includegraphics[width=8cm]{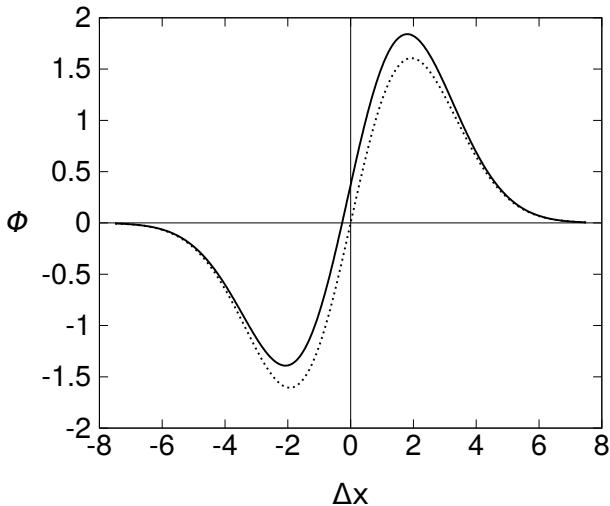}}}
\caption{Comparison of the wavefunctions in the dividing plane
(line) obtained by using ${\mathbf U}_{\perp}$ only (dotted line)
and by using $F+U_i \Delta x_i$ (full line)
in the preexponential factor of the localized wavefunction
in \eqn{excited_state_wavefunction}.}
\label{fig_3}
\end{center}
\end{figure}
It is clear, however, from quantum-mechanical computations that the splitting is not zero,
but can, in fact, even be larger than the splitting in the ground state, as can be seen in
Table IV.

\begin{table}[hbtp]
\begin{center}
\begin{tabular}{ccc} \cline{1-3}
 & $\Delta_1(1,0)$ & $\Delta_1(0,1)$\\
\cline{1-3}
& $1.304(-11)$ & $2.979(-11)$\\
instanton & $1.340(-11)$ & $3.058(-11)$\\
& $1.387(-11)$ & $3.261(-11)$\\
QM & $1.775(-11)$ & $6.531(-11)$\\
\cline{1-3}
$\eta^{(\rm L)}$ & $0.13442$ & $0.99092$\\
$\eta^{(\rm R)}$ & $1.00000$ & $0.00008$\\
\hline
\end{tabular}
\caption{Tunneling splittings in first two vibrationally
excited states ($\Delta_1$) for the potential in \eqn{asymmetric_potential}
obtained using instanton theory.
Displacements, $\eta$ in \eqn{initial_projection}, are given for the left,
$(-\beta, -\beta)$, and the right, $(\beta, -\beta)$, minimum.
QM labels the exact quantum-mechanical results.
The ground-state splitting is $\Delta_0=9.129(-12)$, using the JFI method.
The exact result is $\Delta_0=8.887(-12)$.}
\end{center}
\end{table}

In our treatment, the addition of $F$ term breaks the symmetry of the wavefunction
in the dividing plane, and it moves the node away from the instanton trajectory, while
the maximum of the Gaussian part in \eqn{excited_state_wavefunction} stays on
the trajectory, as shown in Figure 3.
As a result, the integral in Herring formula does not vanish.
Results obtained using our approach are given in Table IV. From the $\eta$ values
in the left minimum, it is clear that in its vicinity, the instanton trajectory rapidly turns
towards the direction of the second (higher) normal mode, while it has to enter the minimum
along the first (lower) mode. As a result of this sharp turn, $F$ value for the left minimum
is not zero and, in the end, gives rise to the non-zero tunneling splitting. Contribution of
the $\mathbf Z$ term in both excited states is quite large compared to its contribution
in the symmetric test case above. This is indicative of the presence of non-negligible
anharmonic effects in this system.
The anharmonicity is also a probable reason for the relatively large
discrepancies between the instanton and the exact quantum-mechanical results (obtained
on the same grid as for the symmetric potential above), where the latter
are $28 \%$ and $100 \%$ higher for the excitation of the first and second vibrational
mode, respectively.
A larger discrepancy in the higher mode could be attributed to its larger energy,
and the larger spread of its wavefunction into the regions away from the instanton path
where anharmonicity is significant.

\subsection{WATER DIMER}
The tunneling splitting pattern of water dimer has been
extensively studied both experimentaly and theoretically
\cite{Coudert1988dimer,Leforestier2012,Wang2018,water},
which makes it a good benchmark system to test our method.
We chose the fully deuterated dimer over the non-deuterated one,
because its vibrational energies are lower. As a consequence,
there are more vibrational excitations which do not exceed
the barrier height, and can be treated with the instanton method.
Analytical potential energy surface MB-pol
\cite{Babin2013MBpol,Babin2014MBpol,Reddy2016MBpol}
was used in all calculations.

\begin{figure} [htbp]
\begin{center}
\rotatebox{0}{ \resizebox{8cm}{!}
{\includegraphics[width=8cm]{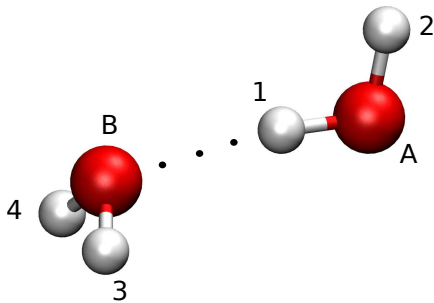}}}
\caption{The minimum energy geometry of the water dimer labeled to represent the reference
version.}
\label{fig_4}
\end{center}
\end{figure}
\begin{table*}[hbtp]
\begin{center}
\begin{tabular}{cccccc} \cline{1-6}
Mode & AT & GI & AI & BT & DE \\
\cline{1-6}
\multirow{2}{*}{1} & $0.99408$ & $0.73315$ & $0.97661$ & $0.00363$ & $0.06834$\\ 
& $0.99413$ & $0.73091$ & $0.97689$ & $0.00205$ & $0.97470$\\
\cline{1-6}
\multirow{2}{*}{2} & $0.08292$ & $0.63492$ & $0.14969$ & $0.00263$ & $0.70957$\\ 
& $0.08307$ & $0.63780$ & $0.14916$ & $0.00070$ & $0.22343$\\
\cline{1-6}
\multirow{2}{*}{3} & $0.07001$ & $0.24363$ & $0.15434$ & $0.99927$ & $0.70112$\\ 
& $0.06920$ & $0.24284$ & $0.15302$ & $0.99927$ & $0.00373$\\
\cline{1-6}
\multirow{2}{*}{4} & $0.00441$ & $0.00276$ & $0.00135$ & $0.03795$ & $0.01621$\\ 
& $0.00449$ & $0.00269$ & $0.00127$ & $0.03792$ & $0.00362$\\
\cline{1-6}
\multirow{2}{*}{5} & $0.00143$ & $0.00163$ & $0.00314$ & $0.00135$ & $0.00032$\\ 
& $0.00145$ & $0.00168$ & $0.00307$ & $0.00122$ & $0.00276$\\
\hline
\end{tabular}
\caption{Left and right displacements $\eta$ in \eqn{initial_projection}
in deuterated water dimer for five instanton pathways
and excitations into lowest five vibrational modes.
Pathways are acceptor tunneling (AT), geared interchange (GI),
antigeared interchange (AI), bifurcation tunneling (BT) and donor exchange (DE).}
\end{center}
\end{table*}

Water dimer, shown labeled in Figure 4, has 8 equivalent symmetry-related and
accessible minima, which correspond to the permutations of hydrogen and
oxygen atoms that do not break the covalent H-O bonds.
Permutations which do break the covalent bonds are considered unfeasable.
These minima are connected by five distinct tunneling rearrangement pathways
\cite{Coudert1988dimer,Watanabe2004dimer,water}.
Acceptor tunneling path (AT) corresponds to the permutation $(34)$.
In the ground state, its effective barrier is relatively low,
$V_{\rm eff}=77$ ${\rm cm}^{-1}$),
so it gives rise to the largest tunneling matrix element. This matrix element
is responsible for the splitting of energy levels into two groups,
whose energy difference is called the acceptor splitting
$\Delta({\rm A}) = 4|h({\rm AT})|$. As seen in Table V, the displacements $\eta$ 
for the AT path lie predominantly along the lowest mode at both minima.
Next contribution to the splitting pattern arises from the geared interchange (GI)
and anti-geared interchange (AI) pathways, which correspond to the (AB)(1324) and
(AB)(14)(23) permutations. These pathways have larger effective barriers in
the ground state, $V_{\rm eff}=188$ ${\rm cm}^{-1}$ and $V_{\rm eff}=227$ ${\rm cm}^{-1}$,
respectively. They cause the energy levels in both groups, formed by
acceptor tunneling, to split into triplets, with the energy width of the lower group
called the lower interchange $\Delta({\rm LI}) = 4|h({\rm GI})+h({\rm AI})|$,
while the upper group energy width is called the upper interchange
$\Delta({\rm UI}) = 4|h({\rm GI})-h({\rm AI})|$.
The AI path is mostly displaced along the lowest mode near minima as well,
but has larger projections onto the second and third mode.
In contrast, the GI path is almost equally displaced along the first and second mode near
minima, while it has to enter the minima along the lowest mode. Finally, the smallest
contribution to the splitting pattern of water dimer arises from
the bifurcation tunneling (BT) and donor exchange (DE) paths, which correspond to
the (12)(34) and (12) permutations, respectively.
These pathways possess the highest effective barriers, $V_{\rm eff}=469$ ${\rm cm}^{-1}$
and $V_{\rm eff}=581$ ${\rm cm}^{-1}$, respectively. They cause the shifts in the energies
of the triplets by the amounts called the lower bifurcation and the upper bifurcation,
$\Delta({\rm LB}) = |h({\rm BT})+4h({\rm DE})|$ and
$\Delta({\rm UB}) = |h({\rm BT})-4h({\rm DE})|$.
Bifurcation tunneling path is displaced mostly along the third mode near minima.
Donor exchange path, on the other hand, is displaced mostly along the lowest mode
near one minimum, while it is displaced mostly along the second and third mode near
the other minimum. Therefore, this path represents a realistic case of
the asymmetric potential which features longitudinal-transversal excitations that
we discussed in the previous subsection on a 2D model potential.

The lowest mode of vibration in the deuterated water dimer corresponds
to donor torsion and has a frequency of $\omega = 84$ ${\rm cm}^{-1}$.
In order to calculate the splitting pattern with the excited donor torsion,
we calculate the matrix elements, $h=-\Delta_1/2$, for all five rearrangement paths.
The AT matrix element, obtained by the instanton method, is 3 times larger
than the experimental value, as seen in Table VI.
Since donor torsion is the longitudinal mode of the AT path and
its excitation frequency is larger than the effective barrier on the path,
this represents a case of over-the-barrier tunneling. The instanton method
is known to overestimate the splittings by a factor of 2$-$3 in such
circumstances \cite{tunnel,Vaillant2019pimd}, as also noted in
the previous subsection.
The sign of the acceptor splitting is found to be opposite to that of
the ground state, indicating that the groups of states associated
with the lower and upper interchange change places. This observation is in
agreement with the experimental measurements \cite{Braly2000}
and the exact quantum-mechanical calculations \cite{Leforestier2012}.

GI and AI matrix elements are found to be in good agreement with
the experimental results \cite{Braly2000} in their absolute values,
but their relative sign appears to be wrong. This results in
the wrong ordering of the LI and UI splittings in magnitude,
as seen in Table VII. We note that the contribution of the $F$ term accounts for 
$86 \%$ and $95 \%$ of the matrix element in \eqn{excited_state_splitting}. 
A large contribution for the AI path is expected, as donor torsion is its
longitudinal mode. However, for the GI path, which lies along a combination
of modes near minima, the contribution of $F$ term is also important.
We presume that the disagreement between the instanton and quantum-mechanical
results of Ref.~\onlinecite{Leforestier2012} is caused by a large
rotation-vibration coupling in the excited mode, which mixes the vibrational
states of $K_a=0$ and $K_a=1$ and is not accounted for in the instanton method.
The values obtained for LI and UI ($0.134$ cm$^{-1}$ and $0.290$ cm$^{-1}$)
are, in fact, in a better agreement with the experimental values \cite{Braly2000}
for $K_a=1$, which are $0.132$ cm$^{-1}$ and $0.257$ cm$^{-1}$, both in
magnitude and in ordering.

Lower and upper bifurcations are underestimated for the first excited
vibrational mode, as can be seen in Table VII.
For the DE path, this represents a longitudinal-transversal excitation,
and it was shown for the model potential above that an underestimate
is expected because of the unaccounted anharmonicities.
However, the difference between the lower and upper bifurcation is not zero,
as it would be in using the theory of Ref.~\onlinecite{Milnikov2005},
and even though it is underestimated, a rough estimate of its value is
obtained. The exact quantum-mechanical calculations
\cite{Leforestier2012} do not report it, probably due to the difficulty in
converging the values with sufficient accuracy.
It is also worth mentioning that the UB and LB change significantly
in the $K_a=1$ rotational state, to $8.906(-4)$ cm$^{-1}$ for UB
and $1.201(-4)$ cm$^{-1}$ for LB.
These values are again in better agreement with those that we computed,
as in the case of the AT path, which provides further indication that
the coupling of the first excited state to rotations plays a significant role.
Finally, the UB and LB are underestimated even in the ground vibrational state,
which suggests the possibility that the BT and DE pathways are poorly described
by the PES, either by too large potential energy barriers,
or by slightly misplaced instanton paths, both of which can have
a drastic effect on the splittings. 

\begin{table*}[hbtp]
\begin{center}
\begin{tabular}{cccccc} \cline{1-6}
Mode & AT & GI & AI & BT & DE \\
\cline{1-6}
GS & $0.766$ & $9.73(-3)$ & $4.88(-4)$ & $1.83(-4)$ & $3.21(-6)$\\
\cline{1-6}
\multirow{6}{*}{1} & $-11.8$ & $-4.58(-2)$ & $1.86(-2)$ & $1.43(-9)$ & $2.95(-6)$\\
& $-11.1$ & $-5.07(-2)$ & $1.83(-2)$ & $-3.96(-5)$ & $1.12(-5)$\\
& $-12.0$ & $-5.30(-2)$ & $1.95(-2)$ & $-3.96(-5)$ & $8.26(-6)$\\
& $(3.953)$ & $(6.643(-2))$ & $(1.561(-2))$ & $(-)$ & $(-)$ \\
& $(3.92)$ & $(6.63(-2))$ & $(1.63(-2))$ & $(-)$ & $(-)$ \\
\cline{1-6}
\multirow{6}{*}{2} & $-0.502$ & $-0.256$ & $4.01(-3)$ & $-6.17(-9)$ & $-5.16(-5)$\\
& $-0.509$ & $-0.254$ & $4.98(-3)$ & $-2.28(-4)$ & $-4.38(-5)$\\
& $-0.457$ & $-0.261$ & $5.18(-3)$ & $-2.28(-4)$ & $-4.78(-5)$\\
& $(0.634)$ & $(0.109)$ & $(1.375(-3))$ & $(-)$ & $(-)$ \\
& $(0.758)$ & $(0.140)$ & $(4.25(-2))$ & $(-)$ & $(-)$ \\\cline{1-6}
\multirow{6}{*}{3} & $1.15$ & $0.147$ & $2.13(-2)$ & $5.47(-3)$ & $3.27(-6)$ \\
& $2.72(-2)$ & $0.141$ & $2.13(-2)$ & $5.42(-3)$ & $2.42(-6)$ \\
& $0.469$ & $0.143$ & $2.21(-2)$ & $5.58(-3)$ & $5.32(-6)$\\
& $(0.442)$ & $(3.033(-2))$ & $(2.427(-3))$ & $(-)$ & $(-)$ \\
& $(0.45)$ & $(2.88(-2))$ & $(1.25(-3))$ & $(-)$ & $(-)$ \\
\cline{1-6}
\multirow{6}{*}{4} & $19.5$ & $2.94(-2)$ & $2.64(-2)$ & $2.17(-3)$ & $-2.43(-4)$ \\
& $8.98$ & $3.20(-2)$ & $2.58(-2)$ & $2.41(-3)$ & $-3.70(-4)$ \\
& $-57.0$ & $0.220$ & $-3.87(-2)$ & $6.45(-2)$ & $4.05(-4)$\\
& $(-)$ & $(-)$ & $(-)$ & $(-)$ & $(-)$ \\
& $(1.23)$ & $(0.173)$ & $(7.75(-2))$ & $(-)$ & $(-)$ \\
\hline
\end{tabular}
\caption{Tunneling matrix elements $-h / {\rm cm}^{-1}$ for different tunneling pathways
in deuterated water dimer $({\rm D}_2{\rm O})_2$ obtained using instanton theory.
Pathways described are acceptor tunneling (AT), geared interchange (GI),
antigeared interchange (AI), bifurcation tunneling (BT) and donor exchange (DE).
The excited-state splittings are, top to bottom, obtained using the expansion of $\exp(-w)$ to
$F$, $F+U_i \Delta x_i$ and $F+U_i \Delta x_i+\frac{1}{2} Z_{ij} \Delta x_i\Delta x_j$
terms, respectively.
The splittings given in parentheses are experimental \cite{Braly2000} (top) and
quantum-mechanical \cite{Leforestier2012} (bottom) results.
(Ground-state (GS) experimental results are from Refs.~\onlinecite{Leforestier2012,Karyakin1993Ddimer}.)}
\end{center}
\end{table*}

\begin{table*}[hbtp]
\begin{center}
\begin{tabular}{cccccc} \cline{1-6}
Mode & A & UI & LI & UB & LB\\
\cline{1-6}
\multirow{3}{*}{GS} & $3.06$ & $3.70(-2)$ & $4.09(-2)$ & $1.70(-4)$ & $1.96(-4)$\\
& $(1.77)$ & $(3.6(-2))$ & $(3.9(-2))$ & $(2.2(-4))$ & $(2.3(-4))$\\
& $(1.78)$ & $(3.6(-2))$ & $(3.8(-2))$ & $(-)$ & $(-)$\\
\cline{1-6}
\multirow{5}{*}{1} & $47.3$ & $0.257$ & $0.109$ & $1.18(-5)$ & $1.18(-5)$\\
& $44.5$ & $0.276$ & $0.129$ & $8.44(-5)$ & $5.15(-6)$\\
& $47.9$ & $0.290$ & $0.134$  & $7.27(-5)$ & $6.57(-6)$\\
& $(15.811)$ & $(0.203)$ & $(0.328)$ & $(8.006(-4))$ & $(1.698(-3))$ \\
& $(15.68)$ & $(0.20)$ & $(0.33)$ & $(-)$ & $(-)$\\
\cline{1-6}
\multirow{5}{*}{2} & $2.01$ & $1.04$ & $1.01$ & $2.06(-4)$ & $2.06(-4)$\\
& $2.04$ & $1.04$ & $1.00$ & $5.31(-5)$ & $4.03(-4)$\\
& $1.83$ & $1.07$ & $1.02$ & $3.67(-5)$ & $4.20(-4)$\\
& $(2.535)$ & $(0.443)$ & $(0.432)$ & $(2.662(-3))$ & $2.635(-3)$\\
& $(3.03)$ & $(0.73)$ & $(0.39)$ & $(-)$ & $(-)$\\
\cline{1-6}
\multirow{5}{*}{3} & $4.60$ & $0.503$ & $0.673$ & $5.46(-3)$ & $5.49(-3)$\\
& $0.109$ & $0.479$ & $0.649$ & $5.41(-3)$ & $5.43(-3)$\\
& $1.88$ & $0.484$ & $0.660$ & $5.56(-3)$ & $5.60(-3)$\\
& $(1.768)$ & $(0.112)$ & $(0.131)$ & $(1.304(-3))$ & $5.174(-3)$\\
& $(1.81)$ & $(0.11)$ & $(0.12)$ & $(-)$ & $(-)$\\
\cline{1-6}
\multirow{5}{*}{4} & $78.1$ & $0.012$ & $0.22$ & $3.14(-3)$ & $1.20(-3)$ \\
& $35.9$ & $0.025$ & $0.231$ & $3.89(-3)$ & $9.35(-4)$ \\
& $228$ & $1.04$ & $0.725$ & $6.29(-2)$ & $6.62(-2)$\\
& $(-)$ & $(-)$ & $(-)$ & $(-)$ & $(-)$\\
& $(4.9)$ & $(0.38)$ & $(1.0)$ & $(-)$ & $(-)$\\
\hline
\end{tabular}
\caption{Acceptor, upper interchange and lower interchange splittings (${\rm cm}^{-1}$) in
deuterated water dimer $({\rm D}_2{\rm O})_2$ obtained using instanton method.
The excited-state splittings are, top to bottom, obtained using the expansion of $\exp(-w)$ to
$F$, $F+U_i \Delta x_i$ and $F+U_i \Delta x_i+\frac{1}{2} Z_{ij} \Delta x_i\Delta x_j$
terms, respectively.
The splittings given in parentheses are experimental \cite{Braly2000} (top) and
quantum-mechanical \cite{Leforestier2012} (bottom) results.
(Ground-state (GS) experimental results are from Refs.~\onlinecite{Leforestier2012,Karyakin1993Ddimer}.)}
\end{center}
\end{table*}

The second mode corresponds to the acceptor twist, with frequency
$\omega = 100$ cm$^{-1}$, while the third mode corresponds to the acceptor wag,
with frequency $\omega = 110$ cm$^{-1}$. However, in quantum-mechanical
calculations \cite{Leforestier2012}, the order of these two motions changes,
and the acceptor wag frequency drops to $82$ cm$^{-1}$, while the acceptor twist
drops to $90$ cm$^{-1}$.
The large deviation of vibrational energies from the harmonic frequencies
is a strong indication of large anharmonic effects in these two vibrational modes.
Furthermore, since their energy difference is very small, it was noticed that
these states interact through a Coriolis perturbation \cite{Braly2000} adding to
the quantitative disagreement with the harmonic analysis.
Nevertheless, the splittings obtained from the second excited mode are in
good agreement with the experimental results. We note that the $F$ term
on the AI path contributes with around $77 \%$ to the matrix element, even though
the displacements near minima along this mode are small.
The overestimation of the GI matrix element can be explained by the fact
that the path has a large projection onto the second mode near minima,
which means that the effective barrier is significantly lowered.
Discrepancy of the AI matrix element can be explained by the inaccuracy of the PES,
since quantum-mechanical results \cite{Leforestier2012} on a similar surface
\cite{Nguyen2018water} also overestimate this matrix element.
Upper and lower bifurcations are again underestimated, probably for the same reasons
as above, namely the inadequate PES and the unaccounted anharmonic effects in
the longitudinal-transversal excitation.

In the case of the third mode excitation, especially interesting is the AT path
for which the contributions of the $F$ term and the $\mathbf U$ term
in the matrix element almost cancel each other out, while the major contribution
arises from the anharmonicity contained in the $\mathbf Z$ term. For this excitation,
both GI and AI matrix elements are overestimated. This can again be attributed to
the rovibrational coupling, since the quantum-mechanical results show
a significant increase in the lower and upper interchange with
the excitation to $K_a = 1$ rotational state \cite{Leforestier2012}.
Upper and lower bifurcations for this excitation show a much better
agreement with the experimental values \cite{Braly2000} than above.

At larger excitation frequencies, the theory breaks down.
A probable cause of this breakdown is the fact that as the frequency increases,
the contribution of the $w$ term to the overall splitting rises significantly.
This is due to the fact that the $F$ contribution depends exponentially
on the frequency of excitation, while the $\eta$ values do not compensate it.
As a result, its contribution becomes comparable to that of $W_0$, while
the WKB approach assumes $\ln{F} << W_0$. A good test of the reliability of
the obtained results is to redo the calculations
with a different value of the initial 'jump' parameter $\varepsilon$.
As the value of $\varepsilon$ is reduced, the results should converge to
the correct value. However, there is a limit to how much $\varepsilon$ can be
reduced, as the propagation from the point too close to the minimum is not stable
\cite{Milnikov2001,Erakovic2020}.
If the results converge before this breakdown, they can be treated as reliable.
Also, as the value of $\varepsilon$ is increased, values of the splittings
should not change by more than a few percent. This is the case for
the excitations in the first three lowest modes. For the fourth excited mode,
if we change $\varepsilon$ from 0.1 $m^{1/2}a_0$ to 1 $m^{1/2}a_0$,
the AT matrix element changes from $8.98$ ${\rm cm}^{-1}$ to $3(+3)$ ${\rm cm}^{-1}$,
which is an indication that the breakdown of theory occured.
Similar behaviour is present for the AI pathway, where the matrix element
changes from $2.58(-2)$ ${\rm cm}^{-1}$ to $0.23$ ${\rm cm}^{-1}$.
The change is not as drastic as in the AT case, but it indicates that the error
bars on our results are very large, which also explains the discrepancies of
results for the LI and UI splittings. Noticeable changes are also present
for the DE pathway (from $-3.70(-4)$ ${\rm cm}^{-1}$ to $-6.85(-4)$ ${\rm cm}^{-1}$),
while the values for other pathways do not change appreciably and
can be considered reliable.

\section{CONCLUSIONS}
We developed a semiclassical theory for calculating tunneling splittings of low-lying
vibrationally excited states based on the instanton method. A WKB wavefunction is
constructed along the instanton path and its harmonic neighborhood for each well, and
inserted into Herring formula to obtain the splitting that matches the JFI result
in the ground state \cite{Erakovic2020}. The excited-state splittings are then obtained
constructing excited-state wavefunctions analogously. The procedure closely follows that
of Ref.~\onlinecite{Milnikov2005}, but uses a more general boundary condition near minima
and does not assume the left-right mirror symmetry of potential along the instanton path.
In our approach, transversal and longitudinal excitations do not require separate
treatments as in Ref.~\onlinecite{Milnikov2005}. This allows us to compute splittings
in the systems where the excited vibrational mode does not line up along the instanton
path near minima, but has both longitudinal and transversal components, or the systems
in which the excited mode is longitudinal at one minimum and transversal at the other.
Both components are propagated simultaneously along the instanton path and cross
interaction is kept in the treatment.
%

The tests on the symmetric double-well model potential showed that a high accuracy
can be expected for low-lying states below the barrier.
It was shown that for transversal modes, even a small longitudinal displacement near
minima can dominate the tunneling splitting.
We also observed that the longitudinal-transversal cross terms improve results.
The tests on the asymmetric model potential showed that we can calculate splitting
estimates for excited longitudinal-transversal modes, albeit with somewhat reduced
accuracy.
Finally, we calculated the tunneling splitting pattern of the deuterated water dimer
in vibrationally excited lowest three modes by computing contributions from
five different rearrangement pathways. This is a particularly challenging
system for treatment with partly harmonic theories. Additionaly, the system
exhibits significant rovibrational couplings, which are, at present, neglected
in our treatment. We could nevertheless obtain reasonable agreement in many cases
in a system which showcases the situations in which the present theory gives
significantly different results from that of Ref.~\onlinecite{Milnikov2005}.

Tunneling splittings in vibrationally excited states require no additional information
about the molecular system. All computational effort is concentrated, as for the
ground-state splittings, in determining the MAP by optimization and
the evaluation of Hessians along the MAP. This allows us to compute and interpret
splitting patterns in many mid-sized molecules using state-of-the-art potentials.
The theory is applied in Cartesian coordinates and requires no modification for
treating different molecular systems. However, tunneling splittings in
vibrational states with higher frequencies, such as the excitations of librational
modes of water trimer \cite{Keutsch2001} and pentamer  \cite{Cole2017pentamer}
that were recently measured, cannot be treated with the theory in the present format.
Also, many small tunneling systems exhibit large rotation-vibration coupling,
which is currently neglected and can affect the splittings. A computationally tractable
theory for calculating splittings in rotationally excited states would also be desirable.
These are some of the immediate challenges remaining in which the future efforts will
certainly be directed in a quest to provide quantitative estimates for splitting
patterns for molecules and clusters that are out of reach to the exact
quantum-mechanical treatments.

\begin{acknowledgments}
This work was supported by Croatian Science Foundation Grant No.~IP-2016-06-1142,
and in part by the QuantiXLie Centre of Excellence, a project
cofinanced by the Croatian Government and European Union through the
European Regional Development Fund -- the Competitiveness and Cohesion
Operational Programme (Grant KK.01.1.1.01.0004).
\end{acknowledgments}

\bigskip

\noindent
{\bf DATA AVAILABILITY}
\\

The data that support the findings of this study are available from
the corresponding author upon reasonable request.

\appendix
\section{Method of Characteristics and local coordinates}
Method of characteristics is a technique for solving
partial differential equations \cite{Hilbert}. It relies on 
locating curves, the characteristics, along which the gradient
of the desired solution is tangential. As a consequence,
the partial differential equation reduces to an ordinary differential
equation. For a non-linear partial differential equation of the form,
\begin{equation}
F(x_1,..., x_N, p_1, ..., p_N, f) = 0,
\label{NLPDF}
\end{equation}
where $p_i=\partial f / \partial x_i$, defining equations of
the characteristics are
\begin{align}
\nonumber
\dfrac{{\rm d} x_i}{{\rm d} \tau}&=\dfrac{\partial F}{\partial p_i}, \\
\nonumber
\dfrac{{\rm d} p_i}{{\rm d} \tau}&=-\dfrac{\partial F}{\partial x_i}-\dfrac{\partial F}{\partial f}p_i, \\
\frac{{\rm d} f}{{\rm d} \tau} &= \frac{\partial F}{\partial p_i} p_i,
\label{non_lin_characteristics_2}
\end{align}
where $\tau$ parametrizes the characteristic.

Hamilton-Jacobi equation is a non-linear partial differential equation for
which $F=\frac{1}{2}p_ip_i-V$, where $p_i=\partial W_0 / \partial x_i$.
Its characteristics are therefore
\begin{align}
\nonumber
\frac{{\rm d} x_i}{{\rm d} \tau} &= p_i, \\
\frac{{\rm d} p_i}{{\rm d} \tau} &= \frac{\partial V}{\partial x_i}.
\label{HJ_characteristics}
\end{align}
The characteristics describe classical trajectories on the inverted PES,
while $\nabla W_0$ is the momentum on the trajectory. The total energy
of the classical motion is $E_{\rm tot} = \frac{1}{2}p_ip_i+(-V) = 0$.
On characteristics, $W_0$ is found by solving
\begin{equation}
\frac{{\rm d} W_0}{{\rm d} \tau} = p_ip_i = 2V.
\label{W_0_on_trajectory}
\end{equation}
The parameter $\tau$ represents time and, as the trajectory approaches minimum,
its value $\tau \to -\infty$. This is numerically problematic, so we reparametrize
characteristics with the arc length distance from the minimum, $S$, using the
transformation in \eqn{arc}.

In order to expand $W_0$ in Taylor series around the characteristic,
it is convenient to define a set of local coordinates $\{S, \Delta {\mathbf x} \}$.
Since coordinate $S$ parametrizes characteristic, it is only defined for
the points lying on it. In order to assign a value $S$ to the point that does not
lie on the characteristic, a point ${\mathbf x}_0(S)$ which does lie on
it is chosen so that
\begin{equation}
\left( x_i - x_{0i}(S) \right) p_{0i} = 0,
\label{local_coordinate_S}
\end{equation}
that is, ${\mathbf x}_0(S)$ is chosen so that the vector connecting it with the point
$\mathbf x$ is orthogonal to the characteristic at ${\mathbf x}_0(S)$. The value
of $S$ which corresponds to ${\mathbf x}_0(S)$ is then assigned to $\mathbf x$.
The orthogonal coordinates $\Delta {\mathbf x}$ are then defined as
$\Delta {\mathbf x} = {\mathbf x} - {\mathbf x}_0(S)$.
Differentiation of \eqn{local_coordinate_S} gives  \cite{Cole2017pentamer}
\begin{equation}
\frac{\partial S}{\partial x_i} = \frac{\frac{p_{0i}}{p_0}}{1-\frac{{\mathbf a}^{\top} \Delta {\mathbf x}}{p_0^2}}
= \frac{p_{0i}}{p_0} \left( 1+\frac{{\mathbf a}^{\top} \Delta {\mathbf x}}{p_0^2}+... \right),
\label{dsdxi}
\end{equation}
where $\mathbf a = \frac{{\rm d} {\mathbf p}_0}{{\rm d} \tau}$ denotes the acceleration.
From the differentiation of Hamilton-Jacobi equation, \eqn{Hamilton-Jacobi},
we obtain 
${\mathbf a} = {\mathbf A} {\mathbf p}_0$. And, finally, the differentiation of
the defining equation of orthogonal coordinates in \eqn{local_coordinate_S} gives
the transformation
\begin{equation}
\frac{\partial \Delta x_i}{\partial x_j} = \delta_{ij}-\frac{p_{0i}p_{0j}}{p_0^2} \left( 1+\frac{{\mathbf a}^{\top} \Delta {\mathbf x}}{p_0^2}+... \right).
\label{ddxidxj}
\end{equation}
\eqnn{dsdxi}{ddxidxj} are used throughout the paper to transform between Cartesian and local coordinates
on the characteristic as $\mathbf{A}$, $\mathbf{U}$, $\mathbf{B}$ and $\mathbf{Z}$ are all given in
differential form.

\section{Wavefunctions near minima}
Near minima $\mathbf{x}_{\rm min}$, the PES can be approximated by a harmonic oscillator potential
\begin{equation}
V = \frac{1}{2} \omega_i^2 q_i^2,
\label{Harmonic_potential}
\end{equation}
where $q_i = V_{ji}(x_j-x_{{\rm min}\ j})$ are normal coordinates,
and $\omega_i$ corresponding harmonic frequencies. Since $\mathbf{A}_0=\mathbf{H}^{1/2}$,
we have
${\mathbf V}^{\top} {\mathbf A}_0 {\mathbf V} = {\mathbf \Omega}$, with
$({\mathbf \Omega})_{ij}=\omega_i \delta_{ij}$.
In the harmonic region near minima, the equations of characteristics,
\eqn{HJ_characteristics}, become
\begin{align}
\nonumber
\frac{{\rm d^2} q_i}{{\rm d} \tau^2} &= \frac{\partial V}{\partial q_i}, \\
\frac{{\rm d^2} q_i}{{\rm d} \tau^2} &= \omega_i^2 q_i.
\label{Harmonic_characteristic_equation}
\end{align}
The trajectory along the characteristic from the minimum to an arbitrary
point ${\mathbf q}_1$ at $\tau=0$ inside the harmonic region is
\begin{equation}
q_i(\tau) = q_{1i} {\rm e}^{\omega_i \tau}.
\label{harmonic_characteristic}
\end{equation}
By considering the tangent vector of the characteristic,
\begin{equation}
t_i = \frac{p_{0i}}{p_0} =
\frac{\omega_i q_{1i} {\rm e}^{\omega_i \tau}}{\sqrt{\omega_j^2 q_{1j}^2 {\rm e}^{2 \omega_j \tau}}},
\label{harmonic_tangent}
\end{equation}
we note that in the limit $\tau \to -\infty$, the tangent becomes
$t_i = \delta_{iM}$, where $M$ denotes the lowest frequency normal mode
for which $q_{1,M} \ne 0$. This means that all characteristics approach
the minimum along the lowest normal mode with a non-zero projection upon
entering the harmonic region.

Function $W_0$ in \eqn{W_0_on_trajectory} can be evaluated in the harmonic region
at the characteristic as
\begin{equation}
W_0(\tau) = \int_{-\infty}^{\tau} \omega_j^2 q_{1j}^2 {\rm e}^{2 \omega_j \tau'} {\rm d} \tau',
\label{Harmonic_W0_tau}
\end{equation}
or, making use of \eqn{harmonic_characteristic}, as
\begin{equation}
W_0(\mathbf q) = \frac{1}{2} \omega_j q_j^2.
\label{Harmonic_W0}
\end{equation}
Furthermore, since in the harmonic region ${\mathbf A} \approx {\mathbf A}_0$,
the ground-state wavefunction corresponds to that of the harmonic oscillator,
\begin{equation}
\phi = {\rm e}^{-\frac{1}{2} \omega_j q_j^2}.
\label{harmonic_ground_state_phi}
\end{equation}
\eqn{harmonic_ground_state_phi} is used to approximate the norm of
the ground-state wavefunction in Herring formula \eqn{herring}.

For vibrationally excitated states, the correct form of the wavefunction
at the minimum is obtained by choosing $(\mathbf{U}_0)_i = V_{i \rm{e}}$,
that is by equating the vector $\mathbf{U}$ with the excited normal mode
at the minimum. The wavefunction then has the form
\begin{equation}
\phi = q_{\rm e} {\rm e}^{-\frac{1}{2} \omega_j q_j^2}.
\label{harmonic_excited_state_phi}
\end{equation}
For a point on the characteristic, which lies in the harmonic
region, $\Delta x_i = 0$, so its form is
\begin{equation}
\phi = F(\varepsilon) {\rm e}^{\frac{1}{2} \omega_j q_j^2(\varepsilon)}.
\label{WKB_harmonic_excited_state_phi}
\end{equation}
Therefore, the initial condition for the $F$ term at $S=\varepsilon$ has to be
\begin{equation}
F(\varepsilon)=q_{\rm e}={\mathbf U}_0^{\top} \left( {\mathbf x}(\varepsilon) - {\mathbf x}(0) \right),
\end{equation}
in order to yield the correct form of the wavefunction in \eqn{harmonic_excited_state_phi}.

\section{Anharmonicity about the instanton path}
The anharmonicity of potential in the directions perpendicular to
the instanton path can be partially accounted for by including
the higher derivatives of the PES along the instanton path, beyond Hessian, 
in the semiclassical treatment of Section III.
We assume below that the third derivative tensor of the PES, with elements
$c_{ijk}=\frac{\partial^3 V}{\partial x_i \partial x_j \partial x_k}$ along
the instanton path has been determined. This allows us to compute
the third derivatives of function $W_0$,
$B_{ijk}=\dfrac{\partial^3 W_0}{\partial x_i \partial x_j \partial x_k}$,
in Taylor expansion \eqn{taylor_W}.
The equation for propagation of tensor $\mathbf B$ is obtained by differentiating
Hamilton-Jacobi equation, \eqn{Hamilton-Jacobi}, three times as,
\begin{align}
\nonumber
&\dfrac{\partial^4 W_0}{\partial x_i \partial x_j \partial x_k \partial x_l}\dfrac{\partial W_0}{\partial x_l}+\dfrac{\partial^3 W_0}{\partial x_i \partial x_j \partial x_l}\dfrac{\partial^2 W_0}{\partial x_l \partial x_k}+ \\
&\dfrac{\partial^3 W_0}{\partial x_i \partial x_l \partial x_k}\dfrac{\partial^2 W_0}{\partial x_l \partial x_j}+\dfrac{\partial^3 W_0}{\partial x_l \partial x_j \partial x_k}\dfrac{\partial^2 W_0}{\partial x_l \partial x_i}=\dfrac{\partial^3 V}{\partial x_i \partial x_j \partial x_k}.
\label{Hamilton-Jacobi_fourth_derivative}
\end{align}
The first term in \eqn{Hamilton-Jacobi_fourth_derivative} represents
a directional derivative of the tensor element $B_{ijk}$ along
the instanton trajectory, while the other terms can be recognized
as tensor elements of $\mathbf B$ and of Hessian $\mathbf A$, which is
determined by solving \eqn{hessian}. \eqn{Hamilton-Jacobi_fourth_derivative}
on the instanton reads
\begin{equation}
p_0B_{ijk}'+B_{ijl}A_{lk}+B_{ilk}A_{lj}+B_{ljk}A_{li}=c_{ijk}.
\label{B_propagation}
\end{equation}

We proceed to determine the initial condition $\mathbf B( \varepsilon)$ 
in the vicinity of the minimum. For that purpose we linearize
\eqn{B_propagation}, following an analogous procedure to that for
$\mathbf A$ in Refs.~\onlinecite{Milnikov2001,Erakovic2020}, as
\begin{align}
\nonumber
{\mathbf B} &= {\mathbf B}^{(0)}+{\mathbf B}^{(1)}S, \\
\nonumber
{\mathbf c} &= {\mathbf c}^{(0)}+{\mathbf c}^{(1)}S, \\
\nonumber
{\mathbf A} &= {\mathbf A}^{(0)}+{\mathbf A}^{(1)}S, \\
p_0 &= p_0^{(1)} S.
\label{B_expansion}
\end{align}
Inserting the above expressions into \eqn{B_propagation} and equating
terms of the same order in $S$ yields equations for ${\mathbf B}^{(0)}$
and ${\mathbf B}^{(1)}$ as
\begin{align}
\nonumber
&B_{ijl}^{(0)}A_{lk}^{(0)}+B_{ilk}^{(0)}A_{lj}^{(0)}+B_{ljk}^{(0)}A_{li}^{(0)}=c_{ijk}^{(0)}, \\
\nonumber
&p_0^{(1)} B_{ijk}^{(1)}+B_{ijl}^{(1)}A_{lk}^{(0)}+B_{ilk}^{(1)}A_{lj}^{(0)}+B_{ljk}^{(1)}A_{li}^{(0)}= \\
&=c_{ijk}^{(1)}-B_{ijl}^{(0)}A_{lk}^{(1)}-B_{ilk}^{(0)}A_{lj}^{(1)}-B_{ljk}^{(0)}A_{li}^{(1)}.
\label{jump_B}
\end{align}
These are solved by transforming to the basis of normal modes, the eigenvectors of ${\mathbf A}^{(0)}$,
using the following relations,
\begin{align}
\nonumber
\omega_i \delta_{ij}=V_{i'i}V_{j'j}A_{i'j'}^{(0)}, \\
\nonumber
\tilde{B}_{ijk}^{(0)}=V_{i'i}V_{j'j}V_{k'k}B_{i'j'k'}^{(0)}, \\
\tilde{c}_{ijk}^{(0)}=V_{i'i}V_{j'j}V_{k'k}c_{i'j'k'}^{(0)}.
\label{normal_modes_B}
\end{align}
Inserting \eqn{normal_modes_B} into \eqn{jump_B} yields equations
\begin{eqnarray}
\tilde{B}_{ijk}^{(0)}&=&\dfrac{\tilde{c}_{ijk}^{(0)}}{\omega_i+\omega_j+\omega_k}, \\
\tilde{B}_{ijk}^{(1)}&=&\dfrac{\tilde{c}_{ijk}^{(1)}-\tilde{B}_{ijl}^{(0)}\tilde{A}_{lk}^{(1)}-\tilde{B}_{ilk}^{(0)}\tilde{A}_{lj}^{(1)}-\tilde{B}_{ljk}^{(0)}\tilde{A}_{li}^{(1)}}{p_0^{(1)}+\omega_i+\omega_j+\omega_k},
\label{B0_B1}
\end{eqnarray}
that are needed to construct $\mathbf B( \varepsilon)$. \eqn{B_propagation} can now
be solved in the interval $[\varepsilon, S]$ using any differential equation solver,
such as the Runge Kutta method \cite{NumRep}.

Tensor $\mathbf B$ cannot be included in the wavefunction of \eqn{WKB_wavefunction}
without the inclusion of fourth derivatives, as the resulting wavefunction would
not be integrable in the dividing plane. However, it is used below to compute
the $\mathbf{Z}$ term in the expansion of $\exp(-w)$, \eqn{w_quadratic}, and
thus indirectly account for a part of anharmonicity.

We first note the following expressions are valid on the instanton path,
\begin{align}
\nonumber
F &= {\rm e}^{-w}, \\
\nonumber
U_i &= \dfrac{\partial}{\partial x_i} {\rm e}^{-w} = -\dfrac{\partial w}{\partial x_i} {\rm e}^{-w}, \\
\nonumber
Z_{ij} &= \dfrac{\partial^2}{\partial x_i \partial x_j} {\rm e}^{-w} = -\dfrac{\partial^2 w}{\partial x_i \partial x_j} {\rm e}^{-w}+\dfrac{\partial w}{\partial x_i} \dfrac{\partial w}{\partial x_j}{\rm e}^{-w}, \\
Z_{ij}p_j &= \left( \dfrac{\partial}{\partial x_j} U_i \right) p_j = p_0 U_i'.
\label{derivative_definition_Z}
\end{align}
In the next step, we differentiate \eqn{w_equation} twice to obtain useful relations
\begin{align}
\nonumber
p_k\dfrac{\partial^2 w}{\partial x_i \partial x_k} {\rm e}^{-w} = &A_{ik}U_k, \\
\nonumber
p_k\dfrac{\partial^3 w}{\partial x_i \partial x_j \partial x_k} {\rm e}^{-w} = &B_{ijk}U_k+A_{ik}Z_{kj}+A_{jk}Z_{ki}+ \\
&\dfrac{\partial w}{\partial x_j} A_{ik}U_k+\dfrac{\partial w}{\partial x_i} A_{jk}U_k.
\label{derivatives_equationw}
\end{align}
Finally, we take the third derivative of $\exp(-w)$ in \eqn{w_quadratic} to arrive at
\begin{align}
\nonumber
\left( \dfrac{\partial}{\partial x_k} Z_{ij} \right) p_k = &-p_k \dfrac{\partial^3 w}{\partial x_i \partial x_j \partial x_k} {\rm e}^{-w}+ \\
\nonumber
&p_k \dfrac{\partial w}{\partial x_k} \dfrac{\partial^2 w}{\partial x_i \partial x_j} {\rm e}^{-w}+p_k \dfrac{\partial w}{\partial x_i} \dfrac{\partial^2 w}{\partial x_k \partial x_j} {\rm e}^{-w}+ \\
&p_k \dfrac{\partial w}{\partial x_j} \dfrac{\partial^2 w}{\partial x_i \partial x_k} {\rm e}^{-w}-p_k \dfrac{\partial w}{\partial x_k} \dfrac{\partial w}{\partial x_i} \dfrac{\partial w}{\partial x_j} {\rm e}^{-w},
\label{Z_expr}
\end{align}
where we insert \eqn{derivatives_equationw} and recognize \eqn{derivative_definition_Z} to obtain
the equation for $\mathbf{Z}$ in the following form, 
\begin{equation}
p_0Z_{ij}'+A_{ik}Z_{kj}+A_{jk}Z_{ki}+B_{ijk}U_k+\omega_{\rm e}Z_{ij} = 0.
\label{Z_propagation}
\end{equation}

This equation is again solved separately in the interval $[0, \varepsilon]$ and $[\varepsilon, S]$,
following the same procedure as for $\mathbf A$ and $\mathbf B$.
All objects are expanded up to linear terms in $S$ and inserted into \eqn{Z_propagation}. By equating
terms of the same order in $S$, we obtain equations for ${\mathbf Z}^{(0)}$ and
${\mathbf Z}^{(1)}$,
\begin{align}
\nonumber
	&{\mathbf A}^{(0)}{\mathbf Z}^{(0)}+{\mathbf Z}^{(0)}{\mathbf A}^{(0)}+\omega_{\rm e} {\mathbf Z}^{(0)}+{\mathbf B}^{(0)}{\mathbf U}^{(0)} = 0, \\
\nonumber
&p_0^{(1)} {\mathbf Z}^{(1)}+{\mathbf A}^{(0)}{\mathbf Z}^{(1)}+{\mathbf Z}^{(1)}{\mathbf A}^{(0)}+\omega_{\rm e} {\mathbf Z}^{(1)}+{\mathbf A}^{(1)}{\mathbf Z}^{(0)}+ \\
&{\mathbf Z}^{(0)}{\mathbf A}^{(1)}+{\mathbf B}^{(1)}{\mathbf U}^{(0)}+{\mathbf B}^{(0)}{\mathbf U}^{(1)} = 0,
\end{align}
where matrices $({\mathbf B}{\mathbf U})_{ij}$ evaluate as $B_{ijk}U_{k}$.
These equations are again solved by transforming to the basis of normal modes, the eigenvectors of
${\mathbf A}^{(0)}$, as
\begin{align}
\nonumber
&\tilde{Z}^{(0)}_{ij} = - \dfrac{(\tilde{B}^{(0)}\tilde{U}^{(0)})_{ij}}{\omega_{\rm e}+\omega_i+\omega_j}, \\
\nonumber
&\tilde{Z}^{(1)}_{ij} = \\
&- \dfrac{(\tilde{B}^{(1)}\tilde{U}^{(0)})_{ij}+(\tilde{B}^{(0)}\tilde{U}^{(1)})_{ij}+(\tilde{A}^{(1)}\tilde{Z}^{(0)})_{ij}+(\tilde{Z}^{(0)}\tilde{A}^{(1)})_{ij}}{p_0^{(1)}+\omega_{\rm e}+\omega_i+\omega_j}.
\end{align}
We are now in the position to compute ${\mathbf Z}(\varepsilon)$, which serves as the initial
condition for the propagation of ${\mathbf Z}$ in the interval $[\varepsilon, S]$ by solving
\eqn{Z_propagation} using, e.g., Runge-Kutta method.

When the $\mathbf{Z}$ term is included in the expansion of $\exp(-w)$, the tunneling splitting formula
assumes the following form
\begin{align}
\nonumber
	\Delta_1 = &\Delta_0 (2 \omega_{\rm e}) \bigg( F^{(\rm L)}F^{(\rm R)} + \frac{1}{2}{\mathbf U}^{(\rm L) \top} \bar{\mathbf A}^{-1} {\mathbf U}^{(\rm R)} \\
	&+\frac{1}{4}F^{(\rm L)} {\rm Tr} \left( {\mathbf Z}^{(\rm R)}\bar{\mathbf A}^{-1} \right)
+\frac{1}{4}F^{(\rm R)} {\rm Tr} \left( {\mathbf Z}^{(\rm L)}\bar{\mathbf A}^{-1} \right) \bigg),
\label{excited_state_splitting_Z}
\end{align}
where terms of the form $Z_{ij}Z_{kl} \Delta x_i\Delta x_j\Delta x_k\Delta x_l$
in the surface integral have been neglected, as their contribution was found
to be negligible.

\section{Invariance of tunneling splittings with respect to the position of the dividing plane}
Invariance of the ground-state tunneling splitting formula can be proved by 
differentiating \eqn{delta_0} with respect to the position of the connection point $S_{\rm cp}$,
where the dividing plane intersects the instanton path,
\begin{align}
\nonumber
\dfrac{\partial \Delta_0}{\partial S_{\rm cp}} = &\dfrac{\Delta_0}{2p_0} \bigg( 2 \dfrac{\partial p_0}{\partial S_{\rm cp}} -\frac{p_0}{{\rm det'} \bar{\mathbf A}} \dfrac{\partial}{\partial S_{\rm cp}}{\rm det'} \bar{\mathbf A}- 2p_0\dfrac{\partial}{\partial S_{\rm cp}} W_1^{(\rm L)} \\
&-2p_0\dfrac{\partial}{\partial S_{\rm cp}} W_1^{(\rm R)} \bigg).
\label{deriv_d0}
\end{align}
The derivative of determinant in \eqn{deriv_d0} is simplified using Jacobi formula (Eq.~(C4) in 
Ref.~\onlinecite{Erakovic2020}), while $W_1^{\rm L/R}$ functions are differentiated in
the upper/lower limit of the integral in \eqn{W1_integral},
\begin{align}
\nonumber
\dfrac{\partial \Delta_0}{\partial S_{\rm cp}} = &\dfrac{\Delta_0}{2p_0} \bigg( 2 \dfrac{\partial p_0}{\partial S_{\rm cp}} -{\rm Tr}\left(\bar{\mathbf A}^{-1} p_0 \dfrac{\partial}{\partial S_{\rm cp}} \bar{\mathbf A} \right)- \\
&{\rm Tr}({\mathbf A}^{\rm{(L)}}-{\mathbf A}_0)+{\rm Tr}({\mathbf A}^{\rm{(R)}}-{\mathbf A}_0) \bigg).
\end{align}
Derivative of $\bar{\mathbf A}$ can be shown to equal
\begin{equation}
\dfrac{\partial}{\partial S_{\rm cp}}\bar{\mathbf A} = \dfrac{1}{2}\bar{\mathbf A} \left( {\mathbf A}^{(\rm R)}-{\mathbf A}^{(\rm L)} \right)+ \dfrac{1}{2}\left( {\mathbf A}^{(\rm R)}-{\mathbf A}^{(\rm L)} \right) \bar{\mathbf A},
\end{equation}
where use has been made of \eqnn{A_bar}{hessian}.
Furthermore, since the tangent $\mathbf t$ is an eigenvector of $\bar{\mathbf A}$ with zero eigenvalue and,
by definition of the pseudoinverse, $\bar{\mathbf A} ^{-1}{\mathbf t} = 0$, we have
${\mathbf P}\bar{\mathbf A}{\mathbf P}=\bar{\mathbf A}$ and ${\mathbf P}\bar{\mathbf A}^{-1}{\mathbf P}=\bar{\mathbf A}^{-1}$,
where ${\mathbf P}={\mathbf I}-{\mathbf t}{\mathbf t}^{\top}$ is the operator that projects out the tangent of the instanton
path. Using the above, one can show that
\begin{align}
\nonumber
{\rm Tr}\left(\bar{\mathbf A}^{-1} p_0 \dfrac{\partial}{\partial S_{\rm cp}} \bar{\mathbf A} \right) 
&= {\rm Tr}\left({\mathbf P} \left( {\mathbf A}^{(\rm R)}-{\mathbf A}^{(\rm L)} \right) {\mathbf P} \right) \\
&= {\rm Tr}\left( {\mathbf A}_{\perp}^{(\rm R)}-{\mathbf A}_{\perp}^{(\rm L)} \right).
\end{align}
Thus, the derivative of the tunneling splitting becomes
\begin{align}
\dfrac{\partial \Delta_0}{\partial S_{\rm cp}} = \dfrac{\Delta_0}{2p_0} \left( 2 \dfrac{\partial p_0}{\partial S_{\rm cp}} +{\rm Tr}({\mathbf A}_{\perp}^{(\rm L)}-{\mathbf A}_{\perp}^{(\rm R)})-{\rm Tr}({\mathbf A}^{(\rm L)}-{\mathbf A}^{(\rm R)}) \right).
\end{align}
Finally, since ${\rm Tr}{\mathbf A}^{(\rm L)}=p_0'+{\rm Tr}{\mathbf A}_{\perp}^{(\rm L)}$ and
${\rm Tr}{\mathbf A}^{(\rm R)}=-p_0'+{\rm Tr}{\mathbf A}_{\perp}^{(\rm R)}$, as shown in Ref.~\onlinecite{Milnikov2001},
we have
\begin{align}
\nonumber
\dfrac{\partial \Delta_0}{\partial S_{\rm cp}} = 
&\dfrac{\Delta_0}{2p_0} \bigg( 2 \dfrac{\partial p_0}{\partial S_{\rm cp}} +{\rm Tr}({\mathbf A}_{\perp}^{(\rm L)}- \\
&{\mathbf A}_{\perp}^{(\rm R)})-{\rm Tr}({\mathbf A}_{\perp}^{(\rm L)}-{\mathbf A}_{\perp}^{(\rm R)})-2 \dfrac{\partial p_0}{\partial S_{\rm cp}} \bigg) = 0,
\end{align}
which proves that the ground-state tunneling splitting does not depend on $S_{\rm cp}$,
the position of the dividing plane.

Similarly, the invariance of the excited-state tunneling splitting on the position of the 
dividing plane is checked by differentiating \eqn{excited_state_splitting}
with respect to $S_{\rm cp}$. If only the $F$ terms are included in the expansion
of $\exp(-w)$, we have
\begin{equation}
\dfrac{\partial \Delta_1}{\partial S_{\rm cp}} = \Delta_0 2 \omega_{\rm e} \left( \dfrac{\partial F^{(\rm L)}}{\partial S_{\rm cp}}F^{(\rm R)}+F^{(\rm L)}\dfrac{\partial F^{(\rm R)}}{\partial S_{\rm cp}} \right),
\end{equation}
which together with \eqn{F_equation} gives
\begin{equation}
\dfrac{\partial \Delta_1}{\partial S_{\rm cp}} = \Delta_0 2 \omega_{\rm e} \left( \dfrac{\omega_{\rm e}}{p_0}F^{(\rm L)}F^{(\rm R)}-\dfrac{\omega_{\rm e}}{p_0}F^{(\rm L)}F^{(\rm R)} \right)= 0.
\end{equation}

If we include the $\mathbf U$ terms in the expansion of $\exp(-w)$, the derivative of the splitting becomes
\begin{align}
\nonumber
\dfrac{\partial \Delta_1}{\partial S_{\rm cp}} = &\Delta_0 \omega_{\rm e} \bigg( \dfrac{\partial}{\partial S_{\rm cp}}{\mathbf U}^{(\rm L) \top}\bar{{\mathbf A}}^{-1}{\mathbf U}^{(\rm R)}+ \\
&{\mathbf U}^{(\rm L) \top}\dfrac{\partial}{\partial S_{\rm cp}}\bar{{\mathbf A}}^{-1}{\mathbf U}^{(\rm R)}+{\mathbf U}^{(\rm L) \top}\bar{{\mathbf A}}^{-1}\dfrac{\partial}{\partial S_{\rm cp}}{\mathbf U}^{(\rm R)} \bigg).
\label{derivative_excited_1}
\end{align}
It can be shown that 
\begin{equation}
p_0\dfrac{\partial}{\partial S_{\rm cp}}\bar{{\mathbf A}}^{-1} = {\mathbf P}{\mathbf A}^{(\rm L)}\bar{{\mathbf A}}^{-1}-\bar{{\mathbf A}}^{-1}{\mathbf A}^{(\rm R)}{\mathbf P},
\end{equation}
which can be used to rewrite \eqn{derivative_excited_1} as
\begin{align}
\nonumber
\dfrac{\partial \Delta_1}{\partial S_{\rm cp}} = &\Delta_0 \dfrac{\omega_{\rm e}}{p_0}\bigg( -({\mathbf U}^{(\rm L)}-{\mathbf P}{\mathbf U}^{(\rm L)})^{\top}{\mathbf A}^{(\rm L)}\bar{{\mathbf A}}^{-1}{\mathbf U}^{(\rm R)}+ \\
&{\mathbf U}^{(\rm L) \top}\bar{{\mathbf A}}^{-1}{\mathbf A}^{(\rm R)}({\mathbf U}^{(\rm R)}-{\mathbf P}{\mathbf U}^{(\rm R)}) \bigg).
\label{derivative_excited_2}
\end{align}
In this form, it is evident that if $\mathbf U$ remains orthogonal to the instanton path, i.e., ${\mathbf P}{\mathbf U}={\mathbf U}$,
the excited-state tunneling splittings become independent on the position of the dividing plane.
If that is not the case, however, \eqn{derivative_excited_1} can be further simplified to
\begin{align}
\nonumber
\dfrac{\partial \Delta_1}{\partial S_{\rm cp}} = &\Delta_0 \dfrac{\omega_{\rm e}}{p_0} \bigg( -\dfrac{\omega_{\rm e}}{p_0}F^{(\rm L)}{\mathbf t}^{\top}{\mathbf A}^{(\rm L)}\bar{{\mathbf A}}^{-1}{\mathbf U}^{(\rm R)} \\
&-\dfrac{\omega_{\rm e}}{p_0} F^{(\rm R)}{\mathbf U}^{(\rm L) \top}\bar{{\mathbf A}}^{-1}{\mathbf A}^{(\rm R)}{\mathbf t} \bigg),
\label{derivative_excited_3}
\end{align}
which does not vanish and the tunneling splitting will, in general,
depend on the position of the dividing plane, as observed in Section IV.

If the same analysis is performed with the $\mathbf Z$ terms,
there arise two factors which cancel out the $\mathbf{U}$ terms. However,
a multitude of other factors also arise, which again cause
the dependence on the position of the dividing plane.
As mentioned above, the root of the problem is that
the expansion of $\exp(-w)$ is inconsistent with the expansion of $W_1$,
and it gives rise to terms of all orders in $\Delta \mathbf x$.
However, in the case of a symmetric potential, all perpendicular components of
the gradients, Hessians and third-order tensors are the same for
the left- and right-localized wavefunctions at the dividing plane in the middle
of the instanton path, while their tangent components differ in sign.
Thus, it is possible to show that the derivative of the $\mathbf Z$ contribution
with respect to $S_{\rm cp}$ vanishes at the middle of the instanton path, and 
numerical tests show that its contribution is minimal there.
Therefore, for symmetric systems, the middle of the path represents
the optimal position of the dividing plane. For the asymmetric paths,
there is no such preferential point on the instanton.
However, good results are obtained by positioning the dividing plane at
the maximum of the barrier, as at this point $p_0$ is the largest,
and the derivatives of the splitting are generally smallest,
which means that, at this point, the splittings are relatively stable.

\bibliography{references,new_references}

\begin{thebibliography}{80}%
\makeatletter
\providecommand \@ifxundefined [1]{%
 \@ifx{#1\undefined}
}%
\providecommand \@ifnum [1]{%
 \ifnum #1\expandafter \@firstoftwo
 \else \expandafter \@secondoftwo
 \fi
}%
\providecommand \@ifx [1]{%
 \ifx #1\expandafter \@firstoftwo
 \else \expandafter \@secondoftwo
 \fi
}%
\providecommand \natexlab [1]{#1}%
\providecommand \enquote  [1]{``#1''}%
\providecommand \bibnamefont  [1]{#1}%
\providecommand \bibfnamefont [1]{#1}%
\providecommand \citenamefont [1]{#1}%
\providecommand \href@noop [0]{\@secondoftwo}%
\providecommand \href [0]{\begingroup \@sanitize@url \@href}%
\providecommand \@href[1]{\@@startlink{#1}\@@href}%
\providecommand \@@href[1]{\endgroup#1\@@endlink}%
\providecommand \@sanitize@url [0]{\catcode `\\12\catcode `\$12\catcode
  `\&12\catcode `\#12\catcode `\^12\catcode `\_12\catcode `\%12\relax}%
\providecommand \@@startlink[1]{}%
\providecommand \@@endlink[0]{}%
\providecommand \url  [0]{\begingroup\@sanitize@url \@url }%
\providecommand \@url [1]{\endgroup\@href {#1}{\urlprefix }}%
\providecommand \urlprefix  [0]{URL }%
\providecommand \Eprint [0]{\href }%
\providecommand \doibase [0]{http://dx.doi.org/}%
\providecommand \selectlanguage [0]{\@gobble}%
\providecommand \bibinfo  [0]{\@secondoftwo}%
\providecommand \bibfield  [0]{\@secondoftwo}%
\providecommand \translation [1]{[#1]}%
\providecommand \BibitemOpen [0]{}%
\providecommand \bibitemStop [0]{}%
\providecommand \bibitemNoStop [0]{.\EOS\space}%
\providecommand \EOS [0]{\spacefactor3000\relax}%
\providecommand \BibitemShut  [1]{\csname bibitem#1\endcsname}%
\let\auto@bib@innerbib\@empty
\bibitem [{\citenamefont {Bell}(1980)}]{BellBook}%
  \BibitemOpen
  \bibfield  {author} {\bibinfo {author} {\bibfnamefont {R.~P.}\ \bibnamefont
  {Bell}},\ }\href@noop {} {\emph {\bibinfo {title} {The Tunnel Effect in
  Chemistry}}}\ (\bibinfo  {publisher} {Chapman and Hall},\ \bibinfo {address}
  {London},\ \bibinfo {year} {1980})\BibitemShut {NoStop}%
\bibitem [{\citenamefont {Coudert}\ and\ \citenamefont
  {Hougen}(1988)}]{Coudert1988dimer}%
  \BibitemOpen
  \bibfield  {author} {\bibinfo {author} {\bibfnamefont {L.~H.}\ \bibnamefont
  {Coudert}}\ and\ \bibinfo {author} {\bibfnamefont {J.~T.}\ \bibnamefont
  {Hougen}},\ }\href {\doibase 10.1016/0022-2852(88)90286-X} {\bibfield
  {journal} {\bibinfo  {journal} {J.~Mol. Spectrosc.}\ }\textbf {\bibinfo
  {volume} {130}},\ \bibinfo {pages} {86 } (\bibinfo {year}
  {1988})}\BibitemShut {NoStop}%
\bibitem [{\citenamefont {Walsh}\ and\ \citenamefont
  {Wales}(1996)}]{Walsh1996rearrangements}%
  \BibitemOpen
  \bibfield  {author} {\bibinfo {author} {\bibfnamefont {T.~R.}\ \bibnamefont
  {Walsh}}\ and\ \bibinfo {author} {\bibfnamefont {D.~J.}\ \bibnamefont
  {Wales}},\ }\href@noop {} {\bibfield  {journal} {\bibinfo  {journal}
  {J.~Chem. Soc. Faraday Trans.}\ }\textbf {\bibinfo {volume} {92}},\ \bibinfo
  {pages} {2505} (\bibinfo {year} {1996})}\BibitemShut {NoStop}%
\bibitem [{\citenamefont {Xu}\ and\ \citenamefont
  {J{\"{a}}ger}(1997)}]{Xu1997}%
  \BibitemOpen
  \bibfield  {author} {\bibinfo {author} {\bibfnamefont {Y.}~\bibnamefont
  {Xu}}\ and\ \bibinfo {author} {\bibfnamefont {W.}~\bibnamefont
  {J{\"{a}}ger}},\ }\href {\doibase 10.1063/1.473808} {\bibfield  {journal}
  {\bibinfo  {journal} {J.~Chem. Phys.}\ }\textbf {\bibinfo {volume} {106}},\
  \bibinfo {pages} {7968} (\bibinfo {year} {1997})}\BibitemShut {NoStop}%
\bibitem [{\citenamefont {Keutsch}\ and\ \citenamefont
  {Saykally}(2001)}]{Keutsch2001water}%
  \BibitemOpen
  \bibfield  {author} {\bibinfo {author} {\bibfnamefont {F.~N.}\ \bibnamefont
  {Keutsch}}\ and\ \bibinfo {author} {\bibfnamefont {R.~J.}\ \bibnamefont
  {Saykally}},\ }\href {\doibase 10.1073/pnas.191266498} {\bibfield  {journal}
  {\bibinfo  {journal} {P. Natl. Acad. Sci. USA}\ }\textbf {\bibinfo {volume}
  {98}},\ \bibinfo {pages} {10533} (\bibinfo {year} {2001})}\BibitemShut
  {NoStop}%
\bibitem [{\citenamefont {Liu}\ \emph {et~al.}(1996)\citenamefont {Liu},
  \citenamefont {Brown}, \citenamefont {Cruzan},\ and\ \citenamefont
  {Saykally}}]{Liu1996pentamer}%
  \BibitemOpen
  \bibfield  {author} {\bibinfo {author} {\bibfnamefont {K.}~\bibnamefont
  {Liu}}, \bibinfo {author} {\bibfnamefont {M.~G.}\ \bibnamefont {Brown}},
  \bibinfo {author} {\bibfnamefont {J.~D.}\ \bibnamefont {Cruzan}}, \ and\
  \bibinfo {author} {\bibfnamefont {R.~J.}\ \bibnamefont {Saykally}},\
  }\href@noop {} {\bibfield  {journal} {\bibinfo  {journal} {Science}\ }\textbf
  {\bibinfo {volume} {271}},\ \bibinfo {pages} {62} (\bibinfo {year}
  {1996})}\BibitemShut {NoStop}%
\bibitem [{\citenamefont {Cvita\v{s}}\ and\ \citenamefont
  {Richardson}(2020)}]{Waterclust}%
  \BibitemOpen
  \bibfield  {author} {\bibinfo {author} {\bibfnamefont {M.~T.}\ \bibnamefont
  {Cvita\v{s}}}\ and\ \bibinfo {author} {\bibfnamefont {J.~O.}\ \bibnamefont
  {Richardson}},\ }in\ \href@noop {} {\emph {\bibinfo {booktitle} {Molecular
  Spectroscopy and Quantum Dynamics}}},\ \bibinfo {editor} {edited by\ \bibinfo
  {editor} {\bibfnamefont {R.}~\bibnamefont {Marquardt}}\ and\ \bibinfo
  {editor} {\bibfnamefont {M.}~\bibnamefont {Quack}}}\ (\bibinfo  {publisher}
  {Elsevier},\ \bibinfo {year} {2020})\ Chap.~\bibinfo {chapter}
  {10}\BibitemShut {NoStop}%
\bibitem [{\citenamefont {Hammer}\ \emph {et~al.}(2009)\citenamefont {Hammer},
  \citenamefont {Coutinho-Neto}, \citenamefont {Viel},\ and\ \citenamefont
  {Manthe}}]{Hammer2009malonaldehyde}%
  \BibitemOpen
  \bibfield  {author} {\bibinfo {author} {\bibfnamefont {T.}~\bibnamefont
  {Hammer}}, \bibinfo {author} {\bibfnamefont {M.~D.}\ \bibnamefont
  {Coutinho-Neto}}, \bibinfo {author} {\bibfnamefont {A.}~\bibnamefont {Viel}},
  \ and\ \bibinfo {author} {\bibfnamefont {U.}~\bibnamefont {Manthe}},\ }\href
  {\doibase 10.1063/1.3272610} {\bibfield  {journal} {\bibinfo  {journal}
  {J.~Chem. Phys.}\ }\textbf {\bibinfo {volume} {131}},\ \bibinfo {pages}
  {224109} (\bibinfo {year} {2009})}\BibitemShut {NoStop}%
\bibitem [{\citenamefont {F{\'{a}}bri}\ \emph {et~al.}(2019)\citenamefont
  {F{\'{a}}bri}, \citenamefont {Marquardt}, \citenamefont
  {Cs{\'{a}}sz{\'{a}}r},\ and\ \citenamefont {Quack}}]{Fabri2019}%
  \BibitemOpen
  \bibfield  {author} {\bibinfo {author} {\bibfnamefont {C.}~\bibnamefont
  {F{\'{a}}bri}}, \bibinfo {author} {\bibfnamefont {R.}~\bibnamefont
  {Marquardt}}, \bibinfo {author} {\bibfnamefont {A.~G.}\ \bibnamefont
  {Cs{\'{a}}sz{\'{a}}r}}, \ and\ \bibinfo {author} {\bibfnamefont
  {M.}~\bibnamefont {Quack}},\ }\href {\doibase 10.1063/1.5063470} {\bibfield
  {journal} {\bibinfo  {journal} {J.~Chem. Phys.}\ }\textbf {\bibinfo {volume}
  {150}},\ \bibinfo {pages} {014102} (\bibinfo {year} {2019})}\BibitemShut
  {NoStop}%
\bibitem [{\citenamefont {Felker}\ and\ \citenamefont
  {Ba{\v{c}}i{\'{c}}}(2019)}]{Felker2019}%
  \BibitemOpen
  \bibfield  {author} {\bibinfo {author} {\bibfnamefont {P.~M.}\ \bibnamefont
  {Felker}}\ and\ \bibinfo {author} {\bibfnamefont {Z.}~\bibnamefont
  {Ba{\v{c}}i{\'{c}}}},\ }\href {\doibase 10.1063/1.5111131} {\bibfield
  {journal} {\bibinfo  {journal} {J.~Chem.~Phys.}\ }\textbf {\bibinfo {volume}
  {151}},\ \bibinfo {pages} {024305} (\bibinfo {year} {2019})}\BibitemShut
  {NoStop}%
\bibitem [{\citenamefont {Ceriotti}\ \emph {et~al.}(2016)\citenamefont
  {Ceriotti}, \citenamefont {Fang}, \citenamefont {Kusalik}, \citenamefont
  {McKenzie}, \citenamefont {Michaelides}, \citenamefont {Morales},\ and\
  \citenamefont {Markland}}]{Ceriotti2016water}%
  \BibitemOpen
  \bibfield  {author} {\bibinfo {author} {\bibfnamefont {M.}~\bibnamefont
  {Ceriotti}}, \bibinfo {author} {\bibfnamefont {W.}~\bibnamefont {Fang}},
  \bibinfo {author} {\bibfnamefont {P.~G.}\ \bibnamefont {Kusalik}}, \bibinfo
  {author} {\bibfnamefont {R.~H.}\ \bibnamefont {McKenzie}}, \bibinfo {author}
  {\bibfnamefont {A.}~\bibnamefont {Michaelides}}, \bibinfo {author}
  {\bibfnamefont {M.~A.}\ \bibnamefont {Morales}}, \ and\ \bibinfo {author}
  {\bibfnamefont {T.~E.}\ \bibnamefont {Markland}},\ }\href {\doibase
  10.1021/acs.chemrev.5b00674} {\bibfield  {journal} {\bibinfo  {journal}
  {Chem. Rev.}\ }\textbf {\bibinfo {volume} {116}},\ \bibinfo {pages} {7529}
  (\bibinfo {year} {2016})}\BibitemShut {NoStop}%
\bibitem [{\citenamefont {Wang}\ \emph {et~al.}(2011)\citenamefont {Wang},
  \citenamefont {Huang}, \citenamefont {Shepler}, \citenamefont {Braams},\ and\
  \citenamefont {Bowman}}]{Wang2011water}%
  \BibitemOpen
  \bibfield  {author} {\bibinfo {author} {\bibfnamefont {Y.}~\bibnamefont
  {Wang}}, \bibinfo {author} {\bibfnamefont {X.}~\bibnamefont {Huang}},
  \bibinfo {author} {\bibfnamefont {B.~C.}\ \bibnamefont {Shepler}}, \bibinfo
  {author} {\bibfnamefont {B.~J.}\ \bibnamefont {Braams}}, \ and\ \bibinfo
  {author} {\bibfnamefont {J.~M.}\ \bibnamefont {Bowman}},\ }\href {\doibase
  10.1063/1.3554905} {\bibfield  {journal} {\bibinfo  {journal} {J.~Chem.
  Phys.}\ }\textbf {\bibinfo {volume} {134}},\ \bibinfo {pages} {094509}
  (\bibinfo {year} {2011})}\BibitemShut {NoStop}%
\bibitem [{\citenamefont {Reddy}\ \emph {et~al.}(2016)\citenamefont {Reddy},
  \citenamefont {Straight}, \citenamefont {Bajaj}, \citenamefont {Huy~Pham},
  \citenamefont {Riera}, \citenamefont {Moberg}, \citenamefont {Morales},
  \citenamefont {Knight}, \citenamefont {G{\"o}tz},\ and\ \citenamefont
  {Paesani}}]{Reddy2016MBpol}%
  \BibitemOpen
  \bibfield  {author} {\bibinfo {author} {\bibfnamefont {S.~K.}\ \bibnamefont
  {Reddy}}, \bibinfo {author} {\bibfnamefont {S.~C.}\ \bibnamefont {Straight}},
  \bibinfo {author} {\bibfnamefont {P.}~\bibnamefont {Bajaj}}, \bibinfo
  {author} {\bibfnamefont {C.}~\bibnamefont {Huy~Pham}}, \bibinfo {author}
  {\bibfnamefont {M.}~\bibnamefont {Riera}}, \bibinfo {author} {\bibfnamefont
  {D.~R.}\ \bibnamefont {Moberg}}, \bibinfo {author} {\bibfnamefont {M.~A.}\
  \bibnamefont {Morales}}, \bibinfo {author} {\bibfnamefont {C.}~\bibnamefont
  {Knight}}, \bibinfo {author} {\bibfnamefont {A.~W.}\ \bibnamefont
  {G{\"o}tz}}, \ and\ \bibinfo {author} {\bibfnamefont {F.}~\bibnamefont
  {Paesani}},\ }\href {\doibase 10.1063/1.4967719} {\bibfield  {journal}
  {\bibinfo  {journal} {J.~Chem. Phys.}\ }\textbf {\bibinfo {volume} {145}},\
  \bibinfo {pages} {194504} (\bibinfo {year} {2016})}\BibitemShut {NoStop}%
\bibitem [{\citenamefont {Richardson}, \citenamefont {Althorpe},\ and\
  \citenamefont {Wales}(2011)}]{water}%
  \BibitemOpen
  \bibfield  {author} {\bibinfo {author} {\bibfnamefont {J.~O.}\ \bibnamefont
  {Richardson}}, \bibinfo {author} {\bibfnamefont {S.~C.}\ \bibnamefont
  {Althorpe}}, \ and\ \bibinfo {author} {\bibfnamefont {D.~J.}\ \bibnamefont
  {Wales}},\ }\href {\doibase 10.1063/1.3640429} {\bibfield  {journal}
  {\bibinfo  {journal} {J.~Chem. Phys.}\ }\textbf {\bibinfo {volume} {135}},\
  \bibinfo {pages} {124109} (\bibinfo {year} {2011})}\BibitemShut {NoStop}%
\bibitem [{\citenamefont {Keutsch}\ \emph {et~al.}(2001)\citenamefont
  {Keutsch}, \citenamefont {Fellers}, \citenamefont {Viant},\ and\
  \citenamefont {Saykally}}]{Keutsch2001}%
  \BibitemOpen
  \bibfield  {author} {\bibinfo {author} {\bibfnamefont {F.~N.}\ \bibnamefont
  {Keutsch}}, \bibinfo {author} {\bibfnamefont {R.~S.}\ \bibnamefont
  {Fellers}}, \bibinfo {author} {\bibfnamefont {M.~R.}\ \bibnamefont {Viant}},
  \ and\ \bibinfo {author} {\bibfnamefont {R.~J.}\ \bibnamefont {Saykally}},\
  }\href {\doibase 10.1063/1.1337052} {\bibfield  {journal} {\bibinfo
  {journal} {J.~Chem.~Phys.}\ }\textbf {\bibinfo {volume} {114}},\ \bibinfo
  {pages} {4005} (\bibinfo {year} {2001})}\BibitemShut {NoStop}%
\bibitem [{\citenamefont {Cole}\ \emph {et~al.}(2017)\citenamefont {Cole},
  \citenamefont {Fellers}, \citenamefont {Viant},\ and\ \citenamefont
  {Saykally}}]{Cole2017pentamer}%
  \BibitemOpen
  \bibfield  {author} {\bibinfo {author} {\bibfnamefont {W.~T.~S.}\
  \bibnamefont {Cole}}, \bibinfo {author} {\bibfnamefont {R.~S.}\ \bibnamefont
  {Fellers}}, \bibinfo {author} {\bibfnamefont {M.~R.}\ \bibnamefont {Viant}},
  \ and\ \bibinfo {author} {\bibfnamefont {R.~J.}\ \bibnamefont {Saykally}},\
  }\href {\doibase 10.1063/1.4973418} {\bibfield  {journal} {\bibinfo
  {journal} {J.~Chem. Phys.}\ }\textbf {\bibinfo {volume} {146}},\ \bibinfo
  {pages} {014306} (\bibinfo {year} {2017})}\BibitemShut {NoStop}%
\bibitem [{\citenamefont {Cvita{\v{s}}}\ and\ \citenamefont
  {Richardson}(2019)}]{Cvitas2019}%
  \BibitemOpen
  \bibfield  {author} {\bibinfo {author} {\bibfnamefont {M.~T.}\ \bibnamefont
  {Cvita{\v{s}}}}\ and\ \bibinfo {author} {\bibfnamefont {J.~O.}\ \bibnamefont
  {Richardson}},\ }\href {\doibase 10.1039/c9cp05561d} {\bibfield  {journal}
  {\bibinfo  {journal} {Phys.~Chem.~Chem.~Phys.}\ }\textbf {\bibinfo {volume}
  {22}},\ \bibinfo {pages} {1035} (\bibinfo {year} {2019})}\BibitemShut
  {NoStop}%
\bibitem [{\citenamefont {Mhin}\ \emph {et~al.}(1994)\citenamefont {Mhin},
  \citenamefont {Kim}, \citenamefont {Lee}, \citenamefont {Lee},\ and\
  \citenamefont {Kim}}]{Mhin1994hexamer}%
  \BibitemOpen
  \bibfield  {author} {\bibinfo {author} {\bibfnamefont {B.~J.}\ \bibnamefont
  {Mhin}}, \bibinfo {author} {\bibfnamefont {J.}~\bibnamefont {Kim}}, \bibinfo
  {author} {\bibfnamefont {S.}~\bibnamefont {Lee}}, \bibinfo {author}
  {\bibfnamefont {J.~Y.}\ \bibnamefont {Lee}}, \ and\ \bibinfo {author}
  {\bibfnamefont {K.~S.}\ \bibnamefont {Kim}},\ }\href@noop {} {\bibfield
  {journal} {\bibinfo  {journal} {J.~Chem. Phys.}\ }\textbf {\bibinfo {volume}
  {100}},\ \bibinfo {pages} {4484} (\bibinfo {year} {1994})}\BibitemShut
  {NoStop}%
\bibitem [{\citenamefont {P{\'e}rez}\ \emph {et~al.}(2012)\citenamefont
  {P{\'e}rez}, \citenamefont {Muckle}, \citenamefont {Zaleski}, \citenamefont
  {Seifert}, \citenamefont {Temelso}, \citenamefont {Shields}, \citenamefont
  {Kisiel},\ and\ \citenamefont {Pate}}]{Perez2012hexamer}%
  \BibitemOpen
  \bibfield  {author} {\bibinfo {author} {\bibfnamefont {C.}~\bibnamefont
  {P{\'e}rez}}, \bibinfo {author} {\bibfnamefont {M.~T.}\ \bibnamefont
  {Muckle}}, \bibinfo {author} {\bibfnamefont {D.~P.}\ \bibnamefont {Zaleski}},
  \bibinfo {author} {\bibfnamefont {N.~A.}\ \bibnamefont {Seifert}}, \bibinfo
  {author} {\bibfnamefont {B.}~\bibnamefont {Temelso}}, \bibinfo {author}
  {\bibfnamefont {G.~C.}\ \bibnamefont {Shields}}, \bibinfo {author}
  {\bibfnamefont {Z.}~\bibnamefont {Kisiel}}, \ and\ \bibinfo {author}
  {\bibfnamefont {B.~H.}\ \bibnamefont {Pate}},\ }\href {\doibase
  10.1126/science.1220574} {\bibfield  {journal} {\bibinfo  {journal}
  {Science}\ }\textbf {\bibinfo {volume} {336}},\ \bibinfo {pages} {897}
  (\bibinfo {year} {2012})}\BibitemShut {NoStop}%
\bibitem [{\citenamefont {Richardson}\ \emph {et~al.}(2016)\citenamefont
  {Richardson}, \citenamefont {P{\'e}rez}, \citenamefont {Lobsiger},
  \citenamefont {Reid}, \citenamefont {Temelso}, \citenamefont {Shields},
  \citenamefont {Kisiel}, \citenamefont {Wales}, \citenamefont {Pate},\ and\
  \citenamefont {Althorpe}}]{hexamerprism}%
  \BibitemOpen
  \bibfield  {author} {\bibinfo {author} {\bibfnamefont {J.~O.}\ \bibnamefont
  {Richardson}}, \bibinfo {author} {\bibfnamefont {C.}~\bibnamefont
  {P{\'e}rez}}, \bibinfo {author} {\bibfnamefont {S.}~\bibnamefont {Lobsiger}},
  \bibinfo {author} {\bibfnamefont {A.~A.}\ \bibnamefont {Reid}}, \bibinfo
  {author} {\bibfnamefont {B.}~\bibnamefont {Temelso}}, \bibinfo {author}
  {\bibfnamefont {G.~C.}\ \bibnamefont {Shields}}, \bibinfo {author}
  {\bibfnamefont {Z.}~\bibnamefont {Kisiel}}, \bibinfo {author} {\bibfnamefont
  {D.~J.}\ \bibnamefont {Wales}}, \bibinfo {author} {\bibfnamefont {B.~H.}\
  \bibnamefont {Pate}}, \ and\ \bibinfo {author} {\bibfnamefont {S.~C.}\
  \bibnamefont {Althorpe}},\ }\href {\doibase 10.1126/science.aae0012}
  {\bibfield  {journal} {\bibinfo  {journal} {Science}\ }\textbf {\bibinfo
  {volume} {351}},\ \bibinfo {pages} {1310} (\bibinfo {year}
  {2016})}\BibitemShut {NoStop}%
\bibitem [{\citenamefont {L{\'{e}}onard}\ \emph {et~al.}(2002)\citenamefont
  {L{\'{e}}onard}, \citenamefont {Handy}, \citenamefont {Carter},\ and\
  \citenamefont {Bowman}}]{Leonard2002}%
  \BibitemOpen
  \bibfield  {author} {\bibinfo {author} {\bibfnamefont {C.}~\bibnamefont
  {L{\'{e}}onard}}, \bibinfo {author} {\bibfnamefont {N.~C.}\ \bibnamefont
  {Handy}}, \bibinfo {author} {\bibfnamefont {S.}~\bibnamefont {Carter}}, \
  and\ \bibinfo {author} {\bibfnamefont {J.~M.}\ \bibnamefont {Bowman}},\
  }\href {\doibase 10.1016/S1386-1425(01)00671-0} {\bibfield  {journal}
  {\bibinfo  {journal} {Spectrochimica Acta Part A}\ }\textbf {\bibinfo
  {volume} {58}},\ \bibinfo {pages} {825} (\bibinfo {year} {2002})}\BibitemShut
  {NoStop}%
\bibitem [{\citenamefont {Neff}\ and\ \citenamefont {Rauhut}(2014)}]{Neff2014}%
  \BibitemOpen
  \bibfield  {author} {\bibinfo {author} {\bibfnamefont {M.}~\bibnamefont
  {Neff}}\ and\ \bibinfo {author} {\bibfnamefont {G.}~\bibnamefont {Rauhut}},\
  }\href {\doibase 10.1016/j.saa.2013.02.033} {\bibfield  {journal} {\bibinfo
  {journal} {Spectrochimica Acta Part A}\ }\textbf {\bibinfo {volume} {119}},\
  \bibinfo {pages} {100} (\bibinfo {year} {2014})}\BibitemShut {NoStop}%
\bibitem [{\citenamefont {{\v{S}}mydke}\ \emph {et~al.}(2019)\citenamefont
  {{\v{S}}mydke}, \citenamefont {F{\'{a}}bri}, \citenamefont {Sarka},\ and\
  \citenamefont {Cs{\'{a}}sz{\'{a}}r}}]{Smydke2019}%
  \BibitemOpen
  \bibfield  {author} {\bibinfo {author} {\bibfnamefont {J.}~\bibnamefont
  {{\v{S}}mydke}}, \bibinfo {author} {\bibfnamefont {C.}~\bibnamefont
  {F{\'{a}}bri}}, \bibinfo {author} {\bibfnamefont {J.}~\bibnamefont {Sarka}},
  \ and\ \bibinfo {author} {\bibfnamefont {A.~G.}\ \bibnamefont
  {Cs{\'{a}}sz{\'{a}}r}},\ }\href {\doibase 10.1039/C8CP04672G} {\bibfield
  {journal} {\bibinfo  {journal} {Phys.~Chem.~Chem.~Phys.}\ }\textbf {\bibinfo
  {volume} {21}},\ \bibinfo {pages} {3453} (\bibinfo {year}
  {2019})}\BibitemShut {NoStop}%
\bibitem [{\citenamefont {Wu}, \citenamefont {Ren},\ and\ \citenamefont
  {Bian}(2016)}]{Wu2016}%
  \BibitemOpen
  \bibfield  {author} {\bibinfo {author} {\bibfnamefont {F.}~\bibnamefont
  {Wu}}, \bibinfo {author} {\bibfnamefont {Y.}~\bibnamefont {Ren}}, \ and\
  \bibinfo {author} {\bibfnamefont {W.}~\bibnamefont {Bian}},\ }\href {\doibase
  10.1063/1.4960789} {\bibfield  {journal} {\bibinfo  {journal}
  {J.~Chem.~Phys.}\ }\textbf {\bibinfo {volume} {145}},\ \bibinfo {pages}
  {074309} (\bibinfo {year} {2016})}\BibitemShut {NoStop}%
\bibitem [{\citenamefont {Leforestier}, \citenamefont {Szalewicz},\ and\
  \citenamefont {van~der Avoird}(2012)}]{Leforestier2012}%
  \BibitemOpen
  \bibfield  {author} {\bibinfo {author} {\bibfnamefont {C.}~\bibnamefont
  {Leforestier}}, \bibinfo {author} {\bibfnamefont {K.}~\bibnamefont
  {Szalewicz}}, \ and\ \bibinfo {author} {\bibfnamefont {A.}~\bibnamefont
  {van~der Avoird}},\ }\href {\doibase 10.1063/1.4722338} {\bibfield  {journal}
  {\bibinfo  {journal} {J.~Chem.~Phys.}\ }\textbf {\bibinfo {volume} {137}},\
  \bibinfo {pages} {014305} (\bibinfo {year} {2012})}\BibitemShut {NoStop}%
\bibitem [{\citenamefont {Wang}\ and\ \citenamefont
  {Carrington}(2018)}]{Wang2018}%
  \BibitemOpen
  \bibfield  {author} {\bibinfo {author} {\bibfnamefont {X.-G.}\ \bibnamefont
  {Wang}}\ and\ \bibinfo {author} {\bibfnamefont {T.}~\bibnamefont
  {Carrington}},\ }\href {\doibase 10.1063/1.5020426} {\bibfield  {journal}
  {\bibinfo  {journal} {J.~Chem.~Phys.}\ }\textbf {\bibinfo {volume} {148}},\
  \bibinfo {pages} {074108} (\bibinfo {year} {2018})}\BibitemShut {NoStop}%
\bibitem [{\citenamefont {Schr{\"o}der}, \citenamefont {Gatti},\ and\
  \citenamefont {Meyer}(2011)}]{Schroeder2011malonaldehyde}%
  \BibitemOpen
  \bibfield  {author} {\bibinfo {author} {\bibfnamefont {M.}~\bibnamefont
  {Schr{\"o}der}}, \bibinfo {author} {\bibfnamefont {F.}~\bibnamefont {Gatti}},
  \ and\ \bibinfo {author} {\bibfnamefont {H.-D.}\ \bibnamefont {Meyer}},\
  }\href {\doibase 10.1063/1.3600343} {\bibfield  {journal} {\bibinfo
  {journal} {J.~Chem. Phys.}\ }\textbf {\bibinfo {volume} {134}},\ \bibinfo
  {pages} {234307} (\bibinfo {year} {2011})}\BibitemShut {NoStop}%
\bibitem [{\citenamefont {Schr{\"{o}}der}\ and\ \citenamefont
  {Meyer}(2014)}]{Schroder2014}%
  \BibitemOpen
  \bibfield  {author} {\bibinfo {author} {\bibfnamefont {M.}~\bibnamefont
  {Schr{\"{o}}der}}\ and\ \bibinfo {author} {\bibfnamefont {H.-D.}\
  \bibnamefont {Meyer}},\ }\href {\doibase 10.1063/1.4890116} {\bibfield
  {journal} {\bibinfo  {journal} {J.~Chem.~Phys.}\ }\textbf {\bibinfo {volume}
  {141}},\ \bibinfo {pages} {034116} (\bibinfo {year} {2014})}\BibitemShut
  {NoStop}%
\bibitem [{\citenamefont {Hammer}\ and\ \citenamefont
  {Manthe}(2011)}]{Hammer2011malonaldehyde}%
  \BibitemOpen
  \bibfield  {author} {\bibinfo {author} {\bibfnamefont {T.}~\bibnamefont
  {Hammer}}\ and\ \bibinfo {author} {\bibfnamefont {U.}~\bibnamefont
  {Manthe}},\ }\href {\doibase 10.1063/1.3598110} {\bibfield  {journal}
  {\bibinfo  {journal} {J.~Chem. Phys.}\ }\textbf {\bibinfo {volume} {134}},\
  \bibinfo {pages} {224305} (\bibinfo {year} {2011})}\BibitemShut {NoStop}%
\bibitem [{\citenamefont {Blume}\ and\ \citenamefont
  {Whaley}(2000)}]{Blume2000}%
  \BibitemOpen
  \bibfield  {author} {\bibinfo {author} {\bibfnamefont {D.}~\bibnamefont
  {Blume}}\ and\ \bibinfo {author} {\bibfnamefont {K.~B.}\ \bibnamefont
  {Whaley}},\ }\href {\doibase 10.1063/1.480788} {\bibfield  {journal}
  {\bibinfo  {journal} {J.~Chem.~Phys.}\ }\textbf {\bibinfo {volume} {112}},\
  \bibinfo {pages} {2218} (\bibinfo {year} {2000})}\BibitemShut {NoStop}%
\bibitem [{\citenamefont {Viel}, \citenamefont {Coutinho-Neto},\ and\
  \citenamefont {Manthe}(2007)}]{Viel2007}%
  \BibitemOpen
  \bibfield  {author} {\bibinfo {author} {\bibfnamefont {A.}~\bibnamefont
  {Viel}}, \bibinfo {author} {\bibfnamefont {M.~D.}\ \bibnamefont
  {Coutinho-Neto}}, \ and\ \bibinfo {author} {\bibfnamefont {U.}~\bibnamefont
  {Manthe}},\ }\href {\doibase 10.1063/1.2406074} {\bibfield  {journal}
  {\bibinfo  {journal} {J.~Chem.~Phys.}\ }\textbf {\bibinfo {volume} {126}},\
  \bibinfo {pages} {024308} (\bibinfo {year} {2007})}\BibitemShut {NoStop}%
\bibitem [{\citenamefont {Wang}\ \emph {et~al.}(2008)\citenamefont {Wang},
  \citenamefont {Braams}, \citenamefont {Bowman}, \citenamefont {Carter},\ and\
  \citenamefont {Tew}}]{Wang2008malonaldehydePES}%
  \BibitemOpen
  \bibfield  {author} {\bibinfo {author} {\bibfnamefont {Y.}~\bibnamefont
  {Wang}}, \bibinfo {author} {\bibfnamefont {B.~J.}\ \bibnamefont {Braams}},
  \bibinfo {author} {\bibfnamefont {J.~M.}\ \bibnamefont {Bowman}}, \bibinfo
  {author} {\bibfnamefont {S.}~\bibnamefont {Carter}}, \ and\ \bibinfo {author}
  {\bibfnamefont {D.~P.}\ \bibnamefont {Tew}},\ }\href {\doibase
  10.1063/1.2937732} {\bibfield  {journal} {\bibinfo  {journal} {J.~Chem.
  Phys.}\ }\textbf {\bibinfo {volume} {128}},\ \bibinfo {pages} {224314}
  (\bibinfo {year} {2008})}\BibitemShut {NoStop}%
\bibitem [{\citenamefont {Vaillant}, \citenamefont {Wales},\ and\ \citenamefont
  {Althorpe}(2019)}]{Vaillant2019pimd}%
  \BibitemOpen
  \bibfield  {author} {\bibinfo {author} {\bibfnamefont {C.~L.}\ \bibnamefont
  {Vaillant}}, \bibinfo {author} {\bibfnamefont {D.~J.}\ \bibnamefont {Wales}},
  \ and\ \bibinfo {author} {\bibfnamefont {S.~C.}\ \bibnamefont {Althorpe}},\
  }\href {\doibase 10.1021/acs.jpclett.9b02951} {\bibfield  {journal} {\bibinfo
   {journal} {J. Phys. Chem. Lett.}\ }\textbf {\bibinfo {volume} {10}},\
  \bibinfo {pages} {7300} (\bibinfo {year} {2019})}\BibitemShut {NoStop}%
\bibitem [{\citenamefont {Nesbitt}\ and\ \citenamefont
  {Dong}(2008)}]{Nesbitt2008}%
  \BibitemOpen
  \bibfield  {author} {\bibinfo {author} {\bibfnamefont {D.~J.}\ \bibnamefont
  {Nesbitt}}\ and\ \bibinfo {author} {\bibfnamefont {F.}~\bibnamefont {Dong}},\
  }\href {\doibase 10.1039/b800880a} {\bibfield  {journal} {\bibinfo  {journal}
  {Phys.~Chem.~Chem.~Phys.}\ }\textbf {\bibinfo {volume} {10}},\ \bibinfo
  {pages} {2113} (\bibinfo {year} {2008})}\BibitemShut {NoStop}%
\bibitem [{\citenamefont {Qu}\ and\ \citenamefont {Bowman}(2016)}]{Qu2016}%
  \BibitemOpen
  \bibfield  {author} {\bibinfo {author} {\bibfnamefont {C.}~\bibnamefont
  {Qu}}\ and\ \bibinfo {author} {\bibfnamefont {J.~M.}\ \bibnamefont
  {Bowman}},\ }\href {\doibase 10.1039/C6CP03073D} {\bibfield  {journal}
  {\bibinfo  {journal} {Phys.~Chem.~Chem.~Phys.}\ }\textbf {\bibinfo {volume}
  {18}},\ \bibinfo {pages} {24835} (\bibinfo {year} {2016})}\BibitemShut
  {NoStop}%
\bibitem [{\citenamefont {Althorpe}\ and\ \citenamefont
  {Clary}(1995)}]{Althorpe1995dimer}%
  \BibitemOpen
  \bibfield  {author} {\bibinfo {author} {\bibfnamefont {S.~C.}\ \bibnamefont
  {Althorpe}}\ and\ \bibinfo {author} {\bibfnamefont {D.~C.}\ \bibnamefont
  {Clary}},\ }\href@noop {} {\bibfield  {journal} {\bibinfo  {journal}
  {J.~Chem. Phys.}\ }\textbf {\bibinfo {volume} {102}},\ \bibinfo {pages}
  {4390} (\bibinfo {year} {1995})}\BibitemShut {NoStop}%
\bibitem [{\citenamefont {Matanovi{\'{c}}}, \citenamefont
  {Do{\v{s}}li{\'{c}}},\ and\ \citenamefont {Johnson}(2008)}]{Matanovic2008}%
  \BibitemOpen
  \bibfield  {author} {\bibinfo {author} {\bibfnamefont {I.}~\bibnamefont
  {Matanovi{\'{c}}}}, \bibinfo {author} {\bibfnamefont {N.}~\bibnamefont
  {Do{\v{s}}li{\'{c}}}}, \ and\ \bibinfo {author} {\bibfnamefont {B.~R.}\
  \bibnamefont {Johnson}},\ }\href {\doibase 10.1063/1.2833978} {\bibfield
  {journal} {\bibinfo  {journal} {J.~Chem.~Phys.}\ }\textbf {\bibinfo {volume}
  {128}},\ \bibinfo {pages} {084103} (\bibinfo {year} {2008})}\BibitemShut
  {NoStop}%
\bibitem [{\citenamefont {Kamarchik}, \citenamefont {Wang},\ and\ \citenamefont
  {Bowman}(2009)}]{Kamarchik2009reduced}%
  \BibitemOpen
  \bibfield  {author} {\bibinfo {author} {\bibfnamefont {E.}~\bibnamefont
  {Kamarchik}}, \bibinfo {author} {\bibfnamefont {Y.}~\bibnamefont {Wang}}, \
  and\ \bibinfo {author} {\bibfnamefont {J.}~\bibnamefont {Bowman}},\
  }\href@noop {} {\bibfield  {journal} {\bibinfo  {journal} {J.~Phys. Chem.~A}\
  }\textbf {\bibinfo {volume} {113}},\ \bibinfo {pages} {7556} (\bibinfo {year}
  {2009})}\BibitemShut {NoStop}%
\bibitem [{\citenamefont {Sewell}, \citenamefont {Guo},\ and\ \citenamefont
  {Thompson}(1995)}]{Sewell1995malonaldehyde}%
  \BibitemOpen
  \bibfield  {author} {\bibinfo {author} {\bibfnamefont {T.~D.}\ \bibnamefont
  {Sewell}}, \bibinfo {author} {\bibfnamefont {Y.}~\bibnamefont {Guo}}, \ and\
  \bibinfo {author} {\bibfnamefont {D.~L.}\ \bibnamefont {Thompson}},\ }\href
  {\doibase 10.1063/1.470166} {\bibfield  {journal} {\bibinfo  {journal}
  {J.~Chem. Phys.}\ }\textbf {\bibinfo {volume} {103}},\ \bibinfo {pages}
  {8557} (\bibinfo {year} {1995})}\BibitemShut {NoStop}%
\bibitem [{\citenamefont {Tautermann}\ \emph {et~al.}(2002)\citenamefont
  {Tautermann}, \citenamefont {Voegele}, \citenamefont {Loerting},\ and\
  \citenamefont {Liedl}}]{Tautermann2002semiclassical}%
  \BibitemOpen
  \bibfield  {author} {\bibinfo {author} {\bibfnamefont {C.~S.}\ \bibnamefont
  {Tautermann}}, \bibinfo {author} {\bibfnamefont {A.~F.}\ \bibnamefont
  {Voegele}}, \bibinfo {author} {\bibfnamefont {T.}~\bibnamefont {Loerting}}, \
  and\ \bibinfo {author} {\bibfnamefont {K.~R.}\ \bibnamefont {Liedl}},\
  }\href@noop {} {\bibfield  {journal} {\bibinfo  {journal} {J.~Chem. Phys.}\
  }\textbf {\bibinfo {volume} {117}},\ \bibinfo {pages} {1967} (\bibinfo {year}
  {2002})}\BibitemShut {NoStop}%
\bibitem [{\citenamefont {Ceotto}(2012)}]{Ceotto2012instanton}%
  \BibitemOpen
  \bibfield  {author} {\bibinfo {author} {\bibfnamefont {M.}~\bibnamefont
  {Ceotto}},\ }\href {\doibase 10.1080/00268976.2012.663943} {\bibfield
  {journal} {\bibinfo  {journal} {Mol. Phys.}\ }\textbf {\bibinfo {volume}
  {110}},\ \bibinfo {pages} {547} (\bibinfo {year} {2012})}\BibitemShut
  {NoStop}%
\bibitem [{\citenamefont {Burd}\ and\ \citenamefont {Clary}(2020)}]{Burd2020}%
  \BibitemOpen
  \bibfield  {author} {\bibinfo {author} {\bibfnamefont {T.~A.~H.}\
  \bibnamefont {Burd}}\ and\ \bibinfo {author} {\bibfnamefont {D.~C.}\
  \bibnamefont {Clary}},\ }\href {\doibase 10.1021/acs.jctc.0c00207} {\bibfield
   {journal} {\bibinfo  {journal} {J.~Chem. Theory Comput.}\ }\textbf {\bibinfo
  {volume} {16}},\ \bibinfo {pages} {3486} (\bibinfo {year}
  {2020})}\BibitemShut {NoStop}%
\bibitem [{\citenamefont {Makri}\ and\ \citenamefont
  {Miller}(1989)}]{Makri1989}%
  \BibitemOpen
  \bibfield  {author} {\bibinfo {author} {\bibfnamefont {N.}~\bibnamefont
  {Makri}}\ and\ \bibinfo {author} {\bibfnamefont {W.~H.}\ \bibnamefont
  {Miller}},\ }\href {\doibase 10.1063/1.456833} {\bibfield  {journal}
  {\bibinfo  {journal} {J.~Chem.~Phys.}\ }\textbf {\bibinfo {volume} {91}},\
  \bibinfo {pages} {4026} (\bibinfo {year} {1989})}\BibitemShut {NoStop}%
\bibitem [{\citenamefont {Coleman}(1977)}]{Uses_of_Instantons}%
  \BibitemOpen
  \bibfield  {author} {\bibinfo {author} {\bibfnamefont {S.}~\bibnamefont
  {Coleman}},\ }in\ \href@noop {} {\emph {\bibinfo {booktitle} {Proc. Int.
  School of Subnuclear Physics}}}\ (\bibinfo {organization} {Erice},\ \bibinfo
  {year} {1977})\ \bibinfo {note} {also in S. Coleman, \emph{Aspects of
  Symmetry}, chapter 7, pp. 265--350 (Cambridge University Press,
  1985)}\BibitemShut {NoStop}%
\bibitem [{\citenamefont {Vainshtein}\ \emph {et~al.}(1982)\citenamefont
  {Vainshtein}, \citenamefont {Zakharov}, \citenamefont {Novikov},\ and\
  \citenamefont {Shifman}}]{ABCofInstantons}%
  \BibitemOpen
  \bibfield  {author} {\bibinfo {author} {\bibfnamefont {A.~I.}\ \bibnamefont
  {Vainshtein}}, \bibinfo {author} {\bibfnamefont {V.~I.}\ \bibnamefont
  {Zakharov}}, \bibinfo {author} {\bibfnamefont {V.~A.}\ \bibnamefont
  {Novikov}}, \ and\ \bibinfo {author} {\bibfnamefont {M.~A.}\ \bibnamefont
  {Shifman}},\ }\href@noop {} {\bibfield  {journal} {\bibinfo  {journal} {Sov.
  Phys. Uspekhi}\ }\textbf {\bibinfo {volume} {25}},\ \bibinfo {pages} {195}
  (\bibinfo {year} {1982})},\ \bibinfo {note} {also in \emph{Instantons in
  Gauge Theories}, edited by M. Shifman, pp. 468 (Singapore: World Scientific,
  1994)}\BibitemShut {NoStop}%
\bibitem [{\citenamefont {Miller}(1975)}]{Miller1975semiclassical}%
  \BibitemOpen
  \bibfield  {author} {\bibinfo {author} {\bibfnamefont {W.~H.}\ \bibnamefont
  {Miller}},\ }\href {\doibase 10.1063/1.430676} {\bibfield  {journal}
  {\bibinfo  {journal} {J.~Chem. Phys.}\ }\textbf {\bibinfo {volume} {62}},\
  \bibinfo {pages} {1899} (\bibinfo {year} {1975})}\BibitemShut {NoStop}%
\bibitem [{\citenamefont {Benderskii}, \citenamefont {Makarov},\ and\
  \citenamefont {Wight}(1994)}]{Benderskii}%
  \BibitemOpen
  \bibfield  {author} {\bibinfo {author} {\bibfnamefont {V.~A.}\ \bibnamefont
  {Benderskii}}, \bibinfo {author} {\bibfnamefont {D.~E.}\ \bibnamefont
  {Makarov}}, \ and\ \bibinfo {author} {\bibfnamefont {C.~A.}\ \bibnamefont
  {Wight}},\ }\href {\doibase 10.1002/9780470141472} {\emph {\bibinfo {title}
  {Chemical Dynamics at Low Temperatures}}},\ \bibinfo {series} {Adv. Chem.
  Phys.}, Vol.~\bibinfo {volume} {88}\ (\bibinfo  {publisher} {Wiley},\
  \bibinfo {address} {New York},\ \bibinfo {year} {1994})\BibitemShut {NoStop}%
\bibitem [{\citenamefont {Siebrand}\ \emph {et~al.}(1999)\citenamefont
  {Siebrand}, \citenamefont {Smedarchina}, \citenamefont {Zgierski},\ and\
  \citenamefont {Fern{\'a}ndez-Ramos}}]{Siebrand1999AIM}%
  \BibitemOpen
  \bibfield  {author} {\bibinfo {author} {\bibfnamefont {W.}~\bibnamefont
  {Siebrand}}, \bibinfo {author} {\bibfnamefont {Z.}~\bibnamefont
  {Smedarchina}}, \bibinfo {author} {\bibfnamefont {M.~Z.}\ \bibnamefont
  {Zgierski}}, \ and\ \bibinfo {author} {\bibfnamefont {A.}~\bibnamefont
  {Fern{\'a}ndez-Ramos}},\ }\href {\doibase 10.1080/014423599229992} {\bibfield
   {journal} {\bibinfo  {journal} {Int. Rev. Phys. Chem.}\ }\textbf {\bibinfo
  {volume} {18}},\ \bibinfo {pages} {5} (\bibinfo {year} {1999})}\BibitemShut
  {NoStop}%
\bibitem [{\citenamefont {Smedarchina}, \citenamefont {Caminati},\ and\
  \citenamefont {Zerbetto}(1995)}]{Smedarchina1995AIM}%
  \BibitemOpen
  \bibfield  {author} {\bibinfo {author} {\bibfnamefont {Z.}~\bibnamefont
  {Smedarchina}}, \bibinfo {author} {\bibfnamefont {W.}~\bibnamefont
  {Caminati}}, \ and\ \bibinfo {author} {\bibfnamefont {F.}~\bibnamefont
  {Zerbetto}},\ }\href@noop {} {\bibfield  {journal} {\bibinfo  {journal}
  {Chemical physics letters}\ }\textbf {\bibinfo {volume} {237}},\ \bibinfo
  {pages} {279} (\bibinfo {year} {1995})}\BibitemShut {NoStop}%
\bibitem [{\citenamefont {Smedarchina}, \citenamefont {Siebrand},\ and\
  \citenamefont {Fern{\'a}ndez-Ramos}(2012)}]{Smedarchina2012rainbow}%
  \BibitemOpen
  \bibfield  {author} {\bibinfo {author} {\bibfnamefont {Z.}~\bibnamefont
  {Smedarchina}}, \bibinfo {author} {\bibfnamefont {W.}~\bibnamefont
  {Siebrand}}, \ and\ \bibinfo {author} {\bibfnamefont {A.}~\bibnamefont
  {Fern{\'a}ndez-Ramos}},\ }\href {\doibase 10.1063/1.4769198} {\bibfield
  {journal} {\bibinfo  {journal} {J.~Chem. Phys.}\ }\textbf {\bibinfo {volume}
  {137}},\ \bibinfo {pages} {224105} (\bibinfo {year} {2012})}\BibitemShut
  {NoStop}%
\bibitem [{\citenamefont {Benderskii}\ \emph
  {et~al.}(1997{\natexlab{a}})\citenamefont {Benderskii}, \citenamefont
  {Vetoshkin}, \citenamefont {Von~Laue},\ and\ \citenamefont
  {Trommsdorff}}]{Benderskii1997excited}%
  \BibitemOpen
  \bibfield  {author} {\bibinfo {author} {\bibfnamefont {V.~A.}\ \bibnamefont
  {Benderskii}}, \bibinfo {author} {\bibfnamefont {E.~V.}\ \bibnamefont
  {Vetoshkin}}, \bibinfo {author} {\bibfnamefont {L.}~\bibnamefont {Von~Laue}},
  \ and\ \bibinfo {author} {\bibfnamefont {H.~P.}\ \bibnamefont
  {Trommsdorff}},\ }\href@noop {} {\bibfield  {journal} {\bibinfo  {journal}
  {Chem. Phys.}\ }\textbf {\bibinfo {volume} {219}},\ \bibinfo {pages} {143}
  (\bibinfo {year} {1997}{\natexlab{a}})}\BibitemShut {NoStop}%
\bibitem [{\citenamefont {Mil'nikov}\ and\ \citenamefont
  {Nakamura}(2001)}]{Milnikov2001}%
  \BibitemOpen
  \bibfield  {author} {\bibinfo {author} {\bibfnamefont {G.~V.}\ \bibnamefont
  {Mil'nikov}}\ and\ \bibinfo {author} {\bibfnamefont {H.}~\bibnamefont
  {Nakamura}},\ }\href {\doibase 10.1063/1.1406532} {\bibfield  {journal}
  {\bibinfo  {journal} {J.~Chem. Phys.}\ }\textbf {\bibinfo {volume} {115}},\
  \bibinfo {pages} {6881} (\bibinfo {year} {2001})}\BibitemShut {NoStop}%
\bibitem [{\citenamefont {Mil'nikov}\ and\ \citenamefont
  {Nakamura}(2005)}]{Milnikov2005}%
  \BibitemOpen
  \bibfield  {author} {\bibinfo {author} {\bibfnamefont {G.~V.}\ \bibnamefont
  {Mil'nikov}}\ and\ \bibinfo {author} {\bibfnamefont {H.}~\bibnamefont
  {Nakamura}},\ }\href@noop {} {\bibfield  {journal} {\bibinfo  {journal}
  {J.~Chem. Phys.}\ }\textbf {\bibinfo {volume} {122}},\ \bibinfo {pages}
  {124311} (\bibinfo {year} {2005})}\BibitemShut {NoStop}%
\bibitem [{\citenamefont {Richardson}\ and\ \citenamefont
  {Althorpe}(2011)}]{tunnel}%
  \BibitemOpen
  \bibfield  {author} {\bibinfo {author} {\bibfnamefont {J.~O.}\ \bibnamefont
  {Richardson}}\ and\ \bibinfo {author} {\bibfnamefont {S.~C.}\ \bibnamefont
  {Althorpe}},\ }\href {\doibase 10.1063/1.3530589} {\bibfield  {journal}
  {\bibinfo  {journal} {J.~Chem. Phys.}\ }\textbf {\bibinfo {volume} {134}},\
  \bibinfo {pages} {054109} (\bibinfo {year} {2011})}\BibitemShut {NoStop}%
\bibitem [{\citenamefont {Vaillant}\ and\ \citenamefont
  {Cvita\v{s}}(2018)}]{Vaillant2018rotation}%
  \BibitemOpen
  \bibfield  {author} {\bibinfo {author} {\bibfnamefont {C.}~\bibnamefont
  {Vaillant}}\ and\ \bibinfo {author} {\bibfnamefont {M.~T.}\ \bibnamefont
  {Cvita\v{s}}},\ }\href {\doibase 10.1039/C8CP04991B (} {\bibfield  {journal}
  {\bibinfo  {journal} {Phys.~Chem.~Chem.~Phys.}\ }\textbf {\bibinfo {volume}
  {20}},\ \bibinfo {pages} {26809} (\bibinfo {year} {2018})}\BibitemShut
  {NoStop}%
\bibitem [{\citenamefont {Richardson}\ \emph {et~al.}(2013)\citenamefont
  {Richardson}, \citenamefont {Wales}, \citenamefont {Althorpe}, \citenamefont
  {McLaughlin}, \citenamefont {Viant}, \citenamefont {Shih},\ and\
  \citenamefont {Saykally}}]{octamer}%
  \BibitemOpen
  \bibfield  {author} {\bibinfo {author} {\bibfnamefont {J.~O.}\ \bibnamefont
  {Richardson}}, \bibinfo {author} {\bibfnamefont {D.~J.}\ \bibnamefont
  {Wales}}, \bibinfo {author} {\bibfnamefont {S.~C.}\ \bibnamefont {Althorpe}},
  \bibinfo {author} {\bibfnamefont {R.~P.}\ \bibnamefont {McLaughlin}},
  \bibinfo {author} {\bibfnamefont {M.~R.}\ \bibnamefont {Viant}}, \bibinfo
  {author} {\bibfnamefont {O.}~\bibnamefont {Shih}}, \ and\ \bibinfo {author}
  {\bibfnamefont {R.~J.}\ \bibnamefont {Saykally}},\ }\href {\doibase
  10.1021/jp311306a} {\bibfield  {journal} {\bibinfo  {journal} {J.~Phys.
  Chem.~A}\ }\textbf {\bibinfo {volume} {117}},\ \bibinfo {pages} {6960 }
  (\bibinfo {year} {2013})}\BibitemShut {NoStop}%
\bibitem [{\citenamefont {Garg}(2000)}]{Garg2000}%
  \BibitemOpen
  \bibfield  {author} {\bibinfo {author} {\bibfnamefont {A.}~\bibnamefont
  {Garg}},\ }\href {\doibase 10.1119/1.19458} {\bibfield  {journal} {\bibinfo
  {journal} {Am. J. Phys.}\ }\textbf {\bibinfo {volume} {68}},\ \bibinfo
  {pages} {430} (\bibinfo {year} {2000})}\BibitemShut {NoStop}%
\bibitem [{\citenamefont {Herring}(1962)}]{Herring1962}%
  \BibitemOpen
  \bibfield  {author} {\bibinfo {author} {\bibfnamefont {C.}~\bibnamefont
  {Herring}},\ }\href {\doibase 10.1103/RevModPhys.34.631} {\bibfield
  {journal} {\bibinfo  {journal} {Rev.~Mod.~Phys.}\ }\textbf {\bibinfo {volume}
  {34}},\ \bibinfo {pages} {631} (\bibinfo {year} {1962})}\BibitemShut
  {NoStop}%
\bibitem [{\citenamefont {Landau}\ and\ \citenamefont
  {Lifshitz}(1965)}]{Landau+Lifshitz}%
  \BibitemOpen
  \bibfield  {author} {\bibinfo {author} {\bibfnamefont {L.~D.}\ \bibnamefont
  {Landau}}\ and\ \bibinfo {author} {\bibfnamefont {E.~M.}\ \bibnamefont
  {Lifshitz}},\ }\href@noop {} {\emph {\bibinfo {title} {Quantum Mechanics:
  Non-Relativistic Theory}}},\ \bibinfo {edition} {2nd}\ ed.\ (\bibinfo
  {publisher} {Pergamon Press},\ \bibinfo {address} {Oxford},\ \bibinfo {year}
  {1965})\BibitemShut {NoStop}%
\bibitem [{\citenamefont {Benderskii}\ \emph
  {et~al.}(1997{\natexlab{b}})\citenamefont {Benderskii}, \citenamefont
  {Vetoshkin}, \citenamefont {von Laue},\ and\ \citenamefont
  {Trommsdorff}}]{Benderskii1997two}%
  \BibitemOpen
  \bibfield  {author} {\bibinfo {author} {\bibfnamefont {V.}~\bibnamefont
  {Benderskii}}, \bibinfo {author} {\bibfnamefont {E.}~\bibnamefont
  {Vetoshkin}}, \bibinfo {author} {\bibfnamefont {L.}~\bibnamefont {von Laue}},
  \ and\ \bibinfo {author} {\bibfnamefont {H.}~\bibnamefont {Trommsdorff}},\
  }\href {\doibase 10.1016/S0301-0104(97)00119-5} {\bibfield  {journal}
  {\bibinfo  {journal} {Chem.~Phys.}\ }\textbf {\bibinfo {volume} {219}},\
  \bibinfo {pages} {143} (\bibinfo {year} {1997}{\natexlab{b}})}\BibitemShut
  {NoStop}%
\bibitem [{\citenamefont {Benderskii}\ \emph {et~al.}(2000)\citenamefont
  {Benderskii}, \citenamefont {Vetoshkin}, \citenamefont {Irgibaeva},\ and\
  \citenamefont {Trommsdorff}}]{Benderskii2000}%
  \BibitemOpen
  \bibfield  {author} {\bibinfo {author} {\bibfnamefont {V.}~\bibnamefont
  {Benderskii}}, \bibinfo {author} {\bibfnamefont {E.}~\bibnamefont
  {Vetoshkin}}, \bibinfo {author} {\bibfnamefont {I.}~\bibnamefont
  {Irgibaeva}}, \ and\ \bibinfo {author} {\bibfnamefont {H.}~\bibnamefont
  {Trommsdorff}},\ }\href {\doibase 10.1016/S0301-0104(00)00319-0} {\bibfield
  {journal} {\bibinfo  {journal} {Chem.~Phys.}\ }\textbf {\bibinfo {volume}
  {262}},\ \bibinfo {pages} {393} (\bibinfo {year} {2000})}\BibitemShut
  {NoStop}%
\bibitem [{\citenamefont {Smedarchina}, \citenamefont {Siebrand},\ and\
  \citenamefont {Zgierski}(1996)}]{Smedarchina1996}%
  \BibitemOpen
  \bibfield  {author} {\bibinfo {author} {\bibfnamefont {Z.}~\bibnamefont
  {Smedarchina}}, \bibinfo {author} {\bibfnamefont {W.}~\bibnamefont
  {Siebrand}}, \ and\ \bibinfo {author} {\bibfnamefont {M.~Z.}\ \bibnamefont
  {Zgierski}},\ }\href {\doibase 10.1063/1.470780} {\bibfield  {journal}
  {\bibinfo  {journal} {J.~Chem. Phys.}\ }\textbf {\bibinfo {volume} {104}},\
  \bibinfo {pages} {1203} (\bibinfo {year} {1996})}\BibitemShut {NoStop}%
\bibitem [{\citenamefont {Fern{\'{a}}ndez-Ramos}\ \emph
  {et~al.}(1998)\citenamefont {Fern{\'{a}}ndez-Ramos}, \citenamefont
  {Smedarchina}, \citenamefont {Zgierski},\ and\ \citenamefont
  {Siebrand}}]{Fernandez-Ramos1998}%
  \BibitemOpen
  \bibfield  {author} {\bibinfo {author} {\bibfnamefont {A.}~\bibnamefont
  {Fern{\'{a}}ndez-Ramos}}, \bibinfo {author} {\bibfnamefont {Z.}~\bibnamefont
  {Smedarchina}}, \bibinfo {author} {\bibfnamefont {M.~Z.}\ \bibnamefont
  {Zgierski}}, \ and\ \bibinfo {author} {\bibfnamefont {W.}~\bibnamefont
  {Siebrand}},\ }\href {\doibase 10.1063/1.476643} {\bibfield  {journal}
  {\bibinfo  {journal} {J.~Chem. Phys.}\ }\textbf {\bibinfo {volume} {109}},\
  \bibinfo {pages} {1004} (\bibinfo {year} {1998})}\BibitemShut {NoStop}%
\bibitem [{\citenamefont {Mil'nikov}, \citenamefont {K{\"u}hn},\ and\
  \citenamefont {Nakamura}(2005)}]{Milnikov2005formic}%
  \BibitemOpen
  \bibfield  {author} {\bibinfo {author} {\bibfnamefont {G.}~\bibnamefont
  {Mil'nikov}}, \bibinfo {author} {\bibfnamefont {O.}~\bibnamefont {K{\"u}hn}},
  \ and\ \bibinfo {author} {\bibfnamefont {H.}~\bibnamefont {Nakamura}},\
  }\href@noop {} {\bibfield  {journal} {\bibinfo  {journal} {J.~Chem. Phys.}\
  }\textbf {\bibinfo {volume} {123}},\ \bibinfo {pages} {074308} (\bibinfo
  {year} {2005})}\BibitemShut {NoStop}%
\bibitem [{\citenamefont {Mil'nikov}, \citenamefont {Ishida},\ and\
  \citenamefont {Nakamura}(2006)}]{Milnikov2006}%
  \BibitemOpen
  \bibfield  {author} {\bibinfo {author} {\bibfnamefont {G.~V.}\ \bibnamefont
  {Mil'nikov}}, \bibinfo {author} {\bibfnamefont {T.}~\bibnamefont {Ishida}}, \
  and\ \bibinfo {author} {\bibfnamefont {H.}~\bibnamefont {Nakamura}},\ }\href
  {\doibase 10.1021/jp055667s} {\bibfield  {journal} {\bibinfo  {journal}
  {J.~Phys.~Chem.~A}\ }\textbf {\bibinfo {volume} {110}},\ \bibinfo {pages}
  {5430} (\bibinfo {year} {2006})}\BibitemShut {NoStop}%
\bibitem [{\citenamefont {Erakovi{\'{c}}}, \citenamefont {Vaillant},\ and\
  \citenamefont {Cvita{\v{s}}}(2020)}]{Erakovic2020}%
  \BibitemOpen
  \bibfield  {author} {\bibinfo {author} {\bibfnamefont {M.}~\bibnamefont
  {Erakovi{\'{c}}}}, \bibinfo {author} {\bibfnamefont {C.~L.}\ \bibnamefont
  {Vaillant}}, \ and\ \bibinfo {author} {\bibfnamefont {M.~T.}\ \bibnamefont
  {Cvita{\v{s}}}},\ }\href {\doibase 10.1063/1.5145278} {\bibfield  {journal}
  {\bibinfo  {journal} {J.~Chem.~Phys.}\ }\textbf {\bibinfo {volume} {152}},\
  \bibinfo {pages} {084111} (\bibinfo {year} {2020})}\BibitemShut {NoStop}%
\bibitem [{\citenamefont {Cvita\v{s}}\ and\ \citenamefont
  {Althorpe}(2016)}]{Cvitas2016instanton}%
  \BibitemOpen
  \bibfield  {author} {\bibinfo {author} {\bibfnamefont {M.~T.}\ \bibnamefont
  {Cvita\v{s}}}\ and\ \bibinfo {author} {\bibfnamefont {S.~C.}\ \bibnamefont
  {Althorpe}},\ }\href {\doibase 10.1021/acs.jctc.5b01073} {\bibfield
  {journal} {\bibinfo  {journal} {J.~Chem. Theory Comput.}\ }\textbf {\bibinfo
  {volume} {12}},\ \bibinfo {pages} {787} (\bibinfo {year} {2016})}\BibitemShut
  {NoStop}%
\bibitem [{\citenamefont {Cvita\v{s}}(2018)}]{Cvitas2018instanton}%
  \BibitemOpen
  \bibfield  {author} {\bibinfo {author} {\bibfnamefont {M.~T.}\ \bibnamefont
  {Cvita\v{s}}},\ }\href {\doibase 10.1021/acs.jctc.7b00881} {\bibfield
  {journal} {\bibinfo  {journal} {J. Chem. Theory Comput.}\ }\textbf {\bibinfo
  {volume} {14}},\ \bibinfo {pages} {1487} (\bibinfo {year}
  {2018})}\BibitemShut {NoStop}%
\bibitem [{\citenamefont {Kleinert}(2009)}]{Kleinert}%
  \BibitemOpen
  \bibfield  {author} {\bibinfo {author} {\bibfnamefont {H.}~\bibnamefont
  {Kleinert}},\ }\href@noop {} {\emph {\bibinfo {title} {Path Integrals in
  Quantum Mechanics, Statistics, Polymer Physics and Financial Markets}}},\
  \bibinfo {edition} {5th}\ ed.\ (\bibinfo  {publisher} {World Scientific},\
  \bibinfo {address} {Singapore},\ \bibinfo {year} {2009})\BibitemShut
  {NoStop}%
\bibitem [{\citenamefont {Benderskii}, \citenamefont {Grebenshchikov},\ and\
  \citenamefont {Mil'nikov}(1995)}]{Benderskii1995excited}%
  \BibitemOpen
  \bibfield  {author} {\bibinfo {author} {\bibfnamefont {V.~A.}\ \bibnamefont
  {Benderskii}}, \bibinfo {author} {\bibfnamefont {S.~Y.}\ \bibnamefont
  {Grebenshchikov}}, \ and\ \bibinfo {author} {\bibfnamefont {G.~V.}\
  \bibnamefont {Mil'nikov}},\ }\href@noop {} {\bibfield  {journal} {\bibinfo
  {journal} {Chem. Phys.}\ }\textbf {\bibinfo {volume} {194}},\ \bibinfo
  {pages} {1} (\bibinfo {year} {1995})}\BibitemShut {NoStop}%
\bibitem [{\citenamefont {Press}\ \emph {et~al.}(2007)\citenamefont {Press},
  \citenamefont {Teukolsky}, \citenamefont {Vetterling},\ and\ \citenamefont
  {Flannery}}]{NumRep}%
  \BibitemOpen
  \bibfield  {author} {\bibinfo {author} {\bibfnamefont {W.~H.}\ \bibnamefont
  {Press}}, \bibinfo {author} {\bibfnamefont {S.~A.}\ \bibnamefont
  {Teukolsky}}, \bibinfo {author} {\bibfnamefont {W.~T.}\ \bibnamefont
  {Vetterling}}, \ and\ \bibinfo {author} {\bibfnamefont {B.~P.}\ \bibnamefont
  {Flannery}},\ }\href@noop {} {\emph {\bibinfo {title} {Numerical Recipes:
  {The} Art of Scientific Computing}}},\ \bibinfo {edition} {3rd}\ ed.\
  (\bibinfo  {publisher} {Cambridge University Press},\ \bibinfo {address}
  {Cambridge},\ \bibinfo {year} {2007})\BibitemShut {NoStop}%
\bibitem [{\citenamefont {Kawatsu}\ and\ \citenamefont
  {Miura}(2014)}]{Kawatsu2014RPI}%
  \BibitemOpen
  \bibfield  {author} {\bibinfo {author} {\bibfnamefont {T.}~\bibnamefont
  {Kawatsu}}\ and\ \bibinfo {author} {\bibfnamefont {S.}~\bibnamefont
  {Miura}},\ }\href {\doibase 10.1063/1.4885437} {\bibfield  {journal}
  {\bibinfo  {journal} {J.~Chem. Phys.}\ }\textbf {\bibinfo {volume} {141}},\
  \bibinfo {pages} {024101} (\bibinfo {year} {2014})}\BibitemShut {NoStop}%
\bibitem [{\citenamefont {Light}, \citenamefont {Hamilton},\ and\ \citenamefont
  {Lill}(1985)}]{Light1985DVR}%
  \BibitemOpen
  \bibfield  {author} {\bibinfo {author} {\bibfnamefont {J.~C.}\ \bibnamefont
  {Light}}, \bibinfo {author} {\bibfnamefont {I.~P.}\ \bibnamefont {Hamilton}},
  \ and\ \bibinfo {author} {\bibfnamefont {J.~V.}\ \bibnamefont {Lill}},\
  }\href@noop {} {\bibfield  {journal} {\bibinfo  {journal} {J.~Chem. Phys.}\
  }\textbf {\bibinfo {volume} {82}},\ \bibinfo {pages} {1400} (\bibinfo {year}
  {1985})}\BibitemShut {NoStop}%
\bibitem [{\citenamefont {Babin}, \citenamefont {Leforestier},\ and\
  \citenamefont {Paesani}(2013)}]{Babin2013MBpol}%
  \BibitemOpen
  \bibfield  {author} {\bibinfo {author} {\bibfnamefont {V.}~\bibnamefont
  {Babin}}, \bibinfo {author} {\bibfnamefont {C.}~\bibnamefont {Leforestier}},
  \ and\ \bibinfo {author} {\bibfnamefont {F.}~\bibnamefont {Paesani}},\ }\href
  {\doibase 10.1021/ct400863t} {\bibfield  {journal} {\bibinfo  {journal}
  {J.~Chem. Theory Comput.}\ }\textbf {\bibinfo {volume} {9}},\ \bibinfo
  {pages} {5395} (\bibinfo {year} {2013})}\BibitemShut {NoStop}%
\bibitem [{\citenamefont {Babin}, \citenamefont {Medders},\ and\ \citenamefont
  {Paesani}(2014)}]{Babin2014MBpol}%
  \BibitemOpen
  \bibfield  {author} {\bibinfo {author} {\bibfnamefont {V.}~\bibnamefont
  {Babin}}, \bibinfo {author} {\bibfnamefont {G.~R.}\ \bibnamefont {Medders}},
  \ and\ \bibinfo {author} {\bibfnamefont {F.}~\bibnamefont {Paesani}},\ }\href
  {\doibase 10.1021/ct500079y} {\bibfield  {journal} {\bibinfo  {journal}
  {J.~Chem. Theory Comput.}\ }\textbf {\bibinfo {volume} {10}},\ \bibinfo
  {pages} {1599} (\bibinfo {year} {2014})}\BibitemShut {NoStop}%
\bibitem [{\citenamefont {Watanabe}, \citenamefont {Taketsugu},\ and\
  \citenamefont {Wales}(2004)}]{Watanabe2004dimer}%
  \BibitemOpen
  \bibfield  {author} {\bibinfo {author} {\bibfnamefont {Y.}~\bibnamefont
  {Watanabe}}, \bibinfo {author} {\bibfnamefont {T.}~\bibnamefont {Taketsugu}},
  \ and\ \bibinfo {author} {\bibfnamefont {D.~J.}\ \bibnamefont {Wales}},\
  }\href@noop {} {\bibfield  {journal} {\bibinfo  {journal} {J.~Chem. Phys.}\
  }\textbf {\bibinfo {volume} {120}},\ \bibinfo {pages} {5993} (\bibinfo {year}
  {2004})}\BibitemShut {NoStop}%
\bibitem [{\citenamefont {Braly}\ \emph {et~al.}(2000)\citenamefont {Braly},
  \citenamefont {Cruzan}, \citenamefont {Liu}, \citenamefont {Fellers},\ and\
  \citenamefont {Saykally}}]{Braly2000}%
  \BibitemOpen
  \bibfield  {author} {\bibinfo {author} {\bibfnamefont {L.~B.}\ \bibnamefont
  {Braly}}, \bibinfo {author} {\bibfnamefont {J.~D.}\ \bibnamefont {Cruzan}},
  \bibinfo {author} {\bibfnamefont {K.}~\bibnamefont {Liu}}, \bibinfo {author}
  {\bibfnamefont {R.~S.}\ \bibnamefont {Fellers}}, \ and\ \bibinfo {author}
  {\bibfnamefont {R.~J.}\ \bibnamefont {Saykally}},\ }\href {\doibase
  10.1063/1.481708} {\bibfield  {journal} {\bibinfo  {journal} {J.~Chem.
  Phys.}\ }\textbf {\bibinfo {volume} {112}},\ \bibinfo {pages} {10293}
  (\bibinfo {year} {2000})}\BibitemShut {NoStop}%
\bibitem [{\citenamefont {Karyakin}, \citenamefont {Fraser},\ and\
  \citenamefont {Suenram}(1993)}]{Karyakin1993Ddimer}%
  \BibitemOpen
  \bibfield  {author} {\bibinfo {author} {\bibfnamefont {E.~N.}\ \bibnamefont
  {Karyakin}}, \bibinfo {author} {\bibfnamefont {G.~T.}\ \bibnamefont
  {Fraser}}, \ and\ \bibinfo {author} {\bibfnamefont {R.~D.}\ \bibnamefont
  {Suenram}},\ }\href@noop {} {\bibfield  {journal} {\bibinfo  {journal} {Mol.
  Phys.}\ }\textbf {\bibinfo {volume} {78}},\ \bibinfo {pages} {1179} (\bibinfo
  {year} {1993})}\BibitemShut {NoStop}%
\bibitem [{\citenamefont {Nguyen}\ \emph {et~al.}(2018)\citenamefont {Nguyen},
  \citenamefont {Sz{\'e}kely}, \citenamefont {Imbalzano}, \citenamefont
  {Behler}, \citenamefont {Cs{\'a}nyi}, \citenamefont {Ceriotti}, \citenamefont
  {G{\"o}tz},\ and\ \citenamefont {Paesani}}]{Nguyen2018water}%
  \BibitemOpen
  \bibfield  {author} {\bibinfo {author} {\bibfnamefont {T.~T.}\ \bibnamefont
  {Nguyen}}, \bibinfo {author} {\bibfnamefont {E.}~\bibnamefont {Sz{\'e}kely}},
  \bibinfo {author} {\bibfnamefont {G.}~\bibnamefont {Imbalzano}}, \bibinfo
  {author} {\bibfnamefont {J.}~\bibnamefont {Behler}}, \bibinfo {author}
  {\bibfnamefont {G.}~\bibnamefont {Cs{\'a}nyi}}, \bibinfo {author}
  {\bibfnamefont {M.}~\bibnamefont {Ceriotti}}, \bibinfo {author}
  {\bibfnamefont {A.~W.}\ \bibnamefont {G{\"o}tz}}, \ and\ \bibinfo {author}
  {\bibfnamefont {F.}~\bibnamefont {Paesani}},\ }\href {\doibase
  10.1063/1.5024577} {\bibfield  {journal} {\bibinfo  {journal} {J.~Chem.
  Phys.}\ }\textbf {\bibinfo {volume} {148}},\ \bibinfo {pages} {241725}
  (\bibinfo {year} {2018})}\BibitemShut {NoStop}%
\bibitem [{\citenamefont {Courant}\ and\ \citenamefont
  {Hilbert}(1962)}]{Hilbert}%
  \BibitemOpen
  \bibfield  {author} {\bibinfo {author} {\bibfnamefont {R.}~\bibnamefont
  {Courant}}\ and\ \bibinfo {author} {\bibfnamefont {D.}~\bibnamefont
  {Hilbert}},\ }\href@noop {} {\emph {\bibinfo {title} {Methods of Mathematical
  Physics, Volume II}}}\ (\bibinfo  {publisher} {Wiley-Interscience},\ \bibinfo
  {year} {1962})\BibitemShut {NoStop}%
\end{thebibliography}%

\end{document}